\documentclass[ aps, showpacs, showkeys, nofootinbib, floatfix, superscriptaddress]{revtex4}

\usepackage{amsfonts}

\usepackage{amssymb}
\usepackage{amsmath}
\usepackage{graphicx}
\usepackage{epstopdf}
\usepackage{multirow}
\usepackage{subfig}
\usepackage{pdfpages}
\usepackage[utf8]{inputenc}

\begin{document}
\title{Observational constraints on the Kerr and its several single-parameter modified spacetimes using quasi-periodic oscillation data}
 \author{Shining Yang}
 \affiliation{Department of Physics, Liaoning Normal University, Dalian 116029, P. R. China}
 \author{Jianbo Lu}
 \email{lvjianbo819@163.com}
 \affiliation{Department of Physics, Liaoning Normal University, Dalian 116029, P. R. China}
 \author{Wenmei Li}
 \affiliation{Department of Physics, Liaoning Normal University, Dalian 116029, P. R. China}
 \author{Mou Xu}
 \affiliation{Department of Physics, Liaoning Normal University, Dalian 116029, P. R. China}
\author{Jingyang Xu}
 \affiliation{Department of Physics, Liaoning Normal University, Dalian 116029, P. R. China}

\begin{abstract}
This paper investigates the dynamical effects of particles moving in the Kerr spacetime and its nine single-parameter modified spacetimes, including Bardeen, Ayon-Beato and Garcia (ABG), Hayward, Kerr-Newman (KN), Kerr-Taub-NUT (KTN), Braneworld Kerr (BK), Kerr-MOG, Kerr-Sen, and Perfect Fluid Dark Matter (PFDM) black holes. Using quasi-periodic oscillation (QPO) observational data, we constrain the free parameters of the ten spacetimes through $\chi^2$ analysis under the relativistic precession model of QPO. We constrain the modification parameters for the nine single-parameter modified spacetimes and provide the spin and mass ranges of three microquasars within the ten spacetime models (including Kerr) at the $68\%$ confidence level (CL).
The results demonstrate that, at the $68 \%$ CL, the QPO data impose stringent constraints on the free parameters, as evidenced by the narrow confidence intervals.
Among them, only the KN spacetime yields a modification parameter constraint spanning both negative and positive values (encompassing the Kerr case at zero). In contrast, all other tested geometries mandate positive-definite parameters at $68 \%$ CL, demonstrating statistical deviation of the Kerr solution.
This highlights the significance of exploring modifications to the Kerr spacetime.
Finally, we evaluate the spacetime models using the Bayes factor and the Akaike Information Criterion (AIC). Based on the current QPO observational data, the Bayesian factor analysis indicates that the ABG, Hayward, KN, BK, and Kerr-MOG spacetime have a slight advantage over the Kerr solution, while the Bardeen, KTN, Kerr-Sen, and PFDM spacetime are somewhat inferior to the Kerr model. In contrast, the AIC analysis shows that the Kerr spacetime remains the optimal model under the current QPO data.

\end{abstract}


\maketitle

\section{$\text{Introduction}$}
The formulation of General Relativity (GR) has profoundly impacted on the development of physics. In weak gravitational fields, the predictions of GR in the solar system and beyond are consistent with most astronomical observations \cite{GR-1}, such as light deflection \cite{GR-2}, gravitational redshift and the correction to the perihelion precession of mercury \cite{GR-3}, etc.. Recently, observations such as gravitational wave signals from binary black hole (BH) mergers \cite{GR-4} and the shadow image of the supermassive black hole $\mathrm{M 87^*}$ \cite{GR-5, GR-6, GR-7,Letter-11} have further strongly supported GR's predictions regarding the existence of BHs.

Although GR has achieved great success, there remain several unresolved physical problems, for example, the accelerated expansion of the universe in its late stages \cite{GR-problem-1, GR-problem-2}, the singularity problem \cite{GR-problem-3}, the dark matter problem \cite{DM}, and others. These issues have prompted in-depth research by physicists into alternative theories.
Modified gravity theories have been proposed to address these problems and have become an important research field in recent years \cite{MG-1, MG-2, MG-3, MG-4, MG-5, MG-6}. By adjusting the components of matter or the gravitational terms in the Einstein-Hilbert action, this approach aims to recover GR in the weak-field limit. However, in strong gravitational fields, other terms such as nonlinear effects gradually dominant, leading to significant differences between the modified gravity theories and GR \cite{MG-aaa-1}.
Furthermore, the problem of the spacetime singularities existed in GR, which cause the breakdown of classical physical laws, has received widespread attention. In this context, phenomenological models based on quantum gravity effects have become one of the research hotspots, including singularity-free gravitational collapse models \cite{NS-1, NS-2}, singularity-free cosmological models \cite{NS-3, NS-4}, and regular black hole models \cite{NS-5, NS-6, NS-7, NS-8, NS-9}, among others.

To date, numerous researches indicate that most celestial objects in the universe exhibit rotational characteristics \cite{a-1, a-2, a-3}. In GR, the unique stationary, axisymmetric, asymptotically flat vacuum solution to Einstein's field equation is represented by the Kerr metric.
This solution describes a non-charged rotating BH through two parameters, the mass $M$  and the spin parameter $a$ \cite{Kerr-1, Kerr-2}.
However, the hypothesis that astrophysical BH candidates are characterized by Kerr spacetime still lacks sufficient evidence \cite{MG-aaa-1}.
Although the predictions of GR regarding the existence of black holes have been confirmed, providing support for its validation in strong gravitational fields, the geometrical structure in strong gravitational fields for GR and the nonlinear behavior of modified gravity theories remain difficult to test and fully understand \cite{Kerr-3}.
To address this, many modified forms of the Kerr BH have been proposed to expand our understanding of rotating black holes. Examples include regular BH models \cite{Bardeen-1, ABG-1, Hayward-1}, BH solutions with a charge \cite{KN-1, KTN-1, BK-1}, BH solutions obtained under different modified gravity theories \cite{MK-1, KS-1}, and BH solutions considering dark matter effects \cite{PFDM-1, PFDM-2} etc.
These modified forms of the Kerr spacetime provide new perspectives for the study of BH geometry and physical properties, opening new research directions in astrophysics. In recent years, astronomical observations, such as quasi-periodic oscillations (QPOs) detected from microquasars \cite{QPO-1, QPO-2, QPO-3, QPO-4, QPO-5}, the shadow of the Sgr A* BH observed by telescopes \cite{shadow-1, shadow-2,Letter-1,Letter-4,Letter-5,Letter-5.5}, gravitational waves \cite{Letter-2,Letter-3,Letter-6} and flare hotspot data observed by gravitational instruments \cite{hot-1}, have become effective ways to test the different gravitational theories.

QPOs are one of the most important tools for testing gravitational theories due to their high precision characteristics \cite{NS-8, NS-9, QPO-1, QPO-2, QPO-3}. QPOs phenomenon can be observed in the X-ray flux from black hole and neutron star X-ray binary systems, and detected as narrow peaks in the power density spectrum \cite{LFQPO-1,QPO-6-2}. Based on the observed frequencies, QPOs are classified into low-frequency (LF) QPOs and high-frequency (HF) QPOs. By analyzing the spectral characteristics of QPOs, scientists can extract physical information about the central object \cite{QPO-6, QPO-8, QPO-9, QPO-10}. Although the generation mechanism of QPOs is not yet fully understood, they are generally believed to be related to relativistic precession and resonance phenomena associated with GR effects \cite{QPO-11, QPO-12, QPO-13}. To investigate these phenomena, various QPO models, including those based on relativistic precession and resonance, have been proposed \cite{QPO-model-1,QPO-model-2,QPO-model-3,QPO-model-4,QPO-model-5,QPO-model-6,QPO-model-7,QPO-model-8}. The construction of these models typically involves the motion of particles around the central object and considers the linear combinations of radial, latitudinal, and orbital frequencies. This paper primarily investigates the dynamical behavior of particles in the Kerr spacetime and nine single-parameter modified versions of the Kerr geometry (Bardeen, Ayon-Beato and García (ABG), Hayward, Kerr-Newman (KN), Kerr-Taub-NUT (KTN), Braneworld Kerr (BK), Kerr-modified theory of gravity (kerr-MOG), Kerr-Sen, and Perfect Fluid Dark Matter (PFDM) black holes), testing these spacetimes using QPO observational data from three microquasars (GRO J1655-40, H1743-322, XTE J1859+226).

The structure of this paper is arranged as follows. The first section is the introduction. The second section provides a brief overview of the Kerr spacetime and its nine single-parameter modified versions. In the third section, we investigate the dynamical behavior of particles in the equatorial plane of the ten aforementioned spacetimes, deriving general expressions for the radial, latitudinal epicyclic frequencies, and orbital angular frequencies of particles undergoing oscillatory motion. In the fourth section, we compare the angular frequencies of particles moving in the Kerr spacetime and its nine single-parameter modified spacetimes with the QPO frequencies observed in three specific microquasars, incorporating the relativistic precession (RP) model. We constrain the modification parameters in the nine single-parameter modified Kerr spacetimes, as well as the mass and spin values of the three microquasars across the ten spacetimes (including Kerr). In the fifth section, we evaluate the ten spacetimes using the Bayes factor and AIC. The sixth section presents the conclusion of the paper.

\section{$\text{Models of stationary axisymmetric spacetime}$}

Let us consider the line element of a circular, stationary, and axisymmetric spacetime written in canonical form \cite{BL}:
\begin{equation}
d s^2=g_{t t} d t^2+2 g_{t \phi} d \phi d t+g_{\phi \phi} d \phi^2+g_{r r} d r^2+g_{\theta \theta} d \theta^2,\label{2.1}
\end{equation}
where there is only one non-vanishing off-diagonal metric coefficient. The non-zero metric components of the Kerr spacetime in Boyer-Lindquist (BL) coordinates are given by \cite{Kerr-4, Kerr-5},
\begin{equation}
\begin{aligned}
g_{t t} &= -\left(1-\frac{2 M r}{\Sigma}\right), \quad g_{t \phi} = -\frac{2 a M r}{\Sigma} \sin ^2 \theta, \quad g_{\theta \theta} = \Sigma, \\
g_{\phi \phi} &= \left(r^2+a^2+\frac{2 M r a^2}{\Sigma} \sin ^2 \theta\right) \sin ^2 \theta, \quad g_{r r} = \frac{\Sigma}{\Delta}, \\
\Sigma &= r^2 + a^2 \cos ^2 \theta, \quad \Delta = r^2 + a^2 - 2 M r,\label{2.2}
\end{aligned}
\end{equation}
where $M$ is the mass of the BH, and $a$ is the spin parameter. Next, we will consider nine stationary, axisymmetric, asymptotically flat spacetime solutions. These solutions are single-parameter modifications of the Kerr spacetime, and when the modification parameter for each model is set to 0, they reduce to the Kerr geometry. These spacetimes encompass not only solutions within the framework of GR (such as rotating regular BH and rotating BH with a charge) but also BH solutions obtained by modified gravitational theories and those derived considering dark matter effects. In this section we use geometrized units.

\subsection{{Rotating regular BHs in GR}}

The regular BH was first introduced as a phenomenological model \cite{Bardeen-1}. In recent studies, spherically symmetric regular BHs have been found to serve as nonsingular exact solutions to the GR field equations coupled to nonlinear electrodynamics \cite{regular-s-1, regular-s-2}. The rotating regular BH, on the other hand, are derived from the spherically symmetric regular BH solution through the non-complexification process \cite{NJ-1}. Next, we will introduce three classical rotating regular BH solutions within the framework of GR.

Bardeen BH is one of the key models of regular black holes. Its construction is based on the theory of nonlinear electrodynamics and is regarded as a potential solution to the black hole singularity problem \cite{Bardeen-1}. By introducing a magnetic monopole charge $b$, the spacetime structure of the Bardeen black hole avoids the occurrence of curvature singularities and satisfies the weak energy condition \cite{Bardeen-1}. The axisymmetric form of this solution, in BL coordinates, has the following metric components  \cite{Bardeen-1, Bardeen-2}
\begin{equation}
\begin{gathered}
g_{t t}=-\left(1-\frac{2 m_1(r) r}{\Sigma}\right),~~ g_{t \phi}=-\frac{2 a m_1(r) r}{\Sigma} \sin ^2 \theta,~~ g_{\theta \theta}=\Sigma \\
g_{\phi \phi}=\left(r^2+a^2+\frac{2 m_1(r) r a^2}{\Sigma} \sin ^2 \theta\right) \sin ^2 \theta,~~ g_{r r}=\frac{\Sigma}{\Delta_1}, \\
\Delta_1=r^2+a^2-2 m_1(r) r,~~\Sigma=r^2+a^2 \cos ^2 \theta,\label{2.3}
\end{gathered}
\end{equation}
$m_1(r)$ represents the mass function, which satisfies $\lim _{r \rightarrow \infty} m_1(r)=M$. In the Bardeen rotating BH, the mass function is given by $m_1(r)=M\left[\frac{r^2}{r^2+b^2}\right]^{3 / 2}$. Obviously, modification parameter $b=0$ in equation (\ref{2.3}) in Bardeen geometry corresponds to the form of Kerr solution in GR.

Ayón-Beato and Garcia (ABG) \cite{ABG-1, ABG-2, ABG-3} also obtained a BH solution free of singularity, referred as the ABG BH. Unlike the Bardeen BH, the ABG BH is a rigorous mathematical solution within the framework of GR coupled with nonlinear electrodynamics. The metric components of the axisymmetric form of the ABG BH are given by \cite{ABG-1, NJ-1}
\begin{equation}
\begin{gathered}
g_{t t}=\left(1-\frac{2 f}{\Sigma}\right),~~ g_{t \phi}=\frac{2 a f \sin ^2 \theta}{\Sigma},~~ g_{\theta \theta}=-\Sigma, \\
g_{\phi \phi}=-\frac{\rho \sin ^2 \theta}{\Sigma},~~ g_{r r}=-\frac{\Sigma}{\Delta_2}, \\
\Delta_2=r^2 F+a^2=r^2-2 f+a^2,~~\Sigma=r^2+a^2 \cos ^2 \theta,\\
2 f=r^2(1-F),~ \rho=\left(r^2+a^2\right)^2-a^2 \Delta_2 \sin ^2 \theta,\label{2.4}
\end{gathered}
\end{equation}
where $~F=1-\frac{2 M r^2}{\left(r^2+Q_1{ }^2\right)^{3 / 2}}+\frac{Q_1{ }^2 r^2}{\left(r^2+Q_1{ }^2\right)^2}$ and $Q_1$ represents the charge. When $Q_1=0$, (\ref{2.4}) is equivalent to classical Kerr solution.

The Hayward BH is a regular black hole model without singularities, proposed by Hayward within the framework of GR \cite{Hayward-1}. This model modifies the classical BH solutions in GR by introducing a mass function. The deviation parameter $l$ in the mass function is used to represent the deviation of the Hayward BH from the Kerr BH \cite{Hayward-1, Hayward-2}. The Hayward metric describes the process of BH formation from an initial vacuum region \cite{Hayward-1, Hayward-3}. In BL coordinates, the metric components of the rotating Hayward BH are given by \cite{Hayward-4}.
\begin{equation}
\begin{gathered}
g_{t t}=-\left(1-\frac{2 r m_2(r)}{\Sigma}\right),~~ g_{t \phi}=-\frac{2 a r m_2(r) \sin ^2 \theta}{\Sigma},~~ g_{\theta \theta}=\Sigma, \\
g_{\phi \phi}=\sin ^2 \theta\left(a^2+r^2+\frac{2 a^2 r m_2(r) \sin ^2 \theta}{\Sigma}\right),~~ g_{r r}=\frac{\Sigma}{\Delta_3}, \\
\Delta_3=r^2-2 r m_2(r)+a^2,~~ \Sigma=r^2+a^2 \cos ^2 \theta. \label{2.5}
\end{gathered}
\end{equation}
The mass function is expressed as $m_2(r)=M \frac{r^{3+\alpha} \Sigma^{-\alpha / 2}}{r^{3+\alpha} \Sigma^{-\alpha / 2}+l^3 r^\beta \Sigma^{-\beta / 2}}$, where the parameters $\alpha$ and $\beta$ are positive real numbers. In the equatorial plane, the mass function is simplified as $m_2(r)=M \frac{r^3}{r^3+l^3}$, and for $l=0$ it corresponds to Kerr solution.

\subsection{{Rotating charged BHs in GR}}

Kerr-Newman (KN) BH is a rotating charged black hole in general relativity, derived from solving the equations for the gravitational and electromagnetic fields to yield a solution for a charged BH. It is characterized by the black hole's mass, charge, and the angular momentum per unit mass. In addition to the charge present in KN BH, there are many kinds of charges such as magnetic, tidal, dyonic, etc. \cite{charges-1,charges-2,charges-3,charges-4,charges-5,charges-6}. Next, we will discuss three black hole solutions with different types of charges, including the KN BH.

The KN BH is an exact solution in general relativity that depicts a charged, rotating BH. It is the charged extension of the Kerr BH and also a solution to the Einstein-Maxwell equations. In the axisymmetric form in the BL coordinate system, the components of its metric are given by  \cite{KN-1}
\begin{equation}
\begin{gathered}
g_{t t}=-\left(1-\frac{2 M r+Q_2^2}{\Sigma}\right),~~ g_{t \phi}=-\frac{2 M a r \sin ^2 \theta}{\Sigma},~~ g_{\theta \theta}=\Sigma, \\
g_{\phi \phi}=\left(r^2+a^2+\frac{2 M a^2 r}{\Sigma}\right) \sin ^2 \theta,~~ g_{r r}=\frac{\Sigma}{\Delta_4}, \\
\Delta_4=r^2-2 M r+a^2+Q_2^2,~~\Sigma=r^2+a^2 \cos ^2 \theta,\label{2.6}
\end{gathered}
\end{equation}
where $Q_2$ represents the charge.

The Kerr-Taub-NUT (KTN) BH is a solution achieved by extending the Kerr BH through the introduction of the Newman-Unti-Tamburino (NUT) parameter $n$ \cite{KTN-1, KTN-2, KTN-3, KTN-4}. The NUT parameter, also called as the gravitational magnetic monopole, represents a unique dual mass effect \cite{KTN-5, KTN-6}. The KTN BH is a stationary, axisymmetric solution to the Einstein field equations, and its metric components in the axisymmetric form in the BL coordinate system are given by \cite{KTN-1}
\begin{equation}
\begin{gathered}
g_{t t}=\frac{-\Delta_5+a^2 \sin ^2 \theta}{\bar{\Sigma}},~~ g_{t \phi}=\frac{A \Delta_5-a B \sin ^2 \theta}{\bar{\Sigma}},~~ g_{\theta \theta}=\bar{\Sigma}, \\
g_{\phi \phi}=\frac{-A^2 \Delta_5+B^2 \sin ^2 \theta}{\bar{\Sigma}},~~ g_{r r}=\frac{\bar{\Sigma}}{\Delta_5}, \\
\Delta_5=r^2-2 M r+a^2-n^2,~~ \bar{\Sigma}=r^2+(n+a \cos \theta)^2, \\
A=a \sin ^2 \theta-2 n \cos \theta,~~ B=r^2+a^2+n^2.\label{2.7}
\end{gathered}
\end{equation}
Obviously, when $n=0$, solution (\ref{2.7}) reduces to the form of Kerr.

The braneworld kerr (BK) BH is an axially symmetric, stationary and asymptotically flat solution of the effective Einstein equations on the brane \cite{BK-1}. The concept of braneworlds, derived from higher-dimensional theories, suggests that the observable universe is confined to a 3-brane, where only the standard-model matter fields (excluding gravity) are present, while gravitational interaction extend into the extra dimensions of space \cite{BK-2}. In this theoretical framework, the effect of the extra dimensions on the gravitational field is described as a "tidal charge" denoted by the parameter $q$, which can take positive or negative values, corresponding to enhanced or weakened gravitational effects, respectively \cite{BK-3, BK-4, BK-5}. The spacetime geometry of the BK BH could be outlined by the metric with BL coordinates as  \cite{BK-1}
\begin{equation}
\begin{gathered}
g_{t t}=-\left(\frac{\Delta_6-a^2 \sin ^2 \theta}{\Sigma}\right), \quad g_{t \phi}=\frac{a \sin ^2 \theta}{\Sigma}\left[\Delta_6-\left(r^2+a^2\right)\right], \quad g_{\theta \theta}=\Sigma, \\
g_{\phi \phi}=\frac{\sin ^2 \theta}{\Sigma}\left[\left(r^2+a^2\right)^2-\Delta_6 a^2 \sin ^2 \theta\right], \quad g_{r r}=\frac{\Sigma}{\Delta_6}, \\
\Delta_6=r^2-2 M r+a^2+q, \Sigma=r^2+a^2 \cos ^2 \theta.\label{2.8}
\end{gathered}
\end{equation}
When $q=0$, the BK solution degenerates into the classical Kerr BH; when $q>0$, it corresponds mathematically to the KN BH; and when $q<0$, it represents a non-standard KN BH with a negative tidal effect \cite{BK-1, BK-5}.

\subsection{{Rotating BHs in modified theories of gravity}}

Researchers have modified classical GR theory by adjusting the matter components or the gravitational term in the Einstein-Hilbert action to explain physical phenomena that cannot be accounted for within the framework of classical GR \cite{GR-problem-1, GR-problem-2, GR-problem-3}. To date, a large number of modified gravity theories have been proposed. Meanwhile, rotating black hole solutions within the modified gravity theories have also been derived. In this section, we will introduce two types of rotating BH solutions obtained within the context of modified gravity theories.

The Kerr-MOG BH, derived by Moffat based on the MOG theory, is a rotating black hole solution \cite{MK-1}. MOG gravity, also called as scalar-tensor-vector gravity (STVG), was proposed by Moffat with the aim of explaining gravitational phenomena in galaxies and the universe without the need for dark matter assumptions \cite{MK-1}. This MOG is based on an action, which consists of the usual Einstein-Hilbert term associated with the metric $g_{\mu \nu}$, a massive vector field $\varphi_\mu$, and three scalar fields that represent the running values of the gravitational constant $G_{\mathrm{K}}$, the vector field mass $\mu$, and its coupling strength $\omega$, respectively \cite{MK-2}. The geometry of the Kerr-MOG BH can be expressed by the line element in the standard BL coordinates \cite{MK-2, MK-3}
\begin{equation}
\begin{gathered}
g_{t t}=-\left(\frac{\Delta_7-a^2 \sin ^2 \theta}{\Sigma}\right), \quad g_{t \phi}=\frac{a \sin ^2 \theta}{\Sigma}\left[\Delta_7-\left(r^2+a^2\right)\right], \quad g_{\theta \theta}=\Sigma, \\
g_{\phi \phi}=\frac{\sin ^2 \theta}{\Sigma}\left[\left(r^2+a^2\right)^2-\Delta_7 a^2 \sin ^2 \theta\right], \quad g_{r r}=\frac{\Sigma}{\Delta_7}, \\
\Delta_7=r^2-2 G_{\mathrm{K}} M r+a^2+\alpha G G_{\mathrm{K}} M^2, \quad \Sigma=r^2+a^2 \cos ^2 \theta,\label{2.9}
\end{gathered}
\end{equation}
where $G_{\mathrm{K}}=G(1+\alpha)$ is the enhanced gravitational constant, where $G=1$ is Newtonian gravitational constant. $\alpha$ is the modification parameter, and when $\alpha=0$ the Kerr-MOG solution reduces to Kerr geometry.

The Kerr-Sen BH, is a black hole solution of the Einstein-Maxwell-dilaton-axion (EMDA) gravity, first initially introduced by Sen in 1992 \cite{KS-1}, arises as a solution to the low energy 4-dimensional effective action of heterotic string theory,  representing a rotating charged black hole. The EMDA gravity consists of the dilaton field, the gauge vector field the metric, and the pseudo-scalar axion field \cite{KS-1,KS-1.1}. The components of its metric, in BL coordinates, are given by \cite{KS-2}
\begin{equation}
\begin{aligned}
g_{t t} & =-\left(1-\frac{2 M r}{\tilde{\Sigma}}\right), g_{t \phi}=-\frac{2 M r a \sin ^2 \theta}{\tilde{\Sigma}}, g_{\theta \theta}=\tilde{\Sigma}, \\
g_{\phi \phi} & =\sin ^2 \theta\left(r(r+Q_3)+a^2+\frac{2 M r a^2 \sin ^2 \theta}{\tilde{\Sigma}}\right), g_{r r}=\frac{\tilde{\Sigma}}{\Delta_8}, \\
\Delta_8 & =r(r+Q_3)-2 M r+a^2, \quad \tilde{\Sigma}=r(r+Q_3)+a^2 \cos ^2 \theta,\label{2.10}
\end{aligned}
\end{equation}
$Q_3=Q^2 / M$ denotes the modification parameter in Kerr-Sen solution relative to Kerr case, where $Q$ represents the electric charge.

\subsection{{Rotating BHs modiffed by dark matter}}

Some astronomical observations suggest the existence of dark matter on galactic and extragalactic scales \cite{DM-1, DM-2, DM-3, DM-4}. Based on this, many physicists have begun to investigate the interaction between dark matter (DM) and black holes \cite{BHDM-1,BHDM-2,BHDM-3,BHDM-4,BHDM-5,BHDM-6,BHDM-7}. Among these, the PFDM model \cite{PFDM-0.1,PFDM-0.2,PFDM-0.3}, which represents dark matter as a perfect fluid, presents a potential explanation for dark matter. In the following, we will introduce the rotating BH solution derived within the PFDM framework.

The spherically symmetric form of the PFDM BH was proposed by Li et al. \cite{PFDM-1}. In this model, the parameter $k$ characterizes the intensity of dark matter, and its introduction alters the radial function of the black hole metric \cite{PFDM-2}. When $k=0$, the model degenerates into the Schwarzschild BH. Based on the spherically symmetric black hole solution, this model was further extended to the axially symmetric rotating form in reference \cite{PFDM-3}. The spacetime metric of rotating BH in PFDM is given by \cite{PFDM-3,PFDM-4}
\begin{equation}
\begin{gathered}
g_{t t}=-\left(1-\frac{2 M r-k r \ln \left(\frac{r}{|k|}\right)}{\Sigma}\right), g_{t \phi}=-\frac{a \sin ^2 \theta\left(2 M r-k r \ln \left(\frac{r}{|k|}\right)\right)}{\Sigma}, g_{\theta \theta}=\Sigma, \\
g_{\phi \phi}=\sin ^2 \theta\left(r^2+a^2+a^2 \sin ^2 \theta \frac{2 M r-k r \ln \left(\frac{r}{|k|}\right)}{\Sigma}\right), g_{r r}=\frac{\Sigma}{\Delta_9}, \\
\Delta_9=r^2-2 M r+a^2+k r \ln \left(\frac{r}{|k|}\right), \quad \Sigma=r^2+a^2 \cos ^2 \theta.\label{2.11}
\end{gathered}
\end{equation}
The values of $k$ can be taken as positive or negative, with different values of $k$ representing various configurations of dark matter distribution. For $k=0$, the equation (\ref{2.11}) becomes Kerr solution in GR.

 \section{$\text{The dynamical effects of particles in some stationary axisymmetric spacetimes}$}

In the spacetime characterized by equation (\ref{2.1}), the geodesic motion of a particle near the central body is governed by $g_{\mu \nu} \dot{x}^\mu \dot{x}^\nu=\epsilon$. When $\epsilon=0$, it corresponds to null geodesics, and when $\epsilon=-1$, it corresponds to timelike geodesics. For a massive particle moving along a timelike geodesic in the equatorial plane, the following relation can be acquired
\begin{equation}
\frac{d t}{d \tau}=\frac{E g_{\phi \phi}+L g_{t \phi}}{g_{t \phi}^2-g_{t t} g_{\phi \phi}}, \label{3.1}
\end{equation}
\begin{equation}
\frac{d \phi}{d \tau}=-\frac{E g_{t \phi}+L g_{t t}}{g_{t \phi}^2-g_{t t} g_{\phi \phi}}, \label{3.2}
\end{equation}
\begin{equation}
V_{e f f}=g_{r r}\left(\frac{d r}{d \tau}\right)^2=-1+\frac{E^2 g_{\phi \phi}+2 E L g_{t \phi}+L^2 g_{t t}}{g_{t \phi}^2-g_{t t} g_{\phi \phi}}.\label{3.3}
\end{equation}
Where $E$ and $L$ are two constants. By setting $V_{\text {eff }}(r)=0$ and ${d V_{\text {eff }}(r)}/{d r}=0$, the expressions for the energy $E$ and angular momentum $L$ of a particle moving on a circular orbit around the central body are written as
\begin{equation}
E=-\frac{g_{t t}+g_{t \phi} \omega_\phi}{\sqrt{-g_{t t}-2 g_{t \phi} \omega_\phi-g_{\phi \phi} \omega_\phi^2}},\label{3.4}
\end{equation}
\begin{equation}
L=\frac{g_{t \phi}+g_{\phi \phi} \omega_\phi}{\sqrt{-g_{t t}-2 g_{t \phi} \omega_\phi-g_{\phi \phi} \omega_\phi^2}},\label{3.5}
\end{equation}
here, $\omega_\phi$ represents the angular velocity of the particle's motion
\begin{equation}
\omega_\phi=\frac{d \phi}{d t}=\frac{-g_{t \phi, r} \pm \sqrt{\left(g_{t \phi, r}\right)^2-g_{t t, r} g_{\phi \phi, r}}}{g_{\phi \phi, r}} .\label{3.6}
\end{equation}
The symbols $+/-$ represent the angular velocity of particles undergoing prograde and retrograde rotations, respectively. To determine the angular frequencies of the radial and latitudinal epicyclic motion of the particle on a stable circular orbit, we consider the geodesic equation \cite{w-1}
\begin{equation}
\frac{d^2 x^\mu}{d p}+\Gamma_{\alpha \beta}^\mu \frac{d x^\alpha}{d p} \frac{d x^\beta}{d p}=0. \label{3.7}
\end{equation}
$p$ is the affine parameter along the geodesic. (\ref{3.7}) describes the equation of motion for a particle moving on a circular orbit around the central body. Next, we consider the particle's motion on the circular orbit, perturbed by small disturbances, and introduce the deviation vector into the geodesic equation, where $Z^\mu(p)$ denotes the undisturbed motion on the circular orbit. Subsequently, for a circular orbit in the equatorial plane, the deviation vector can be expressed as
\begin{equation}
\xi^\mu(p)=x^\mu(p)-Z^\mu(p), \label{3.8}
\end{equation}
and the coordinates of the particle are given by
\begin{equation}
Z^\mu(p)=\left\{t(p), r_0, \pi / 2, \omega_\phi t(p)\right\},\label{3.9}
\end{equation}
$r_0$ represents the radial coordinate position of the particle while it moves on a circular orbit around the central body. By substituting equation (\ref{3.9}) into equation (\ref{3.8}) and expanding $\xi^\mu(p)$ in terms of its powers, and then taking its first-order linear approximation, the deviation equation for $\xi^\mu(p)$ can be determined. This equation represents the small perturbations that occur when the particle moves on a circular orbit around the central body \cite{w-2}
\begin{equation}
\begin{gathered}
\frac{d^2 \xi^\mu}{d t^2}+2 \gamma_\nu^\mu \frac{d^2 \xi^\nu}{d t^2}+\xi^\eta \partial_\eta U^\mu=0 \quad \eta \equiv r, \theta \\
\gamma_\nu^\mu=\Gamma_{\nu \beta}^\mu u^\beta\left(u^0\right)^{-1}, U^\mu=\gamma_\nu^\mu u^\nu\left(u^0\right)^{-1}.\label{3.10}
\end{gathered}
\end{equation}
The quantities $\gamma_\nu^\mu$ and $\partial_\eta U^\mu$ are evaluated on a circular orbit with $r=r_0, \theta=\pi / 2$, and $u^\mu=\dot{Z}^\mu=$ $u^0\left(1,0,0, \omega_\phi\right)$. When $\mu=r$ and $\mu=\theta$, the two decoupled oscillations in the radial and vertical directions can be obtained,
\begin{equation}
\frac{d^2 \xi^\mu}{d t^2}+\omega_r^2 \xi^r=0 ~, ~ \frac{d^2 \xi^\theta}{d t^2}+\omega_\theta^2 \xi^\theta=0.\label{3.11}
\end{equation}
From equation (\ref{3.11}), we have
\begin{equation}
\omega_r^2=\partial_r U^r-4 \gamma_A^r \gamma_r^A~,~\omega_\theta^2=\partial_\theta U^\theta.\label{3.12}
\end{equation}
Expanding equation (\ref{3.12}), the general expressions for $\omega_r^2$ and $\omega_\theta^2$ are expressed as
\begin{equation}
\begin{aligned}
& \omega_r^2=\partial_r \Gamma_{t t}^r+2 \omega_\phi \partial_r \Gamma_{t \phi}^r+\omega_\phi^2 \Gamma_{\phi \phi}^r-4\left(\Gamma_{t t}^r \Gamma_{r t}^t+\Gamma_{t t}^r \Gamma_{r \phi}^t \omega_\phi+\Gamma_{t \phi}^r \Gamma_{r t}^t \omega_\phi\right. \\
& \left.+\Gamma_{t \phi}^r \Gamma_{r \phi}^t \omega_\phi^2+\Gamma_{\phi t}^r \Gamma_{r t}^\phi+\Gamma_{\phi t}^r \Gamma_{r \phi}^\phi \omega_\phi+\Gamma_{\phi \phi}^r \Gamma_{r t}^\phi \omega_\phi+\Gamma_{\phi \phi}^r \Gamma_{r \phi}^\phi \omega_\phi^2\right),\label{3.13}
\end{aligned}
\end{equation}
\begin{equation}
\omega_\theta^2=\partial_\theta \Gamma_{t t}^\theta+2 \omega_\phi \partial_\theta \Gamma_{t \phi}^\theta+\omega_\phi^2 \Gamma_{\phi \phi}^\theta .\label{3.14}
\end{equation}

When $\omega_r^2 \geq 0$ and $\omega_\theta^2 \geq 0$, the perturbation quantities $\xi_r$ and $\xi_\theta$ in equation (\ref{3.11}) exhibit oscillatory behavior with respect to time $t$. Then the particle undergoes harmonic oscillations in the radial and vertical directions along the circular orbit with frequencies $\omega_r$ and $\omega_\theta$, respectively, where the circular motion is stable.
These frequencies are referred to as the epicyclic frequencies. Additionally, $\omega_r^2=0$ corresponds to the position of the innermost stable circular orbit. When $\omega_r^2<0$ or $\omega_\theta^2<0$, the solutions $\xi_r$ or $\xi_\theta$ to equation (\ref{3.11}) grow exponentially with time $t$, indicating that small perturbations in the corresponding direction will deviate from the circular orbit exponentially, leading to instability in the motion \cite{w-1, w-3}.
By substituting equations (\ref{2.2})-(\ref{2.11}) into (\ref{3.6}), (\ref{3.13})-(\ref{3.14}), we obtain the expressions for the orbital frequency, as well as the radial and latitudinal epicyclic frequencies of particles moving in circular orbits in the equatorial plane in the Kerr spacetime and its nine classes of single-parameter modified spacetimes. The concrete frequency expressions for particle motion in the ten types of spacetimes are derived and summarized in Appendix A.
\begin{figure}[ht]
  \includegraphics[width=7.8cm]{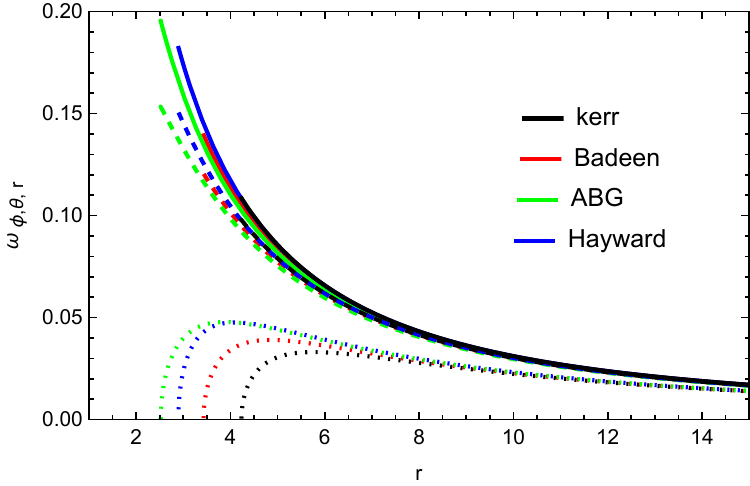}
  \includegraphics[width=7.8cm]{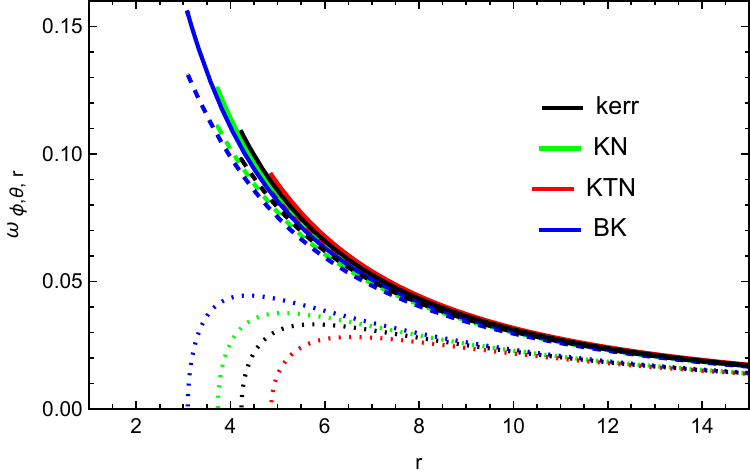}
  \includegraphics[width=7.8cm]{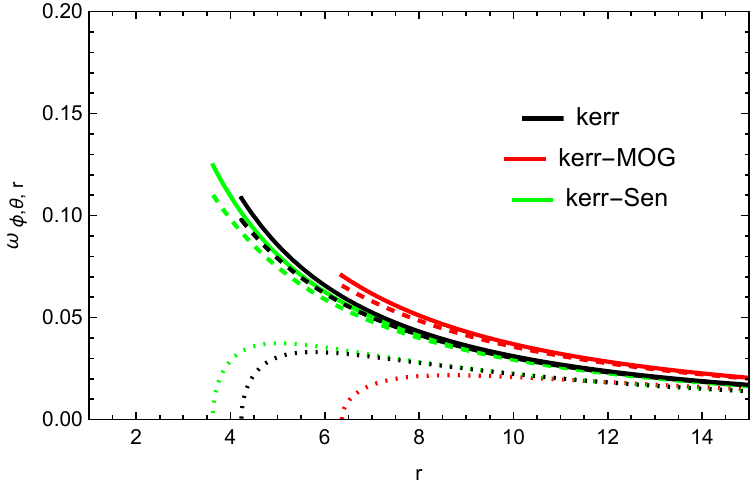}
  \includegraphics[width=7.8cm]{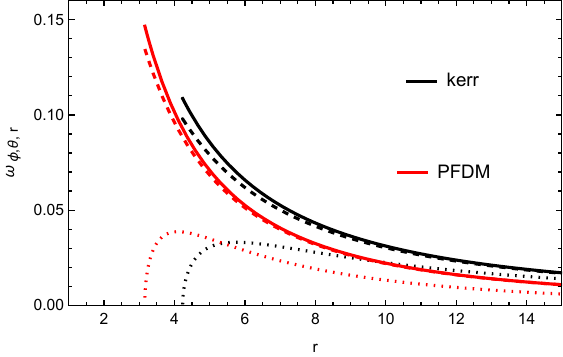}
  \caption{With a spin parameter $a=0.5$, in the Kerr and its nine types of single-parameter modified spacetimes (set each modification parameter to 0.5), the frequencies of the test particle, $\omega_r$ (dotted line), $\omega_\theta$ (dashed line), and $\omega_\phi$ (solid line), are plotted as functions of the radial coordinate $r$.}
  \label{fig:1}
\end{figure}

In Figure \ref{fig:1}, we present the variation of the test particle's frequencies $\omega_r$ (dotted line), $\omega_\theta$ (dashed line), and $\omega_\phi$ (solid line) with respect to the radial coordinate $r$ in the Kerr spacetime, as well as in the nine types of single-parameter modified spacetimes of Kerr.
In the plot, we take spin parameter $a=0.5$, and only the range $r>r_{\text {ISCO }}$ is considered. Additionally, to visually illustrate the impact of the modification parameters on the test particle's frequencies, we set the modification parameter for all nine types of modified Kerr spacetimes to 0.5.
From the figure, it is observed that as the particle approaches the BH, its radial frequency distribution in the non-Kerr spacetimes significantly deviates from that in the Kerr spacetime. However, as the particle moves away from the black hole, the radial frequency distributions across the different spacetimes gradually converge.
One knows that the particle's epicyclic and orbital motions around the central body are theoretically regarded as one possible origin of the observed QPO phenomenon. Researchers focus on using the angular frequencies of the particle's epicyclic motion and orbital motion to construct corresponding theoretical models, and by combining these with the observed QPO phenomena, conduct in-depth studies of the astrophysical QPO behavior. To facilitate the subsequent use of QPO observational data in constraining various models, we introduce
\begin{equation}
\nu=\frac{1}{2 \pi} \frac{c^3}{G M} \omega_{r, \theta, \phi},\label{3.16}
\end{equation}
where $c$ is the speed of light, $G$ is the Newtonian gravitational constant, and $M$ is the mass of the celestial body.

\section{$\text{Data fitting and observational constraints on models of stationary axisymmetric spacetime}$}

In this section, we will use the QPO data observed in microquasars to constrain the modification parameters in nine types of modified spacetimes, as well as the spin and mass parameters of microquasars in ten types of spacetimes (including the Kerr spacetime). We selecte four sets of QPO observation data observed from three microquasars, which are displayed in Table \ref{table-1}. These three microquasars are labeled as: GRO J1655-40 $\left(p_1\right)$, H1743-322 $(p_2)$, and XTE J1859+226 $\left(p_3\right)$. The data in Table \ref{table-1} include the HFQPO double peak frequencies and the LFQPO frequency for the three microquasars \cite{microquasars-1,microquasars-2,microquasars-3,microquasars-4,microquasars-5,microquasars-6,microquasars-7,microquasars-8}.

\begin{table}[ht]
\begin{center}
\begin{tabular}{|l|l|l|l|l|l|}
\hline Source & $\nu_{1 u}[\mathrm{~Hz}]$ & $\nu_{1 l}[\mathrm{~Hz}]$ & $\nu_{1 c}[\mathrm{~Hz}]$ & $\nu_{2 u}[\mathrm{~Hz}]$ & $\nu_{2 c}[\mathrm{~Hz}]$ \\
\hline GRO J1655-40 ($p_1$) & $441 \pm 2$ & $298 \pm 4$ & $17.3 \pm 0.1$ & $451 \pm 5$ & $18.3 \pm 0.1$ \\
\hline H1743-322 ($p_2$) & $240 \pm 3$ & $165_{-5.0}^{+9.0}$ & $9.44 \pm 0.02$ & --- & --- \\
\hline XTE J1859+226 ($p_3$) & $227.5{ }_{-2.4}^{+2.1}$ & $128.6_{-1.8}^{+1.6}$ & $3.65 \pm 0.01$ & --- & --- \\
\hline
\end{tabular}
\end{center}
\caption{\label{table-1}
 The four sets of observational data from the three microquasars \cite{microquasars-1,microquasars-2,microquasars-3,microquasars-4,microquasars-5,microquasars-6,microquasars-7,microquasars-8}. }
\end{table}

In the study of QPO phenomena, the construction of related models typically involves linear combinations of the radial, latitudinal epicyclic motion frequencies, and orbital frequencies of a particle around the central body. Overall, QPO models can be broadly classified into four categories: the tidal precession model \cite{QPO-model-9}, the disk oscillation model \cite{QPO-model-10, QPO-model-11, QPO-model-12}, the relativistic precession model \cite{QPO-model-13, QPO-model-14}, and the resonance model \cite{QPO-model-1}. Considering the QPO data observed in the three aforementioned microquasars, to date, only the RP model has been able to simultaneously explain the effects of both HFQPO and LFQPO \cite{LFQPO-1, LFQPO-2}. Therefore, this paper will employ the RP model to analyze the QPO observational data presented in Table \ref{table-1}.

The RP model, proposed by Stella and Vietri \cite{QPO-model-3}, is a kinematic model. In the RP model, it is believed that QPOs are generated by the local motion of plasma in the accretion disk surrounding the central body \cite{microquasars-1}. In this model, the high-frequency component of the HFQPO double peak is represented by the orbital frequency of the test particle ($\nu_u=\nu_\phi$), while the low-frequency component is represented by a linear combination of the particle's orbital frequency and the radial epicyclic frequency ($\nu_l=\nu_\phi+\nu_r$, also referred to as the periastron precession frequency). The nodal precession frequency ($\nu_c=\nu_\phi-\nu_\theta$) corresponds to LFQPO. It is assumed that the four sets of QPO resuls in Table \ref{table-1} (including both high and low frequency observational data) originate from different circular orbital positions with radii $r_1 / M, r_1^{\prime} / M, r_2 / M, r_3 / M$. As shown in equations (\ref{3.6}) and (\ref{3.13})-(\ref{3.14}), the theoretical model contains seven free parameters, in addition to the resonance positions: the modification parameters $\Delta^*\left(\Delta^*=b^*, Q_1^*, l^*, Q_2^*, n^*, q^*, \alpha^*, Q_3^*, k^*\right)$, the spin parameters of the three microquasars $a_{p}^*$ and their masses $M_p^*$. Here, for convenience, we define the dimensionless quantities: $r^*=r / M, \Delta^*=$ $\Delta / M, a_p^*=a_p / M, M_p^*=M_p / M_{\odot}$ (where $p=1,2,3$ ). Next, using the RP model, we perform a $\chi^2$ analysis with the following equation

\begin{equation}
\begin{gathered}
\chi^2=\frac{\left\{\nu_{2 u, 1}-\nu_{2 u}\left(\Delta^*, a_1^*, r_1^{\prime *}, M_1^*\right)\right\}^2}{\sigma_{\nu_{2 u, 1}}^2}+\frac{\left\{\nu_{2 c, 1}-\nu_{2 c}\left(\Delta^*, a_1^*, r_1^{\prime *}, M_1^*\right)\right\}^2}{\sigma_{\nu_{2 c, 1}}^2}+\sum_{p=1}^3 \left(\frac{\left\{\nu_{1 u, p}-\nu_{1 u}\left(\Delta^*, a_p^*, r_p^*, M_p^*\right)\right\}^2}{\sigma_{\nu_{1 u, p}}^2}+ \right.\\
\left. \frac{\left\{\nu_{1 l, p}-\nu_{1 l}\left(\Delta^*, a_p^*, r_p^*, M_p^*\right)\right\}^2}{\sigma_{\nu_{1 l, p}}^2}+\frac{\left\{\nu_{1 c, p}-\nu_{1 c}\left(\Delta^*, a_p^*, r_p^*, M_p^*\right)\right\}^2}{\sigma_{\nu_{1 c, p}}}\right) .  \label{4.1}
\end{gathered}
\end{equation}

\begin{figure}[ht]
\includegraphics[width=5.8cm]{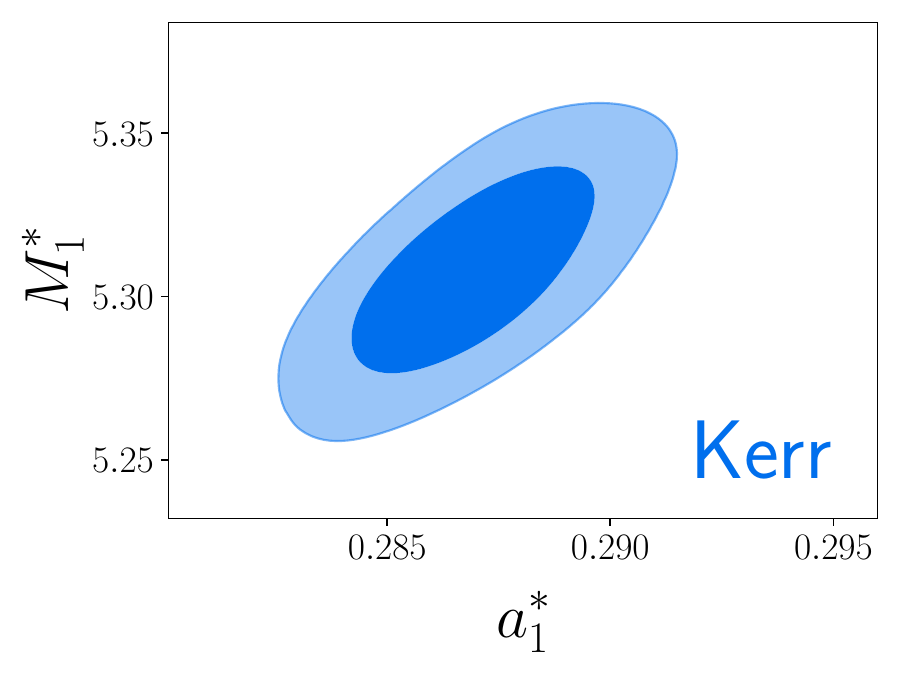}
\includegraphics[width=5.8cm]{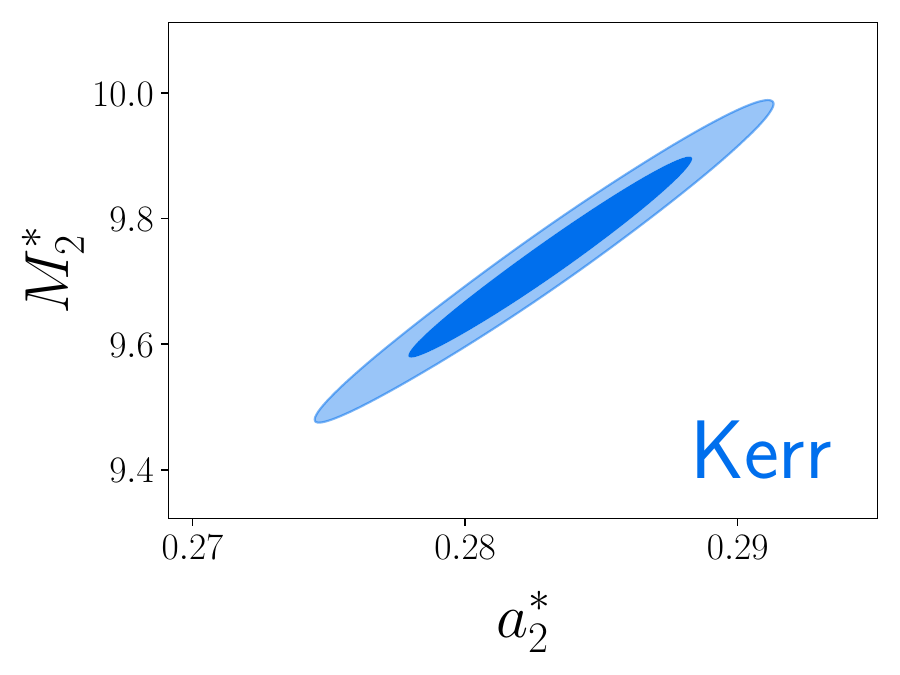}
\includegraphics[width=5.8cm]{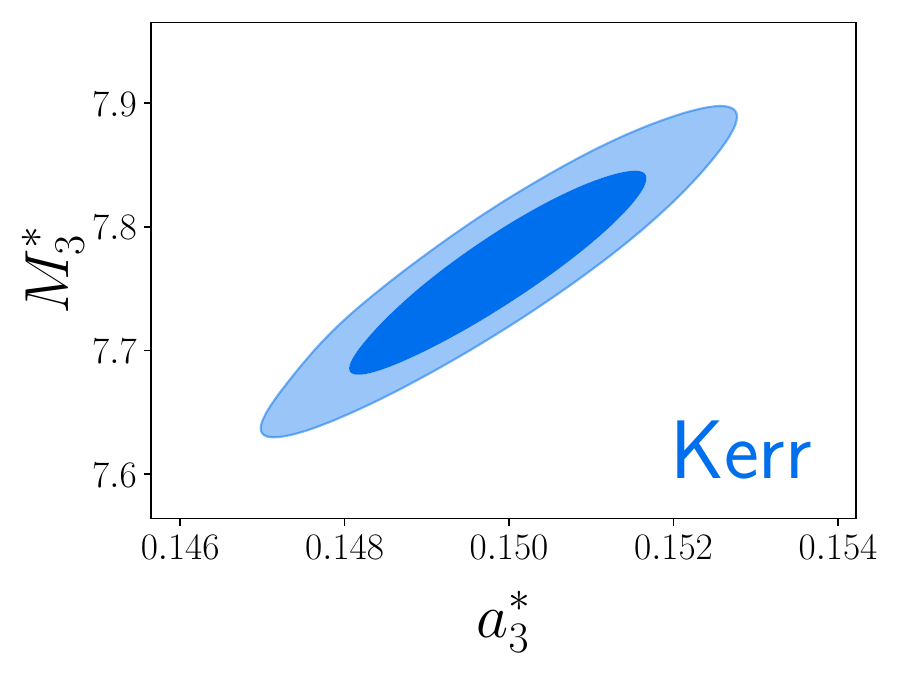}
\caption{The parameter plots of the spin parameters $a_p^*$ and masses $M_p^*$ of the three microquasars in the Kerr spacetime within the $68 \%$ and $95 \%$ CL.}
    \label{fig:2}
\end{figure}

On the basis of the constraint results from the QPO data, we summarize in Table \ref{table-2} the best-fit values of the modification parameters in the nine types of modified spacetimes, as well as the spin values $a_p^*$ and masses $M_p^*$ of the three microquasars in ten types of spacetimes within the $68 \%$ confidence level (CL), along with the $\chi_{\min }^2$ values and marginal likelihood values. Additionally, the best-fit values of the circular orbital radii (i.e., potential resonance positions) associated with the four sets of QPO data are listed in Table \ref{table-3}. In Figure \ref{fig:2}, we show the parameter plots of the spin $a_p^*$ and mass $M_p^*$ of the three microquasars in the Kerr spacetime within the $68 \%$ and $95 \%$ CL. In Figures \ref{fig:3} and \ref{fig:4}, we show the parameter plots of the modification parameters, as well as the spin parameters $a_p^*$ and masses $M_p^*$ of the three microquasars in the nine modified Kerr spacetimes, within the $68 \%$ and $95 \%$ CL.

\begin{table}[ht]
\begin{center}
\begin{tabular}{|l|l|l|l|l|l|l|}
\hline
\multicolumn{2}{|c|}{} & Kerr & Bardeen ($b^*$) & ABG ($Q_1{ }^*$) & Hayward ($l^*$) & $\mathrm{KN}(Q_2{ }^*)$ \\
\hline
\multicolumn{2}{|c|}{$\Delta^*$} & --- & $0.229_{-0.034}^{+0.045}$ & $0.360_{-0.011}^{+0.012}$ & $0.295 \pm 0.019$ & $-0.004 \pm 0.078$ \\
\hline
\multirow{3}{*}{$a_p^*$} & GRO J1655-40 & $0.287_{-0.002}^{+0.002}$ & $0.279 \pm 0.002$ & $0.254 \pm 0.002$ & $0.267 \pm 0.003$ & $0.287 \pm 0.002$ \\
\cline{2-7}
                         & H1743-322 & $0.283_{-0.003}^{+0.003}$ & $0.274 \pm 0.005$ & $0.251_{-0.004}^{+0.003}$ & $0.264_{-0.006}^{+0.005}$ & $0.282 \pm 0.005$ \\
\cline{2-7}
                         & XTE J1859+226 & $0.150 \pm 0.001$ & $0.146 \pm 0.004$ & $0.135_{-0.001}^{+0.001}$ & $0.141 \pm 0.005$ & $0.149 \pm 0.004$ \\
\hline
\multirow{3}{*}{$M_p^*$} & GRO J1655-40 & $5.307_{-0.021}^{+0.024}$ & $5.425 \pm 0.026$ & $5.796_{-0.030}^{+0.027}$ & $5.671 \pm 0.028$ & $5.302 \pm 0.025$ \\
\cline{2-7}
                         & H1743-322 & $9.745_{-0.082}^{+0.092}$ & $10.000 \pm 0.110$ & $10.620_{-0.140}^{+0.120}$ & $10.440_{-0.130}^{+0.120}$ & $9.800_{-0.120}^{+0.100}$ \\
\cline{2-7}
                         & XTE J1859+226 & $7.774 \pm 0.060$ & $8.009 \pm 0.063$ & $8.509 \pm 0.073$ & $8.367_{-0.070}^{+0.059}$ & $7.823 \pm 0.062$ \\
\hline
\multicolumn{2}{|c|}{$\chi_{\text {min }}^2$} & 0.976 & 0.572 & 0.428 & 0.442 & 0.429 \\
\hline
\multicolumn{2}{|c|}{Marginal likelihood} & 0.070 & 0.068 & 0.074 & 0.072 & 0.078 \\
\hline\hline
\multicolumn{2}{|c|}{} & KTN ($n^*$) & BK ($q{ }^*$) & Kerr-MOG ($\alpha^*$) & Kerr-Sen ($Q_3^*$) & PFDM ($k^*$) \\
\hline
\multicolumn{2}{|c|}{$\Delta^*$} & $0.257_{-0.025}^{+0.054}$ & $0.152 \pm 0.019$ & $0.795 \pm 0.011$ & $0.361_{-0.024}^{+0.028}$ & $0.019_{-0.003}^{+0.002}$ \\
\hline
\multirow{3}{*}{$a_p^*$} & GRO J1655-40 & $0.311_{-0.005}^{+0.006}$ & $0.270 \pm 0.002$ & $0.422 \pm 0.003$ & $0.272 \pm 0.002$ & $0.278 \pm 0.002$ \\
\cline{2-7}
                         & H 1743-322 & $0.305_{-0.008}^{+0.009}$ & $0.266 \pm 0.005$ & $0.415 \pm 0.008$ & $0.267_{-0.004}^{+0.003}$ & $0.2471_{-0.004}^{+0.003}$ \\
\cline{2-7}
                         & XTE J1859+226 & $0.170_{-0.007}^{+0.007}$ & $0.141 \pm 0.004$ & $0.223 \pm 0.006$ & $0.142 \pm 0.001$ & $0.144 \pm 0.001$ \\
\hline
\multirow{3}{*}{$M_p^*$} & GRO J1655-40 & $5.196 \pm 0.029$ & $5.556 \pm 0.027$ & $3.411 \pm 0.018$ & $5.515 \pm 0.027$ & $5.532 \pm 0.031$ \\
\cline{2-7}
                         & H1743-322 & $9.550 \pm 0.110$ & $10.160 \pm 0.120$ & $6.288 \pm 0.073$ & $10.240_{-0.130}^{+0.120}$ & $10.140 \pm 0.120$ \\
\cline{2-7}
                         & XTE J1859+226 & $7.716 \pm 0.080$ & $8.254 \pm 0.070$ & $5.064 \pm 0.045$ & $8.173 \pm 0.069$ & $8.267 \pm 0.075$ \\
\hline
\multicolumn{2}{|c|}{$\chi_{\text {min }}^2$} & 0.424 & 0.497 & 0.480 & 0.699 & 0.712 \\
\hline
\multicolumn{2}{|c|}{Marginal likelihood} & 0.068 & 0.072 & 0.075 & 0.065 & 0.064 \\
\hline
\end{tabular}
\end{center}
\caption{\label{table-2}
 The best-fit ranges of the modification parameters in the nine modified versions of Kerr spacetime, as well as the spin values $a_p^*$ and masses $M_p^*$ of the three microquasars within the $68 \%$ CL under ten types of rotating spacetimes, along with the $\chi_{\min }^2$ values and marginal likelihood values.}
\end{table}

\begin{table}[ht]
\setlength{\tabcolsep}{7pt}  
\begin{center}
\begin{tabular}{|l|l|l|l|l|l|l|l|l|l|l|}
\hline
& Kerr & Bardeen & ABG & Hayward & KN & KTN & BK & Kerr-MOG & Kerr-Sen & PFDM \\
\hline
$r_1 / M$ & 5.685 & 5.592 & 5.283 & 5.441 & 5.684 & 5.796 & 5.466 & 9.008 & 5.498 & 5.448 \\
\hline
$r_1^{\prime} / M$ & 5.582 & 5.485 & 5.182 & 5.339 & 5.577 & 5.691 & 5.366 & 8.834 & 5.400 & 5.347 \\
\hline
$r_2 / M$ & 5.662 & 5.550 & 5.264 & 5.411 & 5.635 & 5.762 & 5.449 & 8.940 & 5.442 & 5.425 \\
\hline
$r_3 / M$ & 6.895 & 6.765 & 6.427 & 6.586 & 6.866 & 6.975 & 6.592 & 10.904 & 6.638 & 6.542 \\
\hline
\end{tabular}
\end{center}
\caption{\label{table-3}
 The best-fit values of the circular orbital radii of particle motion (potential resonance positions) in the Kerr spacetime and nine types of modified spacetimes of Kerr.}
\end{table}

\begin{figure}[htbp]
    \renewcommand{\thefigure}{3}
    \centering

    \setlength{\tabcolsep}{2pt}

    \begin{tabular}{cc}
        \subfloat[]{
            \begin{minipage}[b]{0.45\linewidth}
                \centering
                \includegraphics[width=2.6cm]{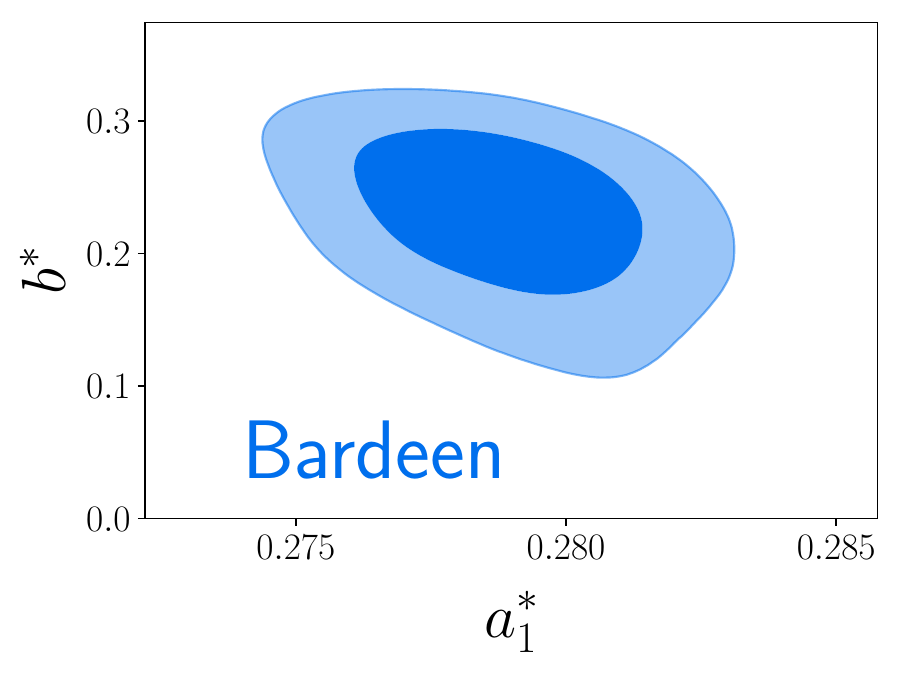}  
                \includegraphics[width=2.6cm]{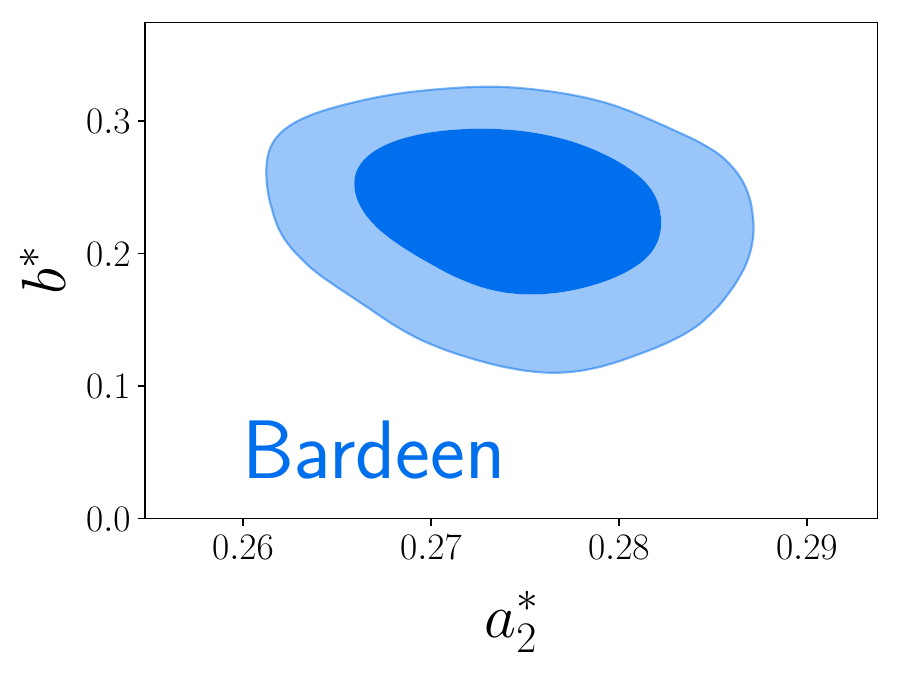}
                \includegraphics[width=2.6cm]{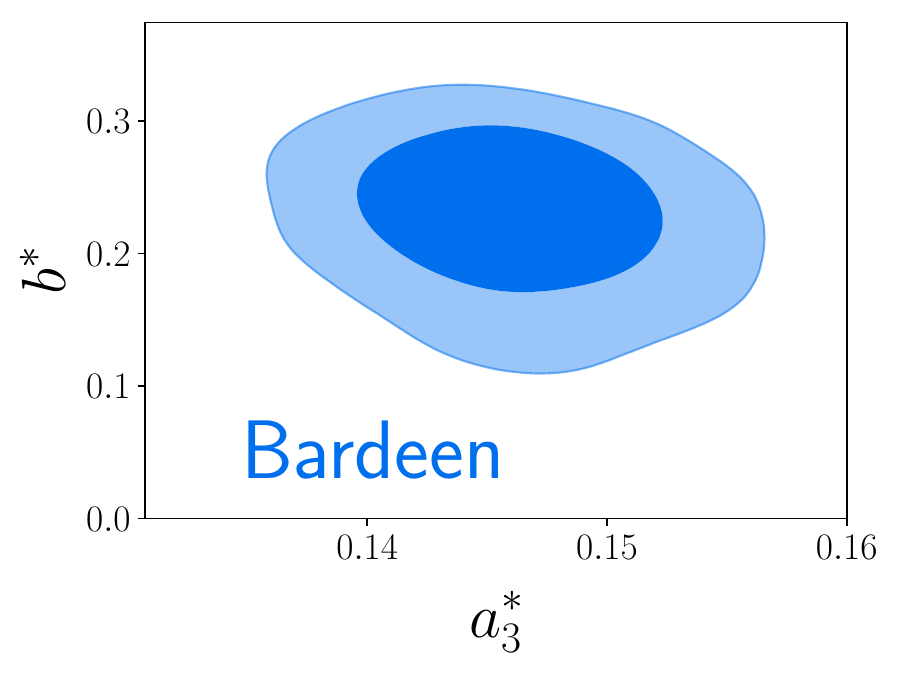}
            \end{minipage}
        }
        & 
        \subfloat[]{
            \begin{minipage}[b]{0.45\linewidth}
                \centering
                \includegraphics[width=2.6cm]{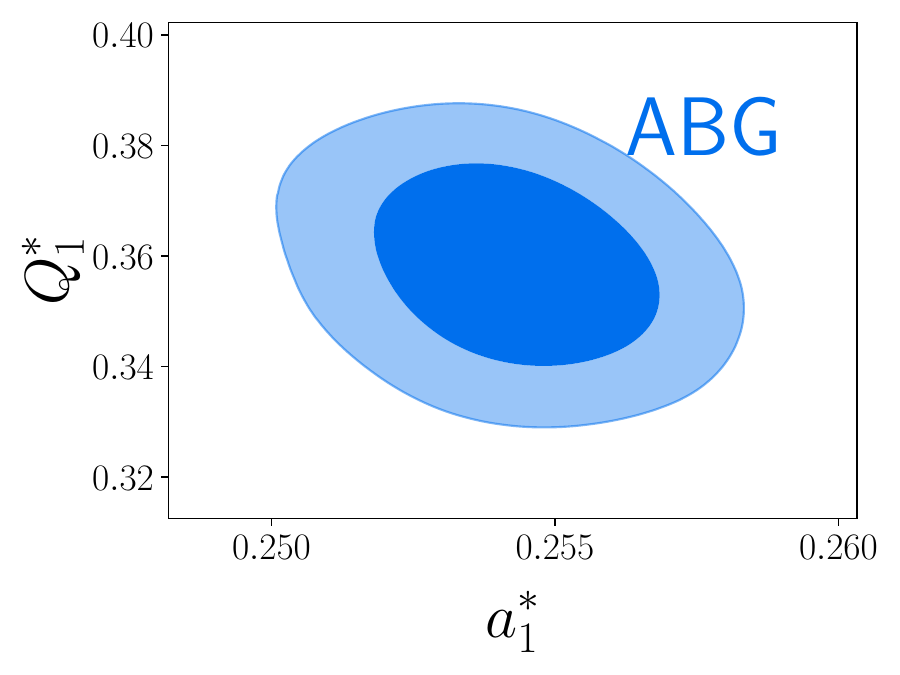}
                \includegraphics[width=2.6cm]{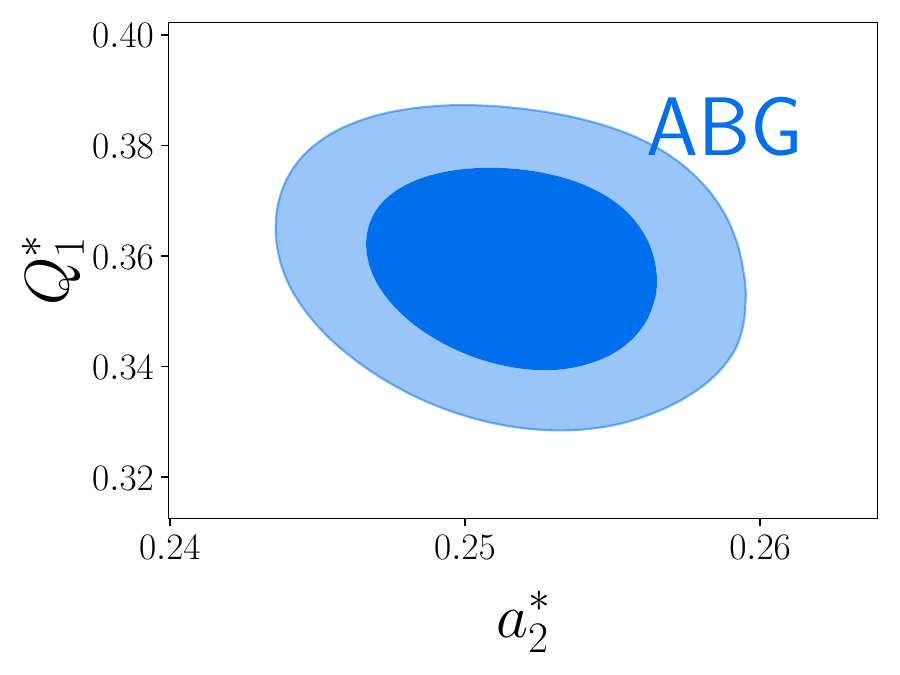}
                \includegraphics[width=2.6cm]{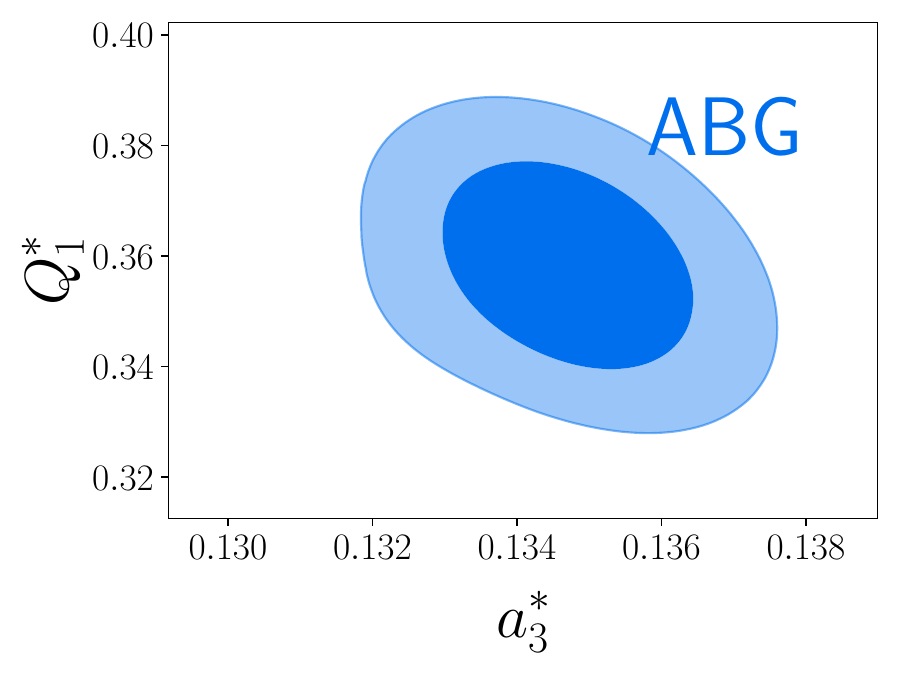}
            \end{minipage}
        }
        \\[1pt]
        \subfloat[]{
            \begin{minipage}[b]{0.45\linewidth}
                \centering
                \includegraphics[width=2.6cm]{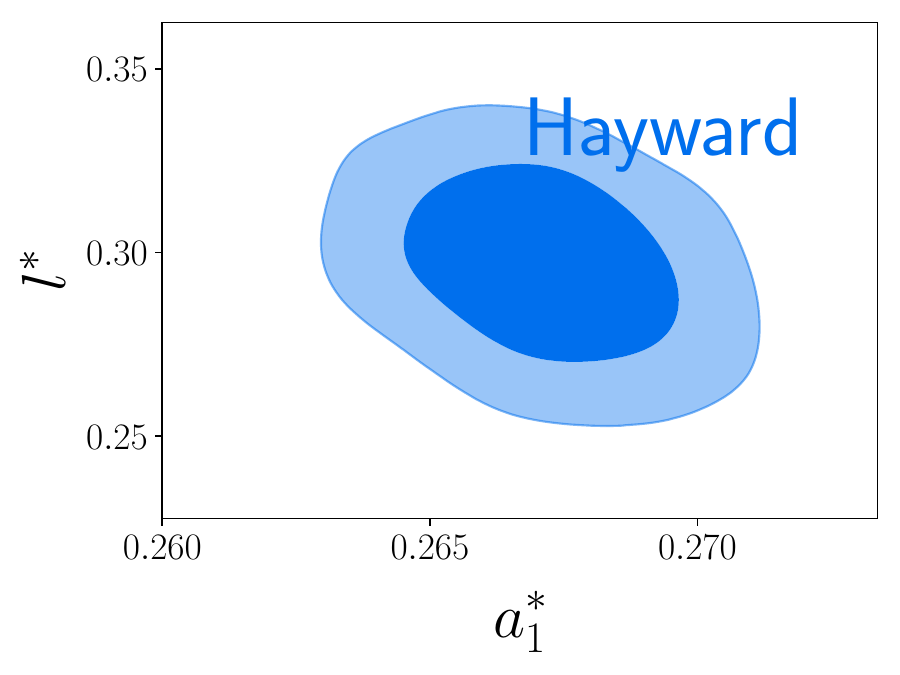}
                \includegraphics[width=2.6cm]{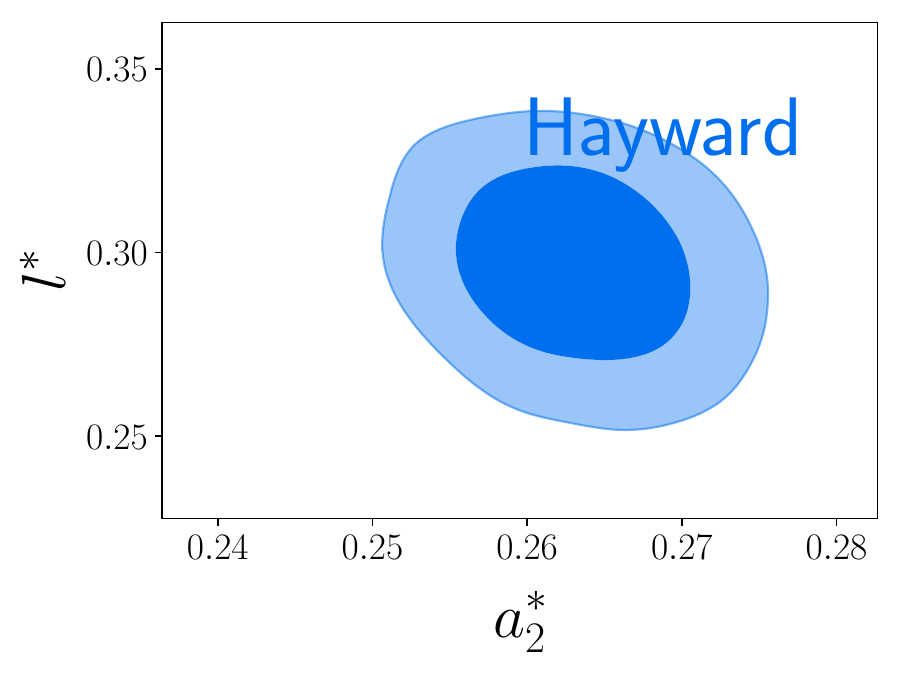}
                \includegraphics[width=2.6cm]{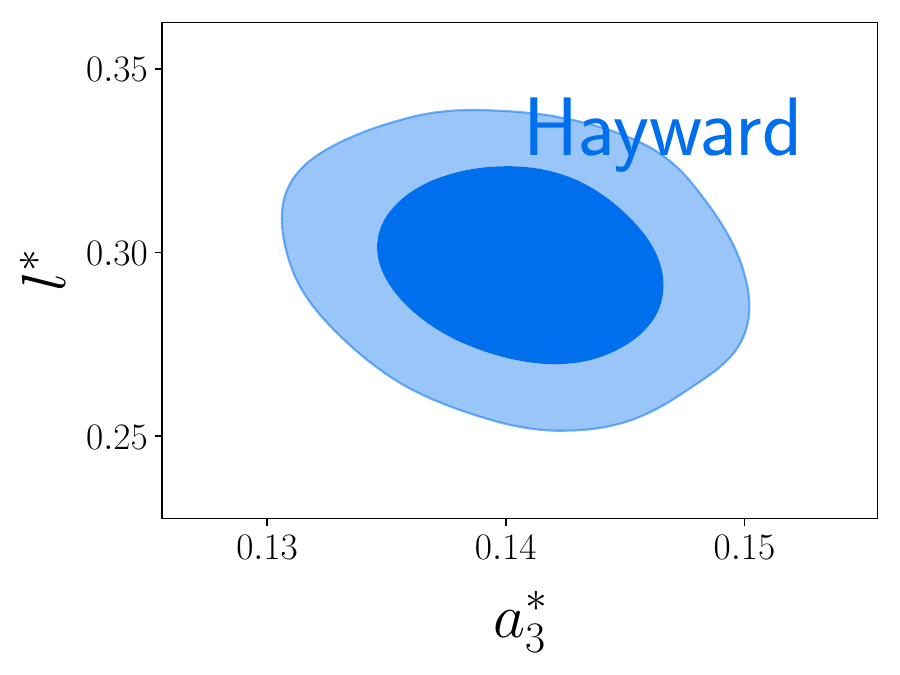}
            \end{minipage}
        }
        & 
        \subfloat[]{
            \begin{minipage}[b]{0.45\linewidth}
                \centering
                \includegraphics[width=2.6cm]{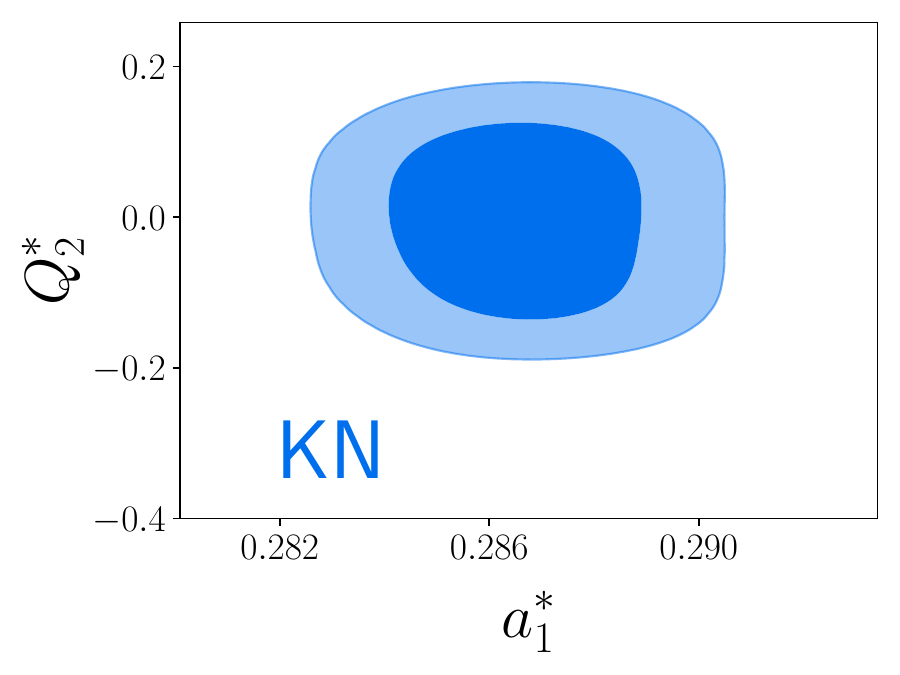}
                \includegraphics[width=2.6cm]{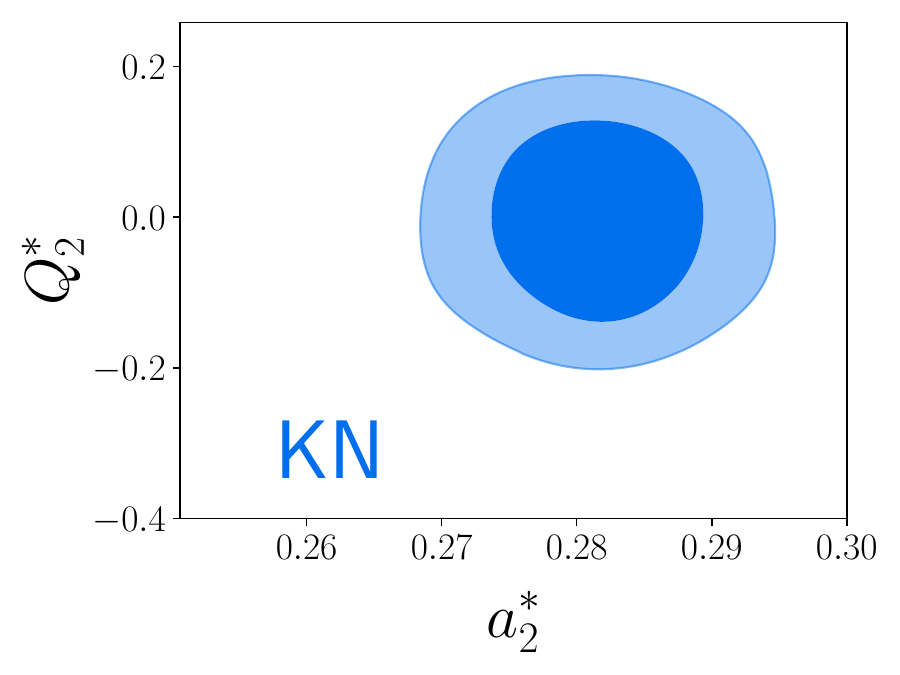}
                \includegraphics[width=2.6cm]{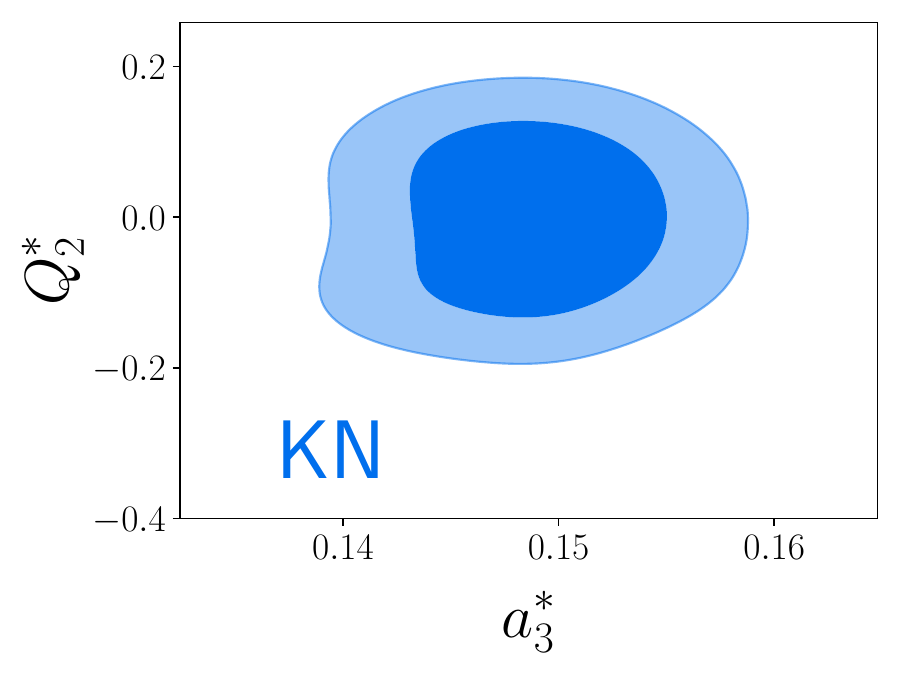}
            \end{minipage}
        }
        \\[1pt]
        \subfloat[]{
            \begin{minipage}[b]{0.45\linewidth}
                \centering
                \includegraphics[width=2.6cm]{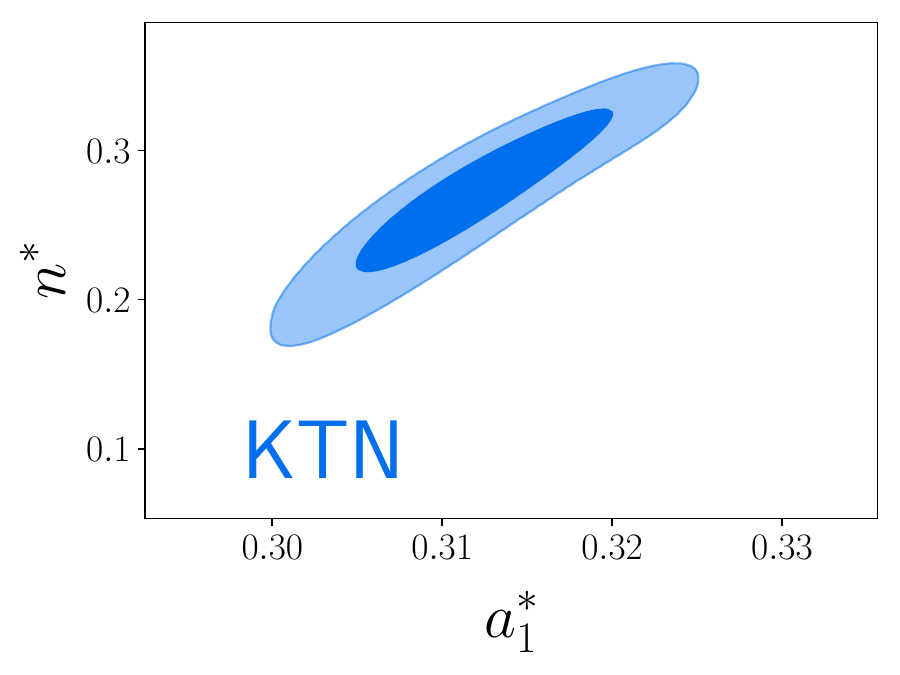}
                \includegraphics[width=2.6cm]{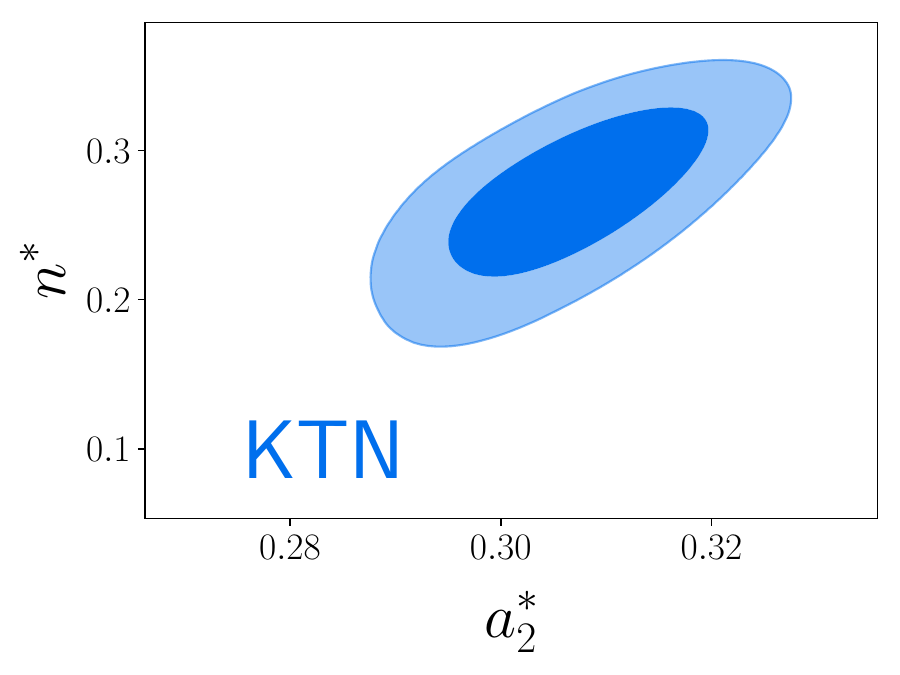}
                \includegraphics[width=2.6cm]{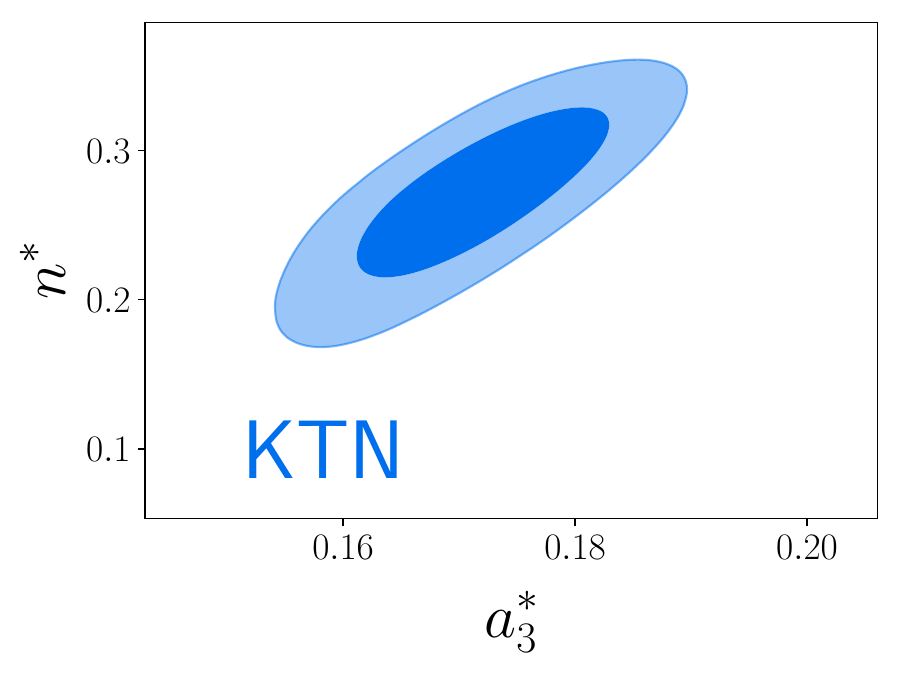}
            \end{minipage}
        }
        & 
        \subfloat[]{
            \begin{minipage}[b]{0.45\linewidth}
                \centering
                \includegraphics[width=2.6cm]{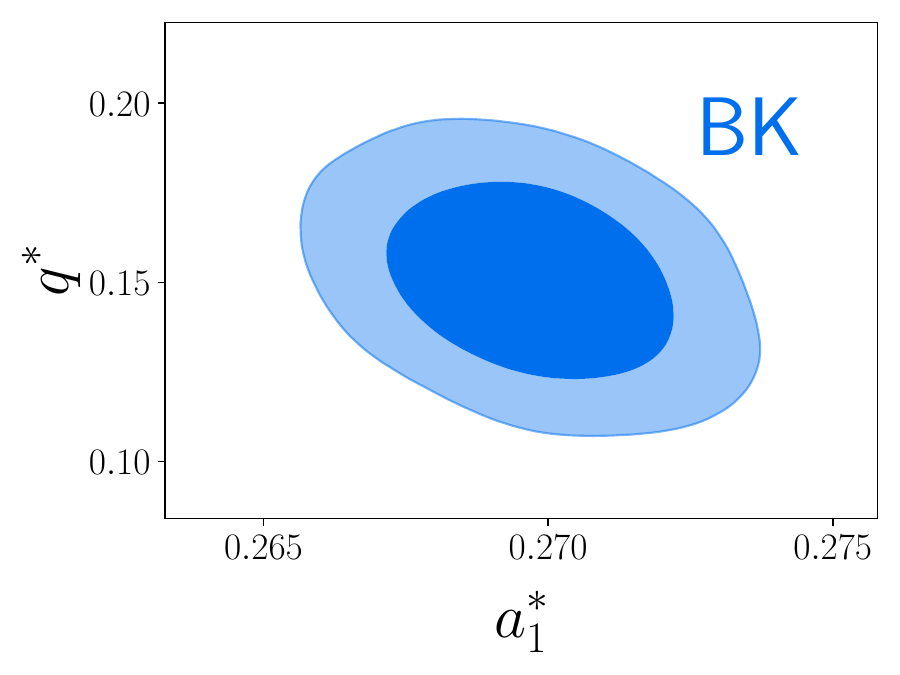}
                \includegraphics[width=2.6cm]{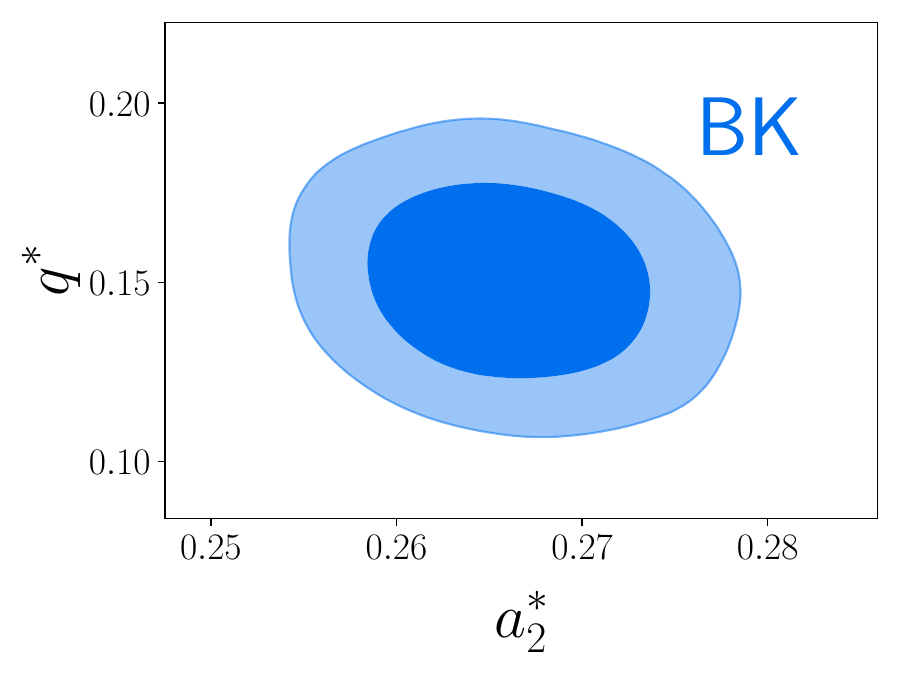}
                \includegraphics[width=2.6cm]{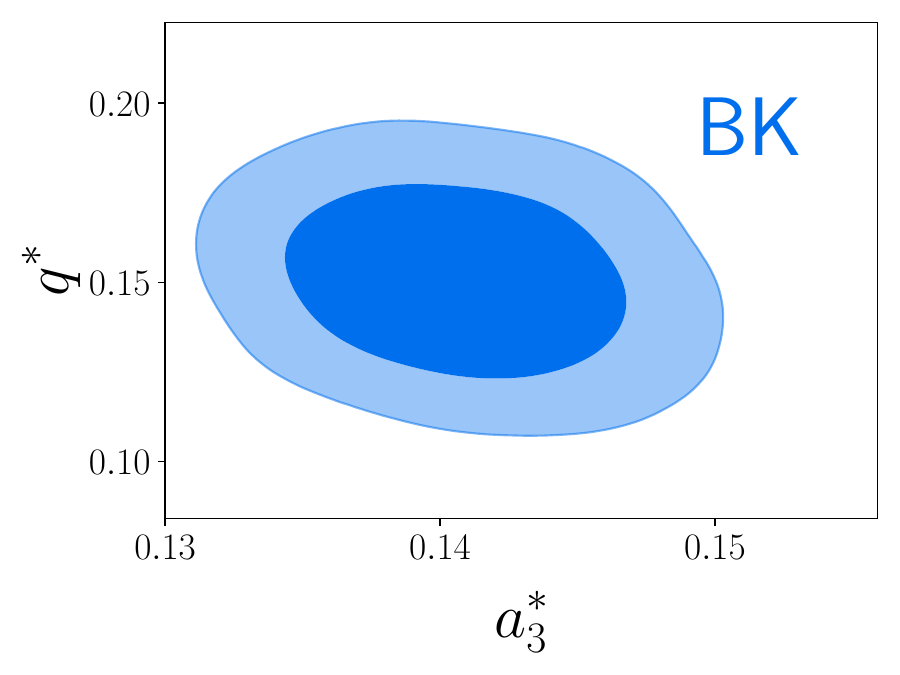}
            \end{minipage}
        }
        \\[4pt]
        \subfloat[]{
            \begin{minipage}[b]{0.45\linewidth}
                \centering
                \includegraphics[width=2.6cm]{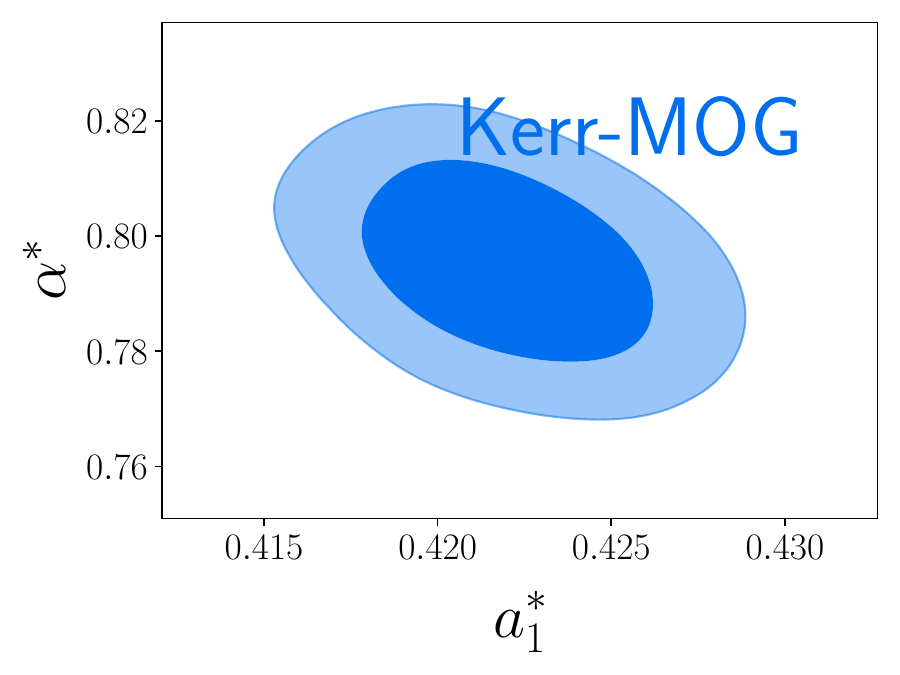}
                \includegraphics[width=2.6cm]{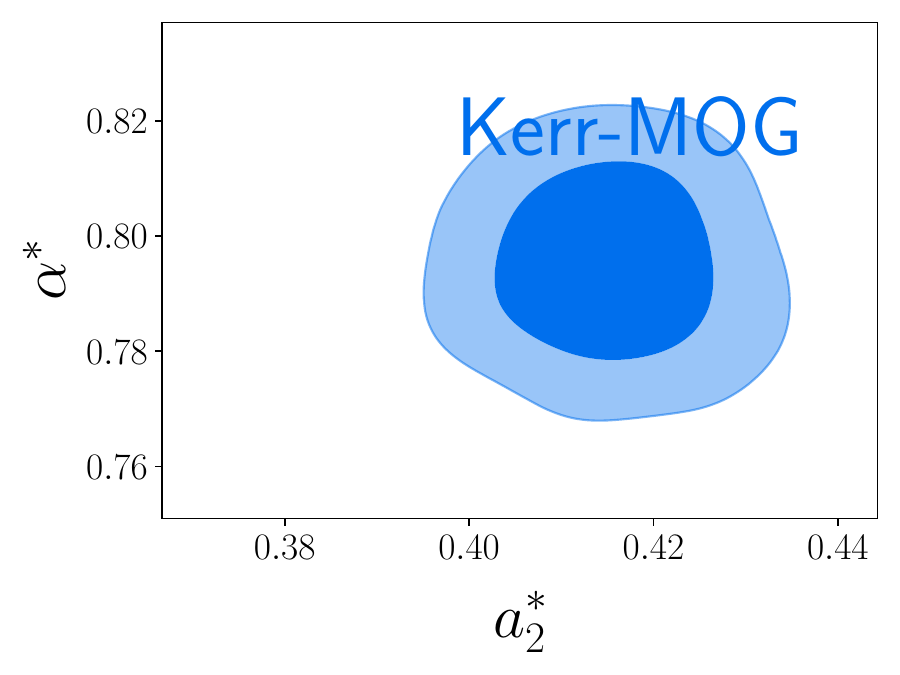}
                \includegraphics[width=2.6cm]{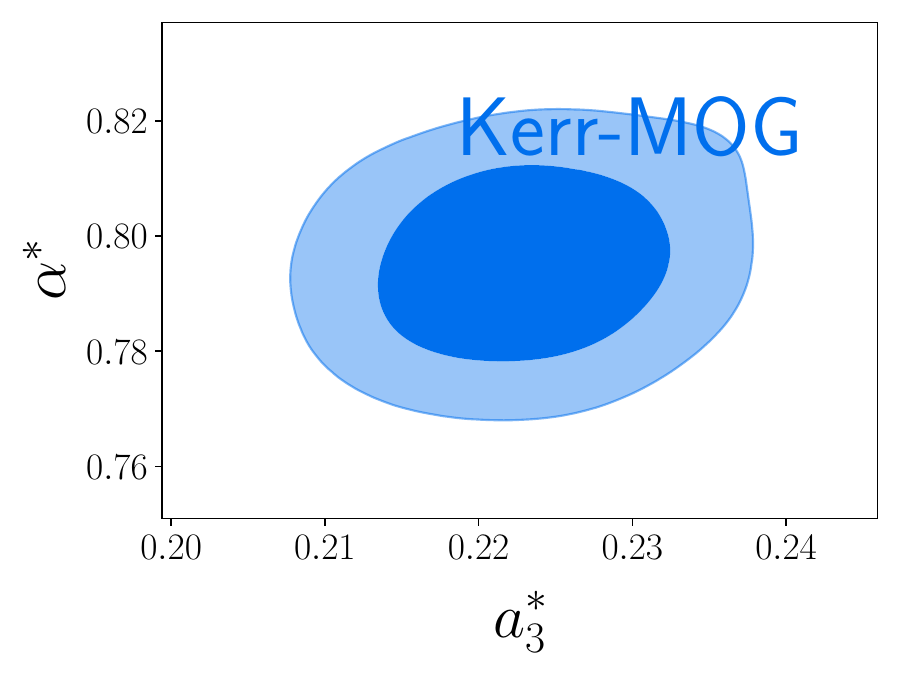}
            \end{minipage}
        }
        & 
        \subfloat[]{
            \begin{minipage}[b]{0.45\linewidth}
                \centering
                \includegraphics[width=2.6cm]{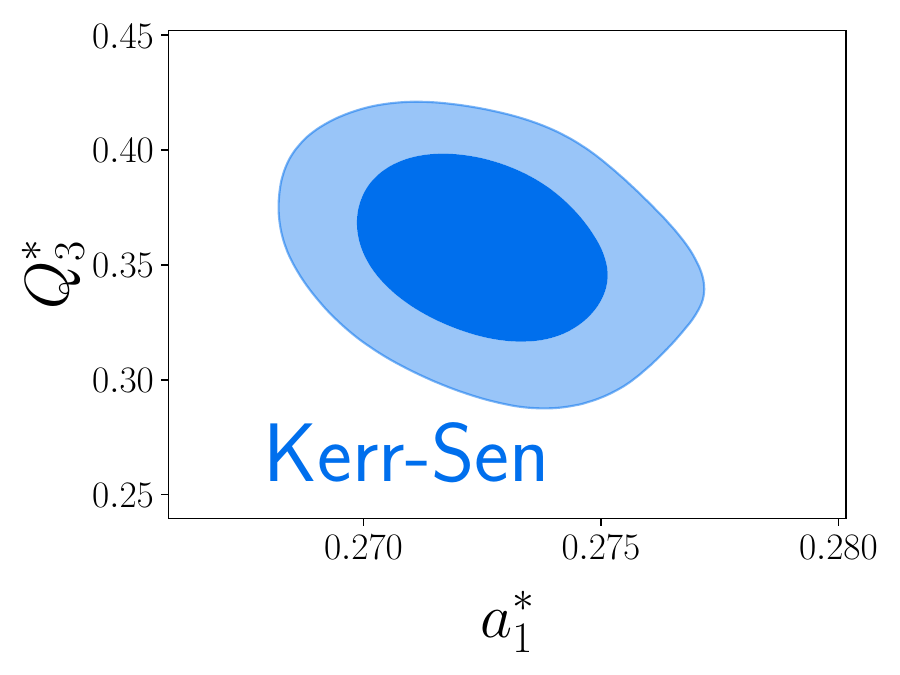}
                \includegraphics[width=2.6cm]{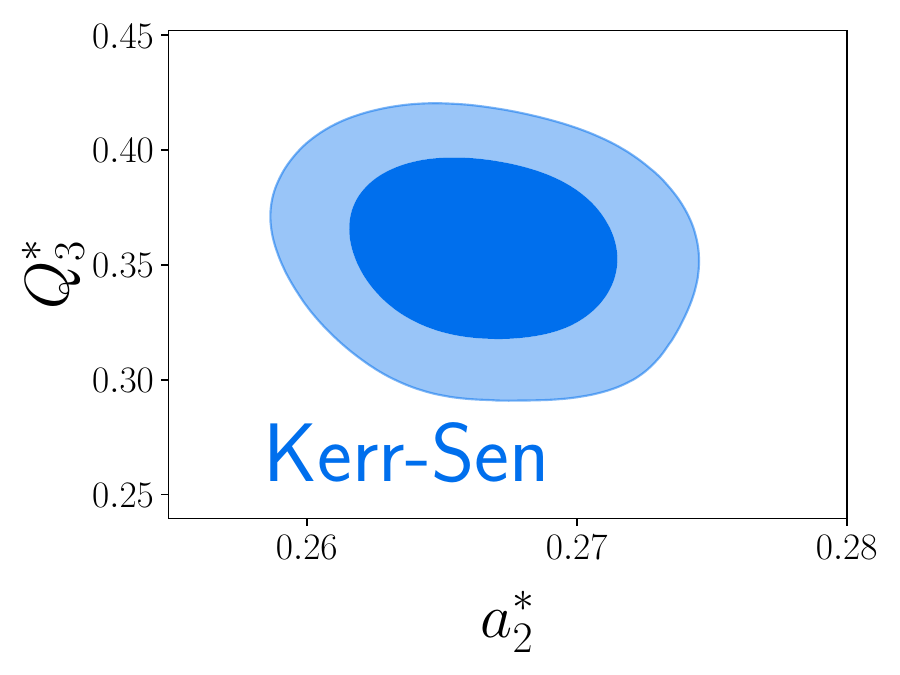}
                \includegraphics[width=2.6cm]{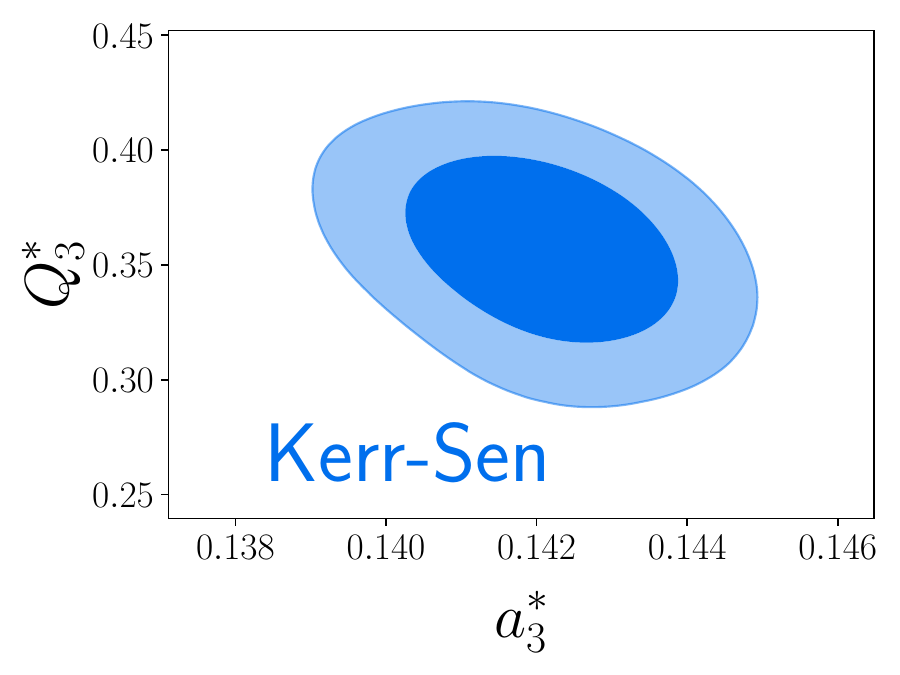}
            \end{minipage}
        }
        \\[1pt]
        \multicolumn{2}{c}{
            \subfloat[]{
                \begin{minipage}[b]{0.45\linewidth}
                    \centering
                    \includegraphics[width=2.6cm]{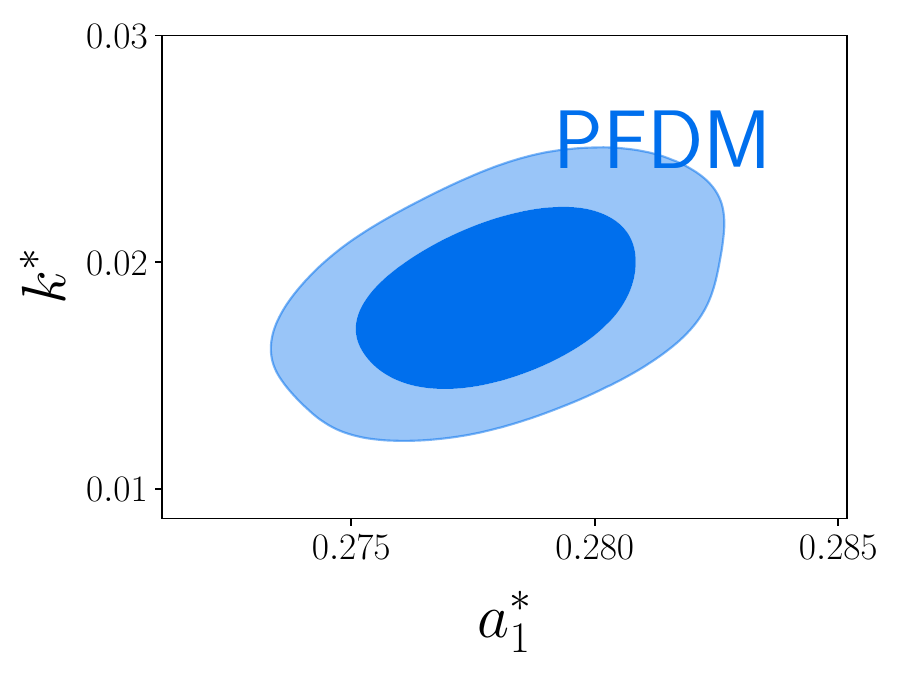}
                    \includegraphics[width=2.6cm]{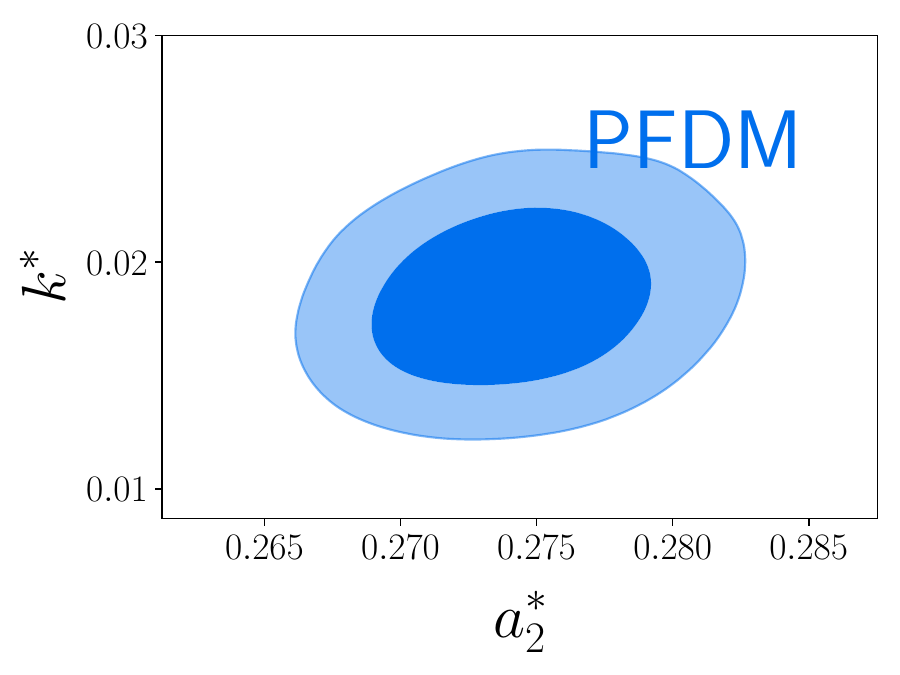}
                    \includegraphics[width=2.6cm]{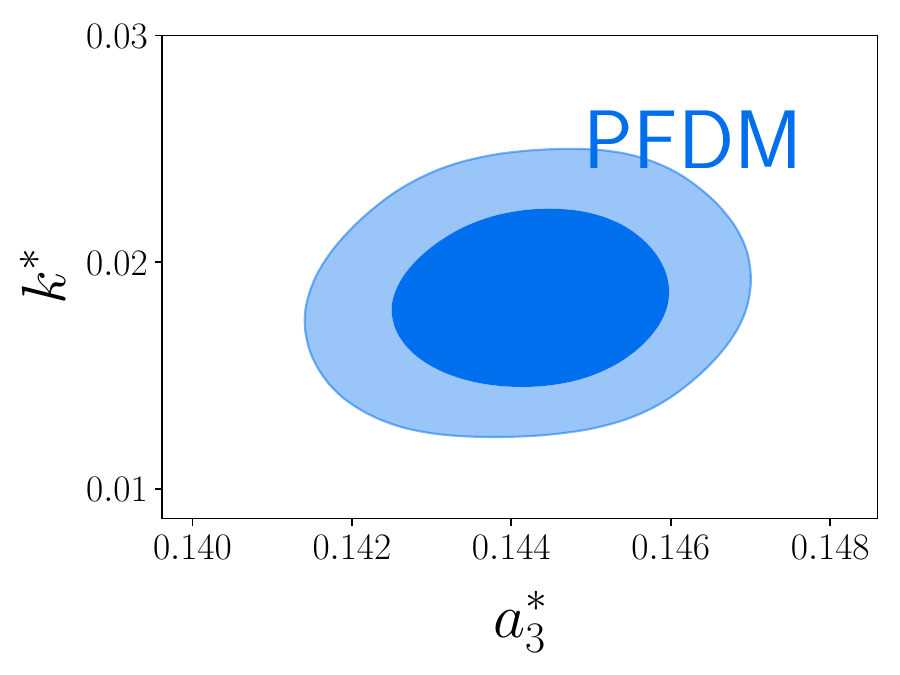}
                \end{minipage}
            }
        }
    \end{tabular}

    \caption{The parameter plots of the modification parameters and the spin parameters $a_p^*$ of the three microquasars within the $68\%$ and $95\%$ CL under the single-parameter modified spacetimes of Kerr.}
    \label{fig:3}
\end{figure}

\begin{figure}[htbp]
    \renewcommand{\thefigure}{4}
    \centering
    \setlength{\tabcolsep}{2pt}

    \begin{tabular}{cc}
        \subfloat[]{
            \begin{minipage}[b]{0.45\linewidth}
                \centering
                \includegraphics[width=2.6cm]{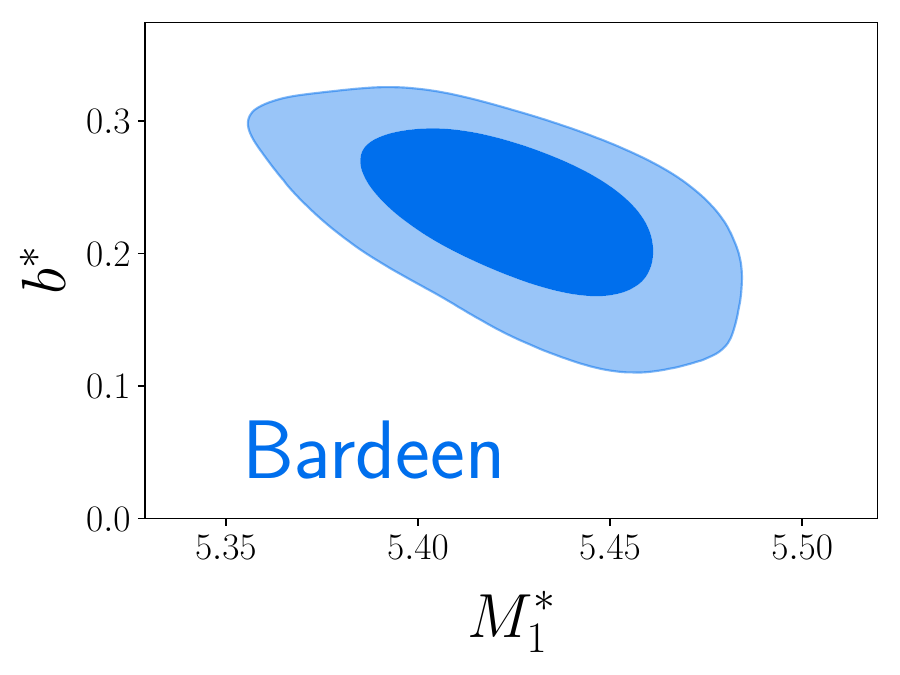}
                \includegraphics[width=2.6cm]{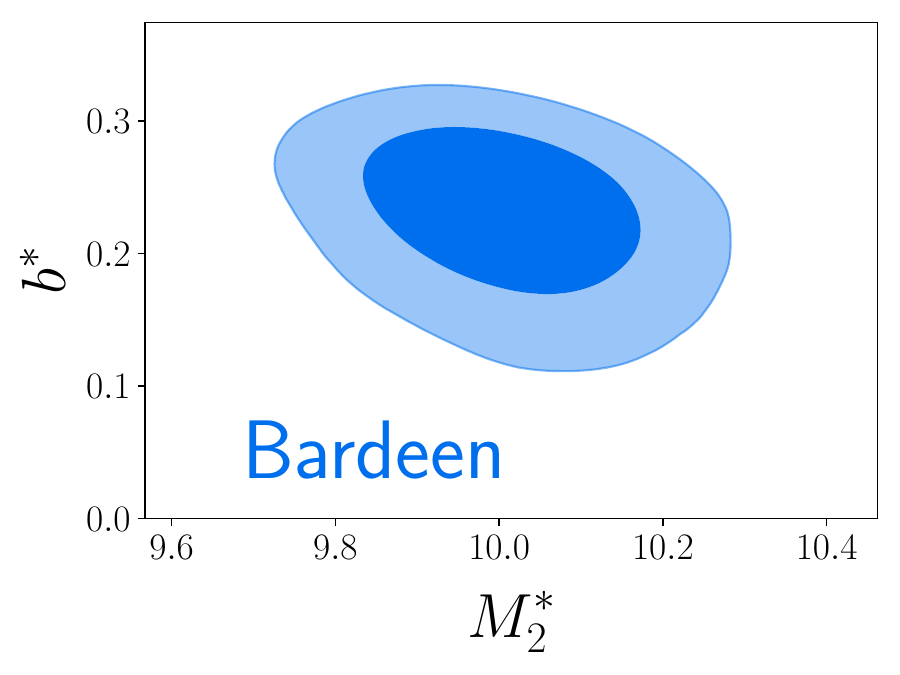}
                \includegraphics[width=2.6cm]{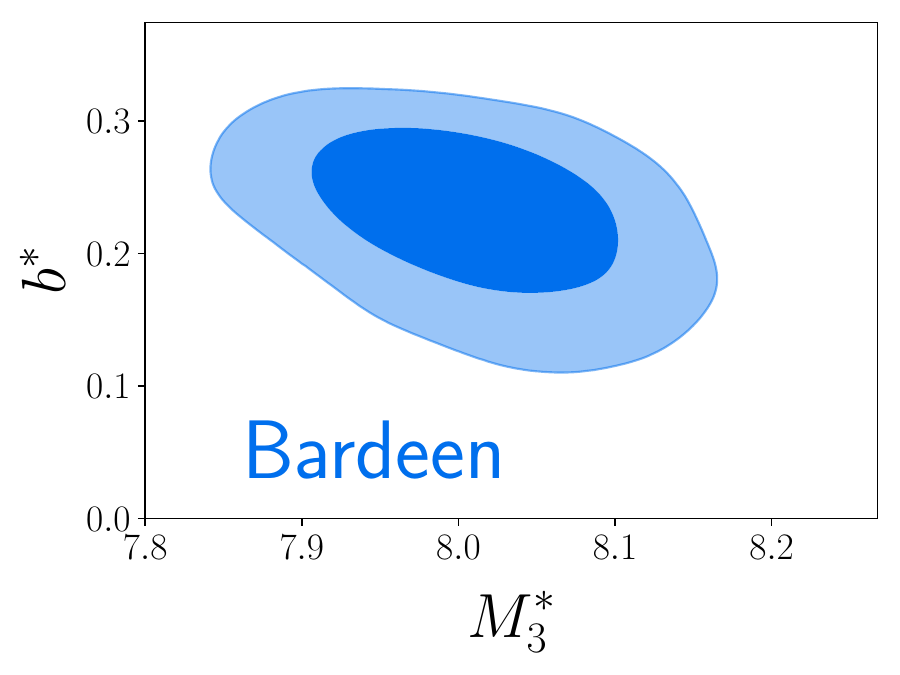}
            \end{minipage}
        }
        &
        \subfloat[]{
            \begin{minipage}[b]{0.45\linewidth}
                \centering
                \includegraphics[width=2.6cm]{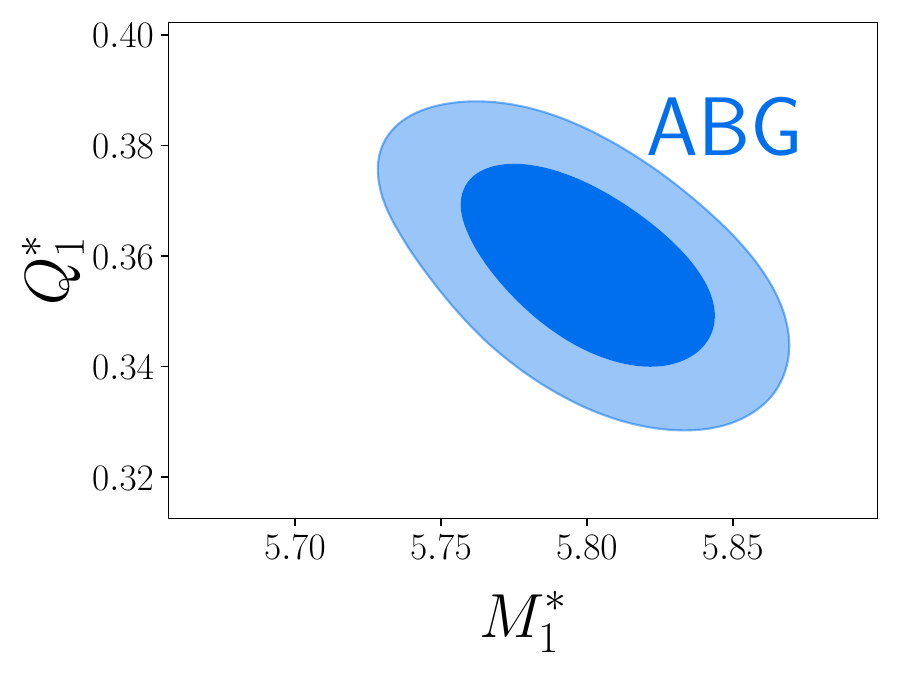}
                \includegraphics[width=2.6cm]{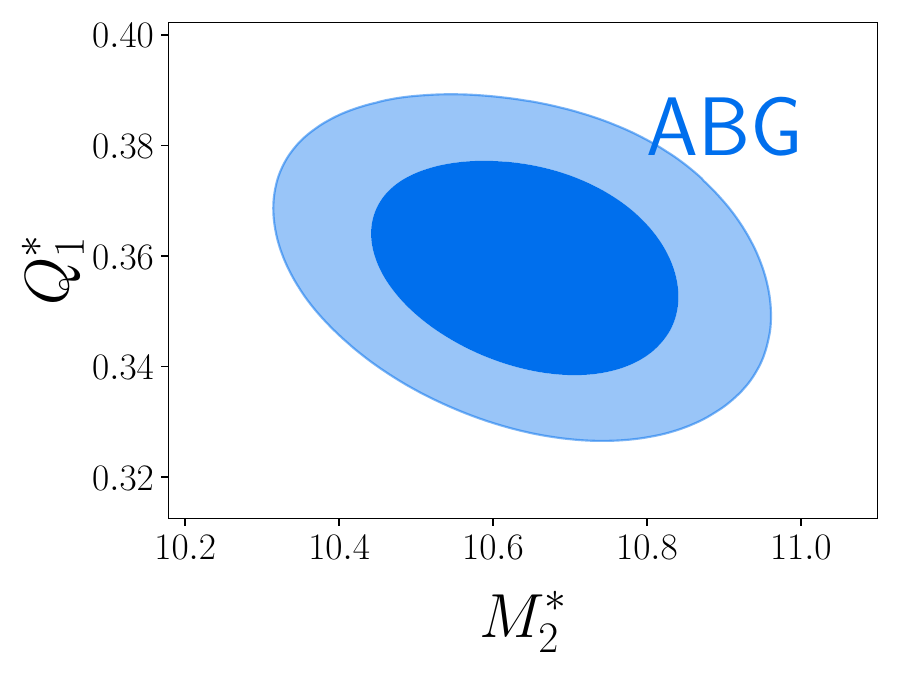}
                \includegraphics[width=2.6cm]{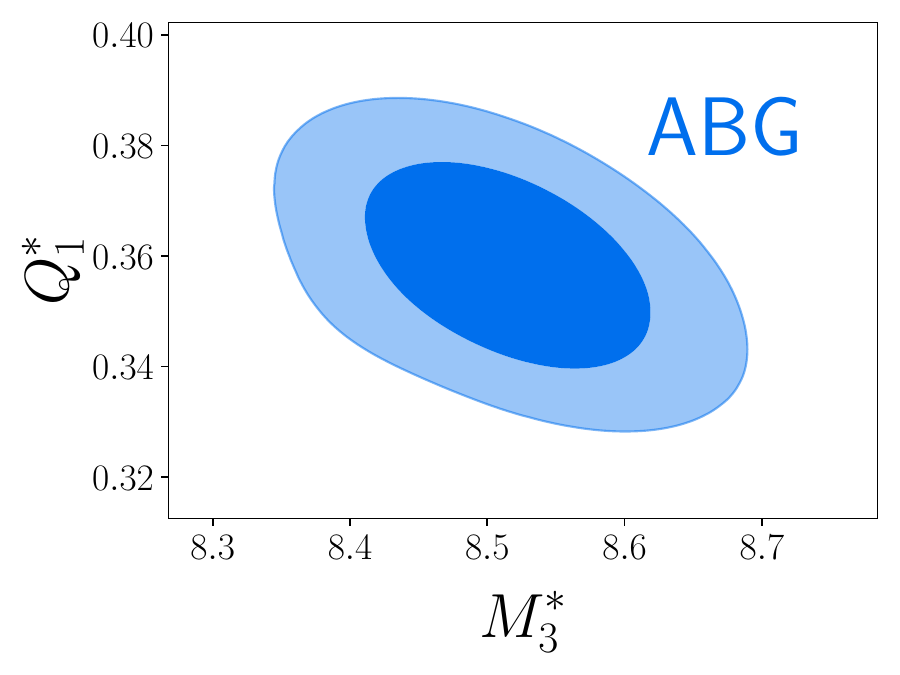}
            \end{minipage}
        }
        \\[1pt]
        \subfloat[]{
            \begin{minipage}[b]{0.45\linewidth}
                \centering
                \includegraphics[width=2.6cm]{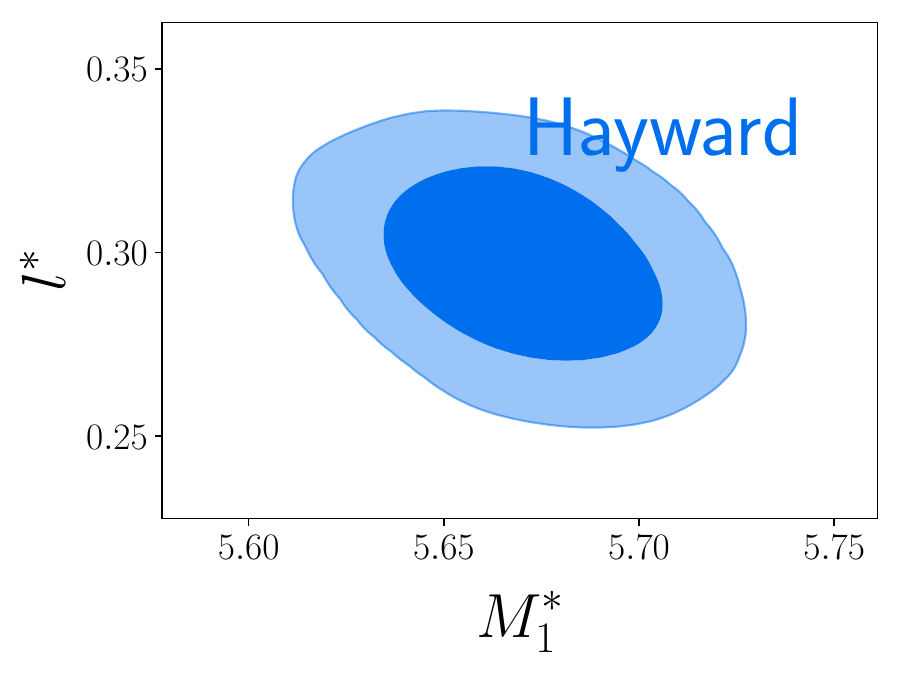}
                \includegraphics[width=2.6cm]{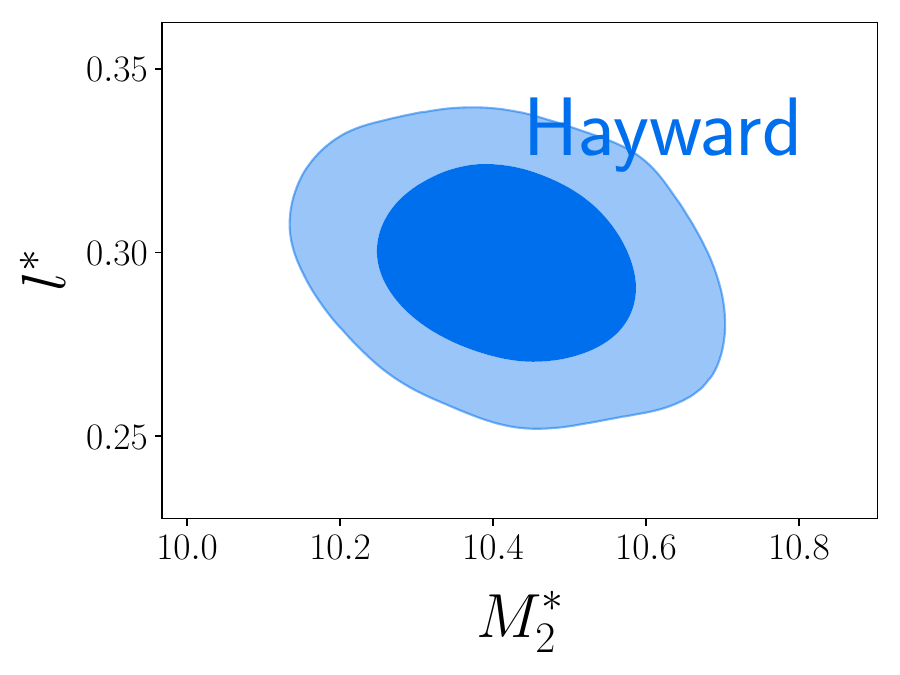}
                \includegraphics[width=2.6cm]{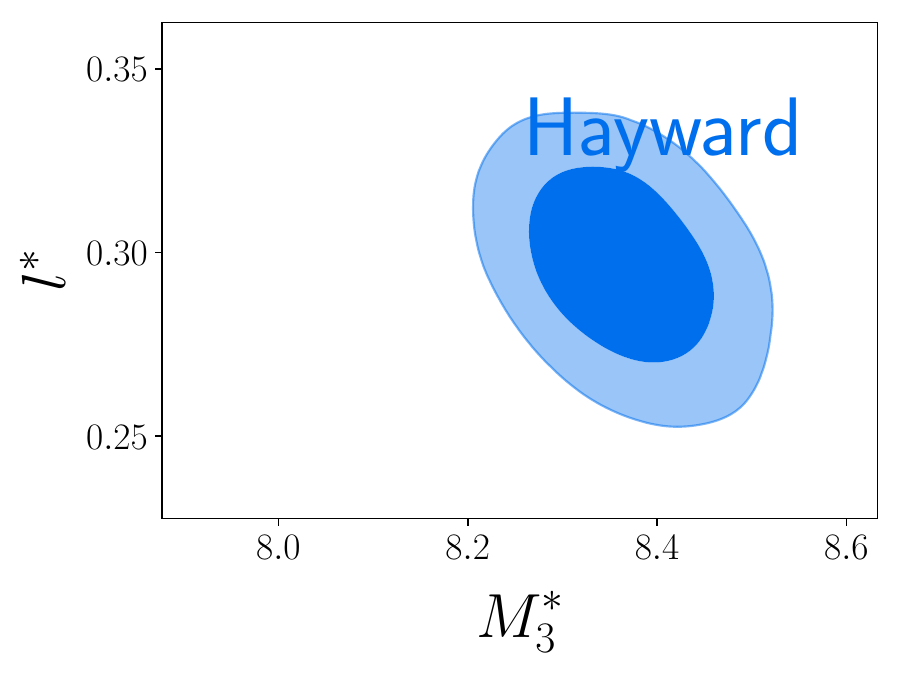}
            \end{minipage}
        }
        &
        \subfloat[]{
            \begin{minipage}[b]{0.45\linewidth}
                \centering
                \includegraphics[width=2.6cm]{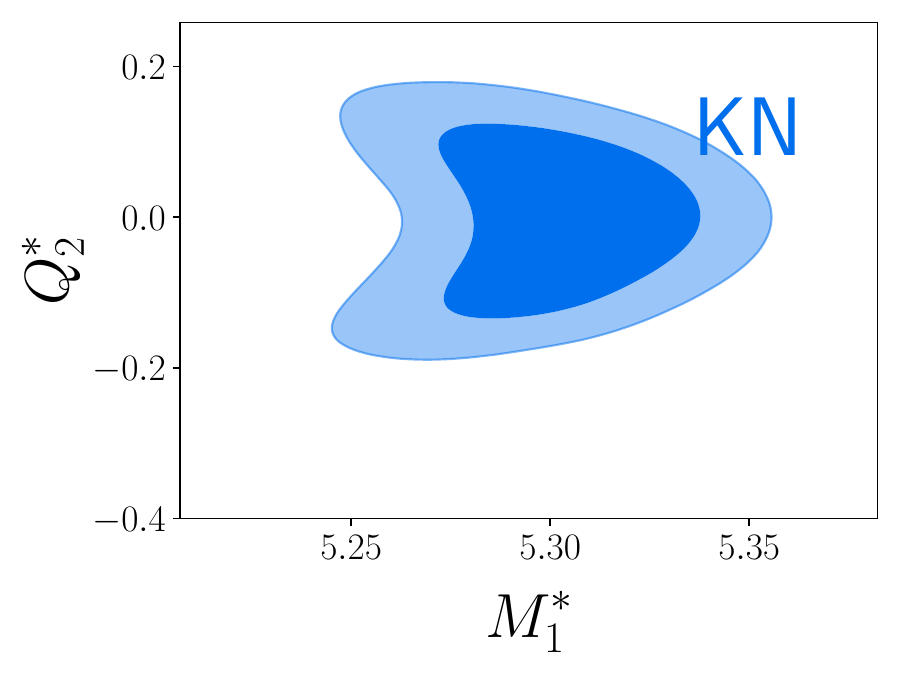}
                \includegraphics[width=2.6cm]{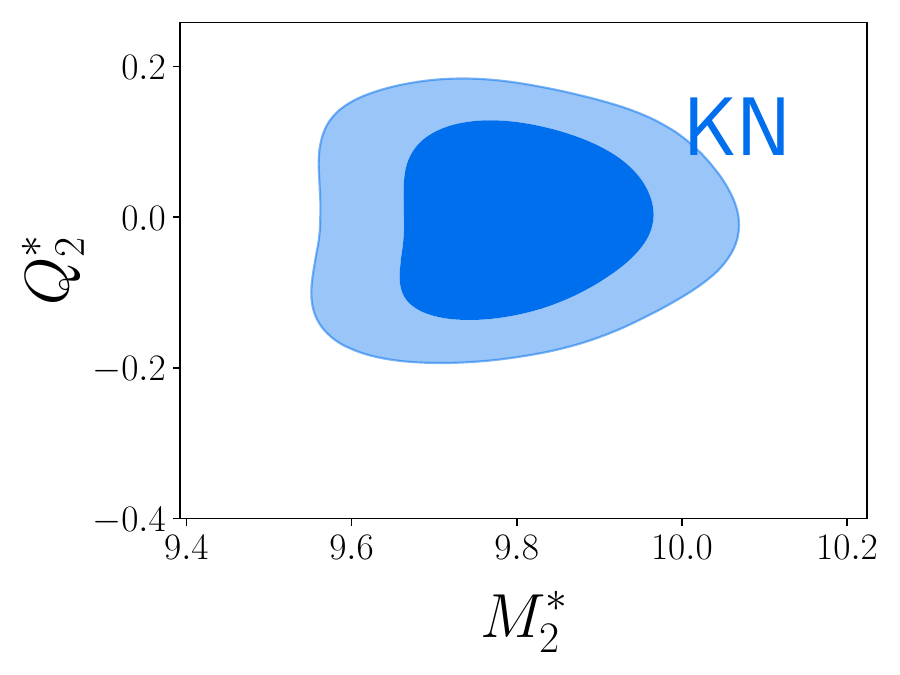}
                \includegraphics[width=2.6cm]{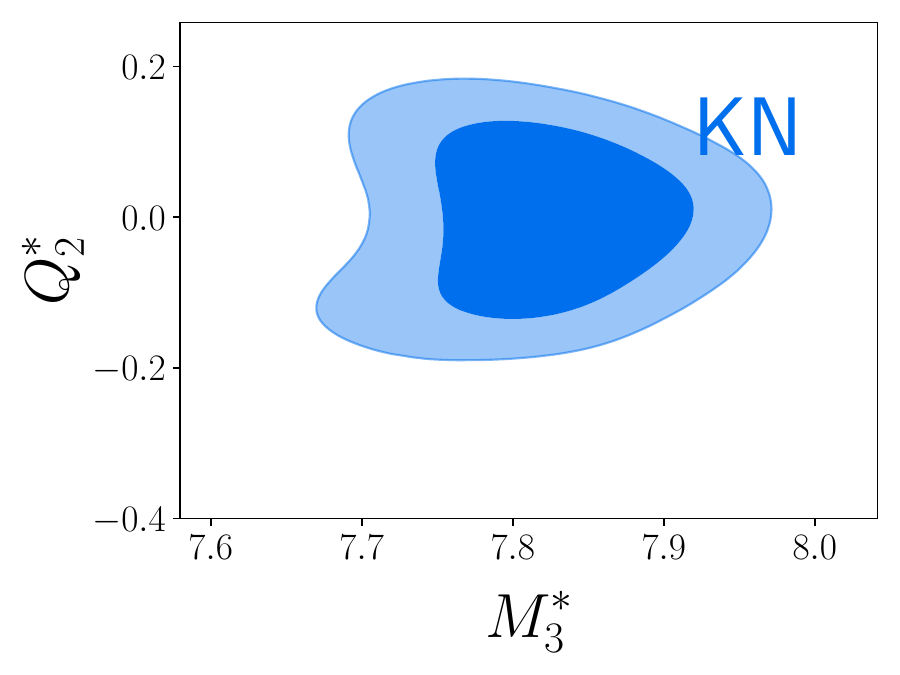}
            \end{minipage}
        }
        \\[1pt]
        \subfloat[]{
            \begin{minipage}[b]{0.45\linewidth}
                \centering
                \includegraphics[width=2.6cm]{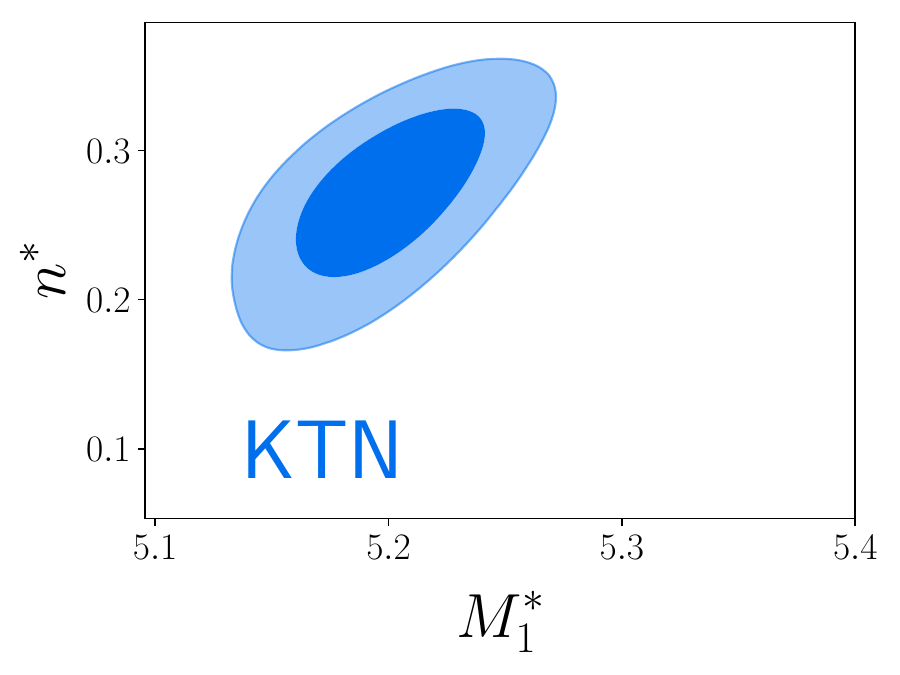}
                \includegraphics[width=2.6cm]{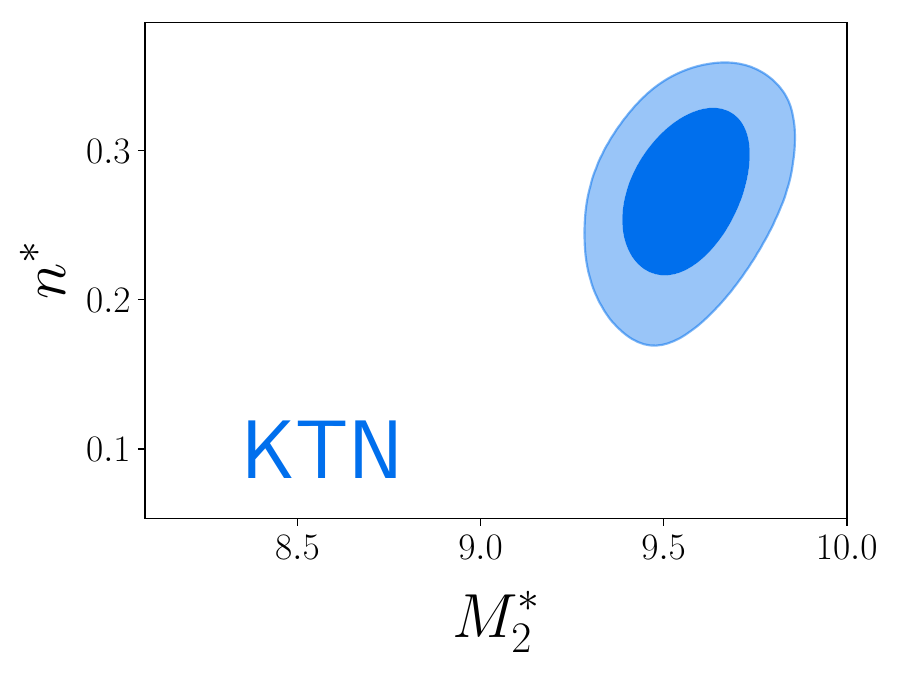}
                \includegraphics[width=2.6cm]{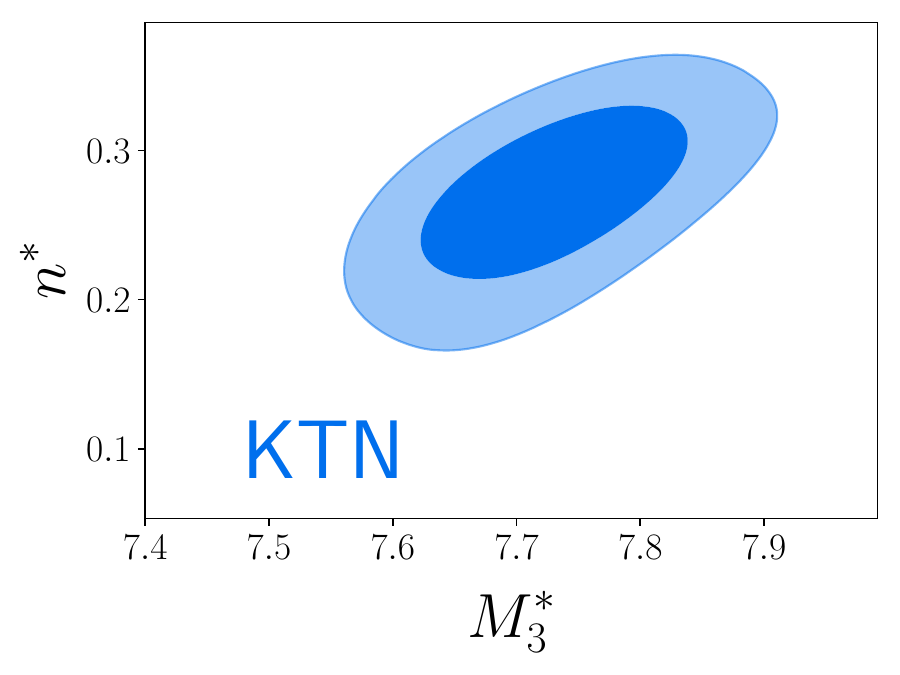}
            \end{minipage}
        }
        &
        \subfloat[]{
            \begin{minipage}[b]{0.45\linewidth}
                \centering
                \includegraphics[width=2.6cm]{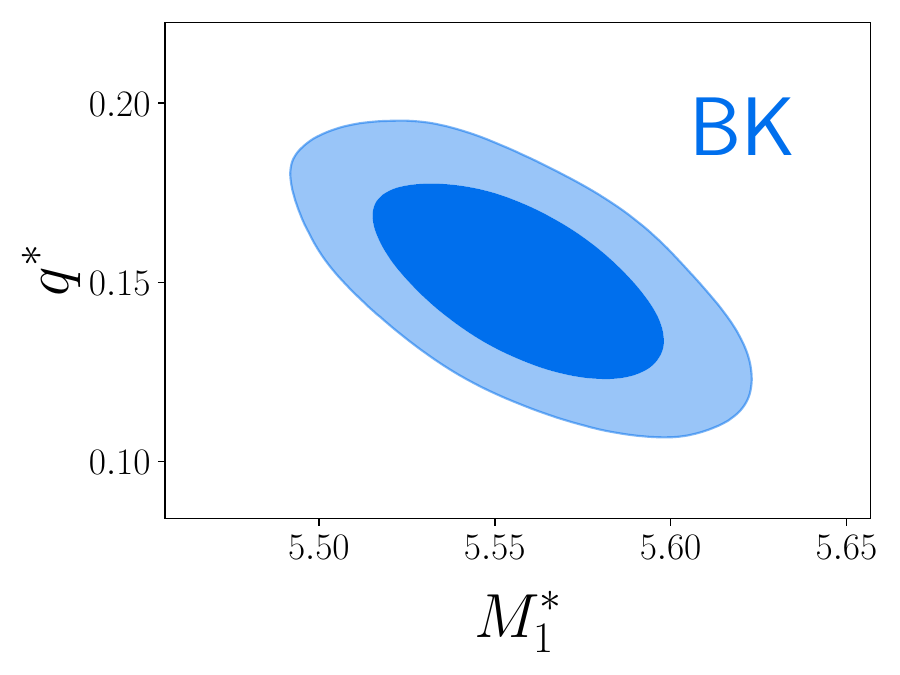}
                \includegraphics[width=2.6cm]{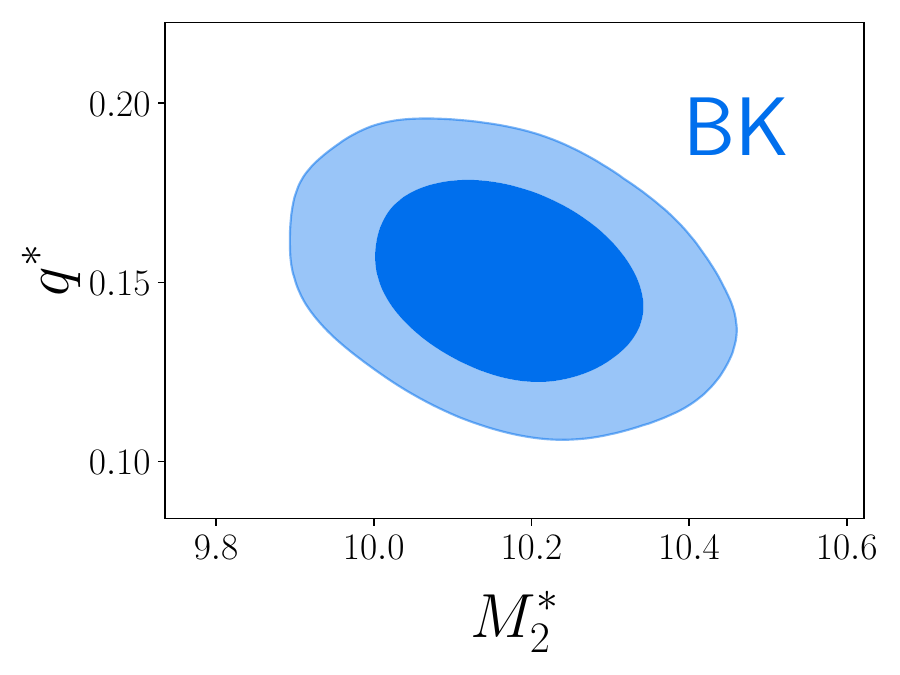}
                \includegraphics[width=2.6cm]{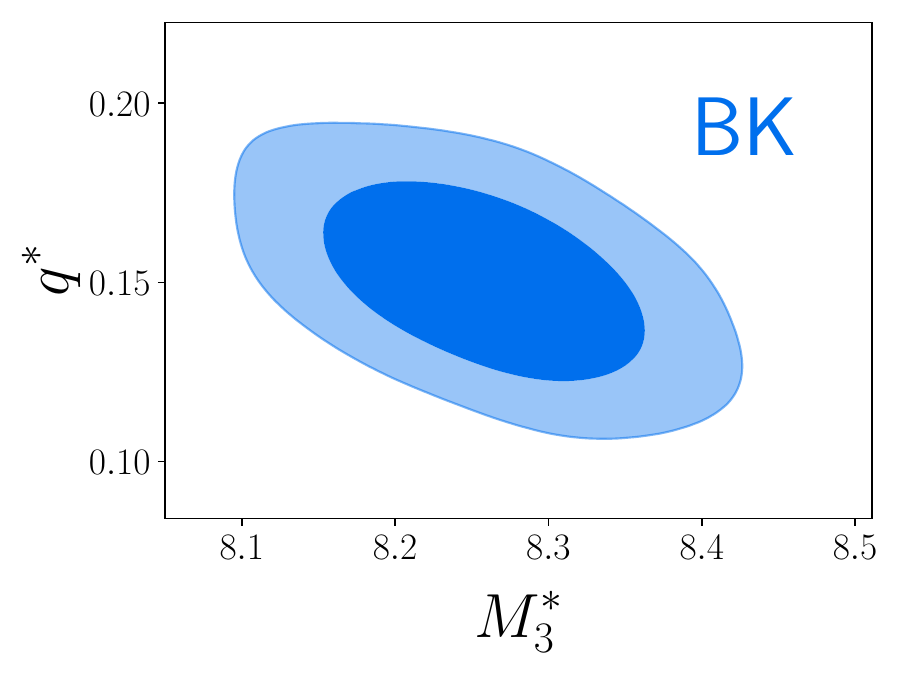}
            \end{minipage}
        }
        \\[1pt]
        \subfloat[]{
            \begin{minipage}[b]{0.45\linewidth}
                \centering
                \includegraphics[width=2.6cm]{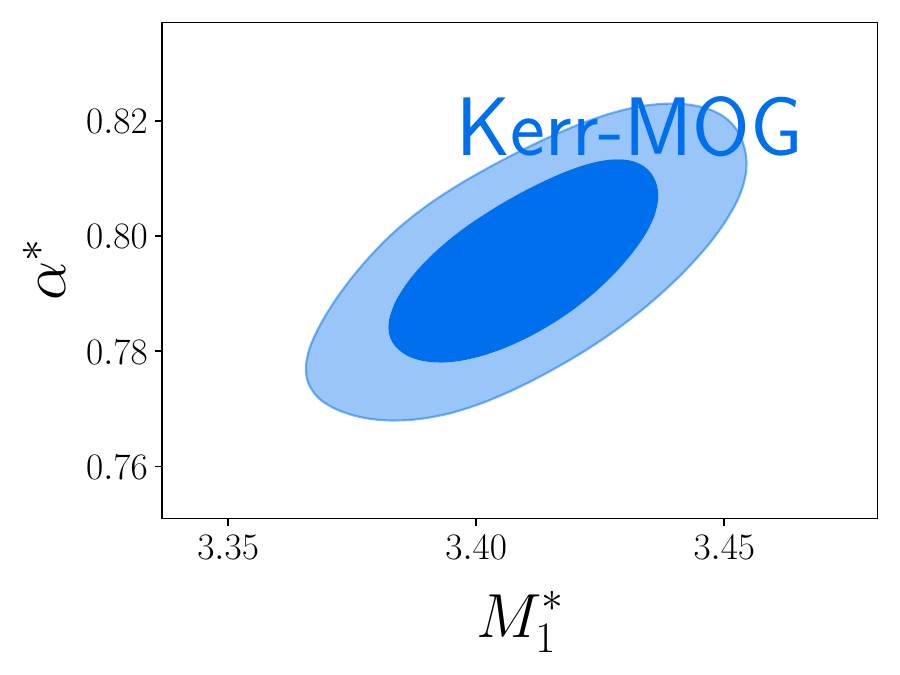}
                \includegraphics[width=2.6cm]{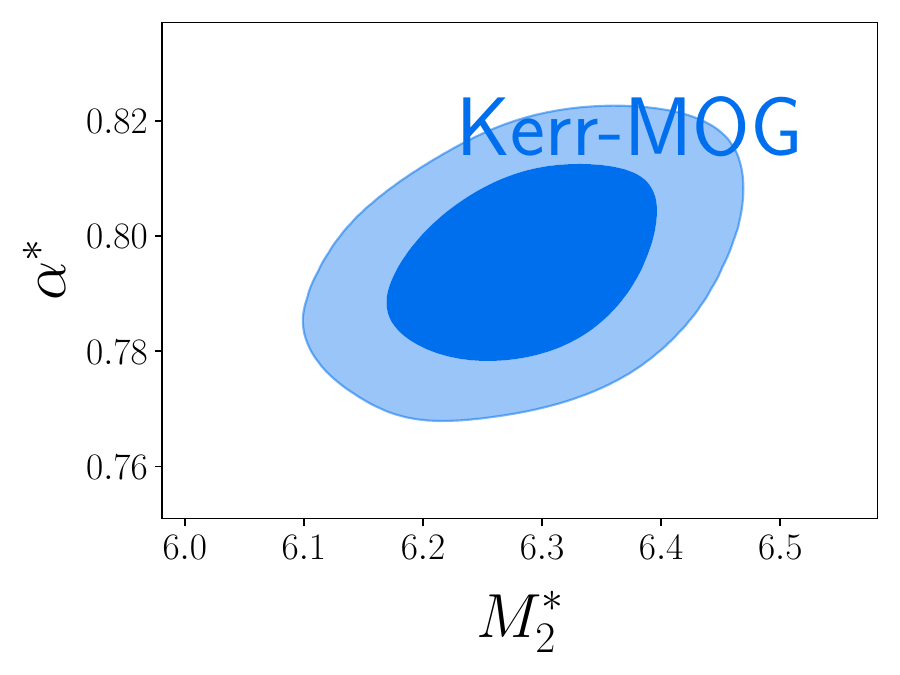}
                \includegraphics[width=2.6cm]{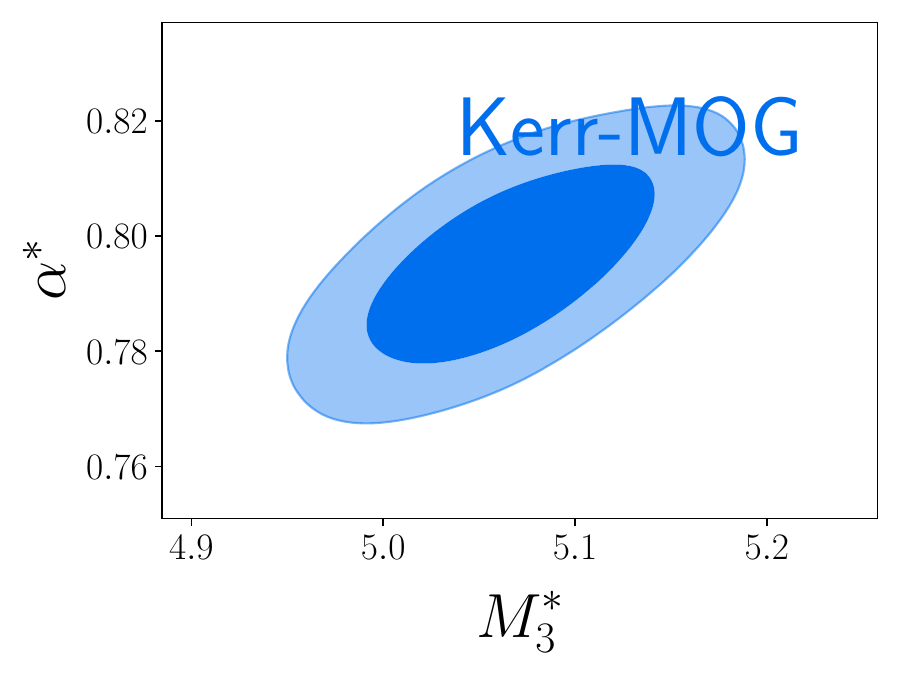}
            \end{minipage}
        }
        &
        \subfloat[]{
            \begin{minipage}[b]{0.45\linewidth}
                \centering
                \includegraphics[width=2.6cm]{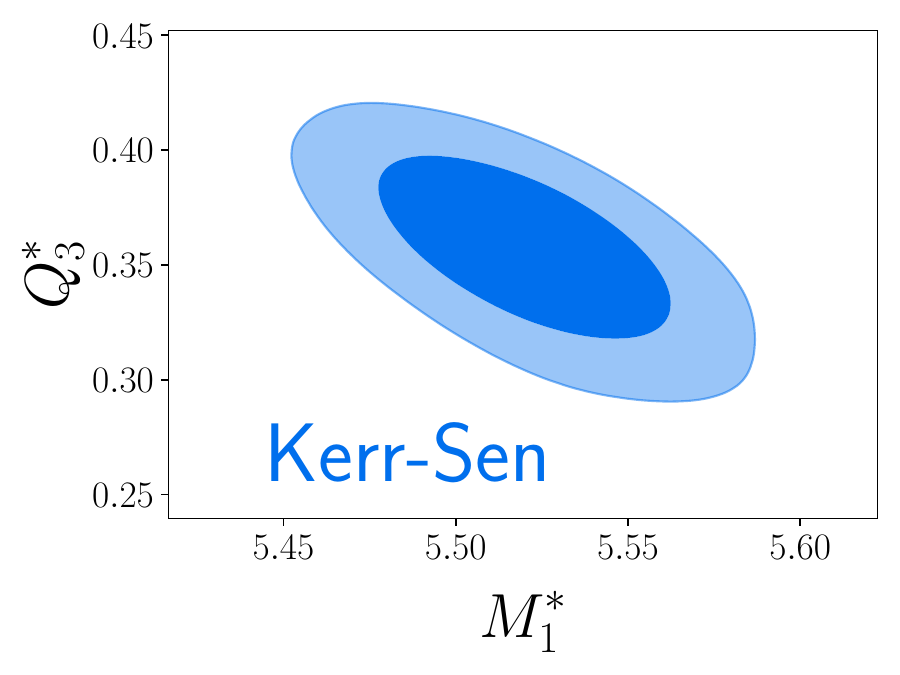}
                \includegraphics[width=2.6cm]{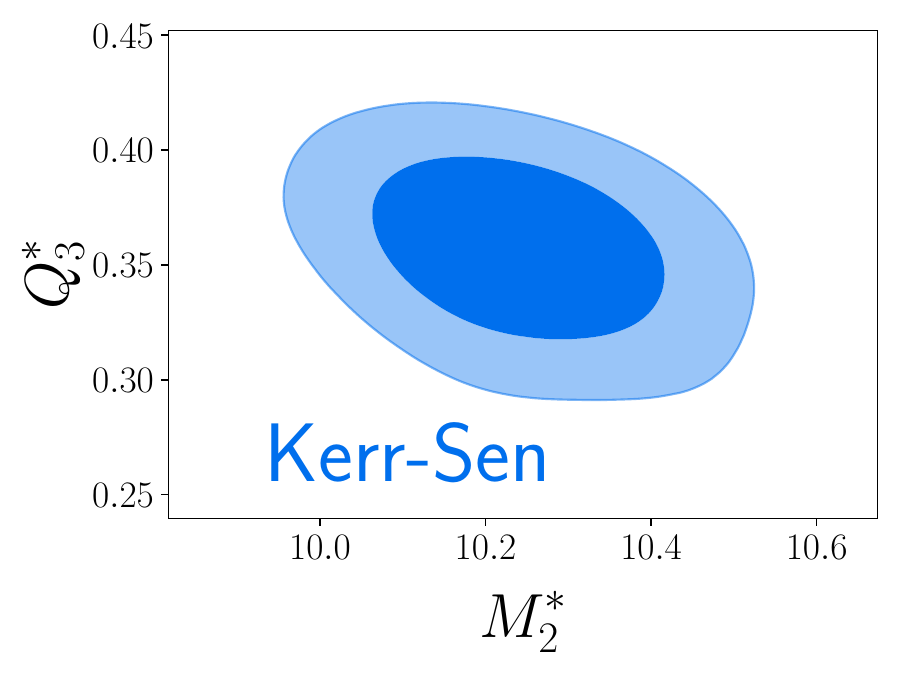}
                \includegraphics[width=2.6cm]{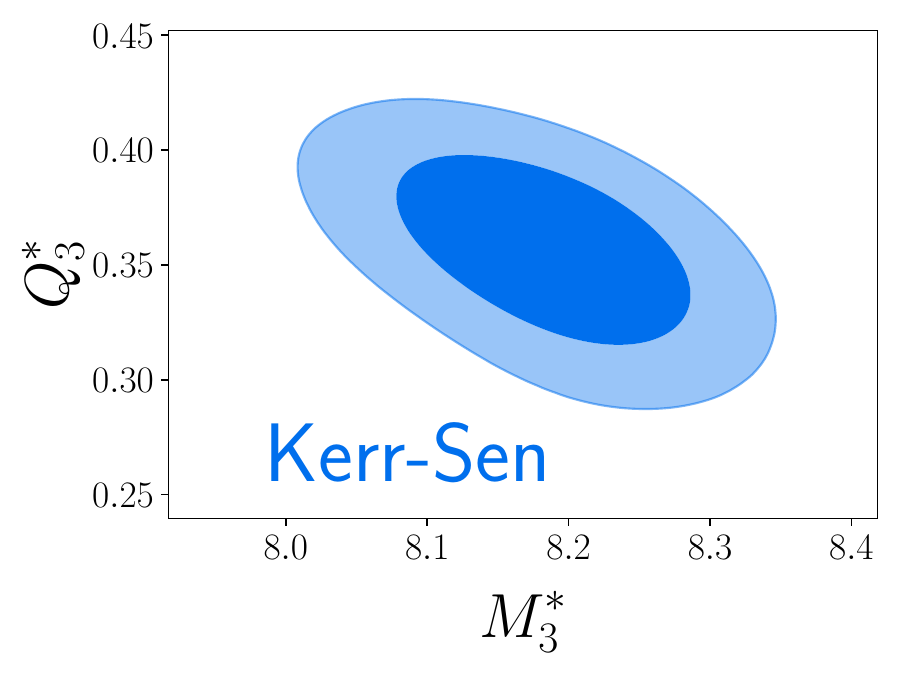}
            \end{minipage}
        }
        \\[1pt]
        \multicolumn{2}{c}{
            \subfloat[]{
                \begin{minipage}[b]{0.45\linewidth}
                    \centering
                    \includegraphics[width=2.6cm]{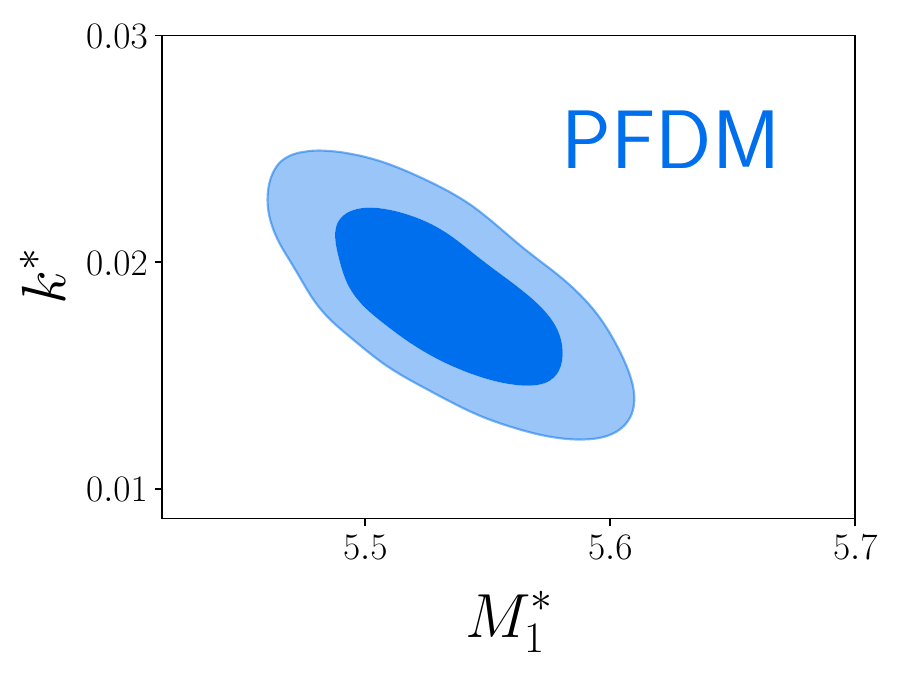}
                    \includegraphics[width=2.6cm]{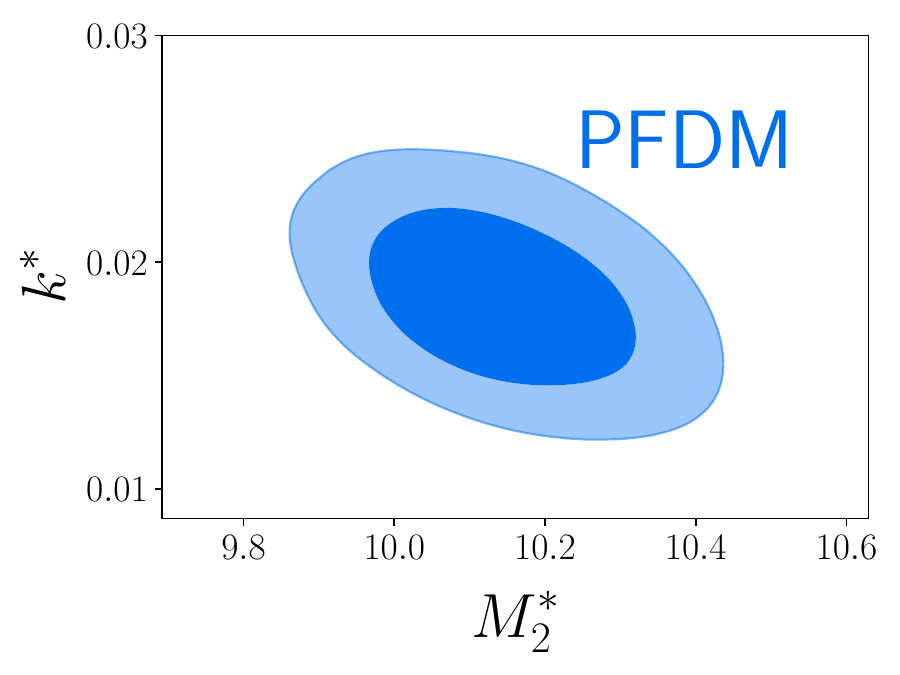}
                    \includegraphics[width=2.6cm]{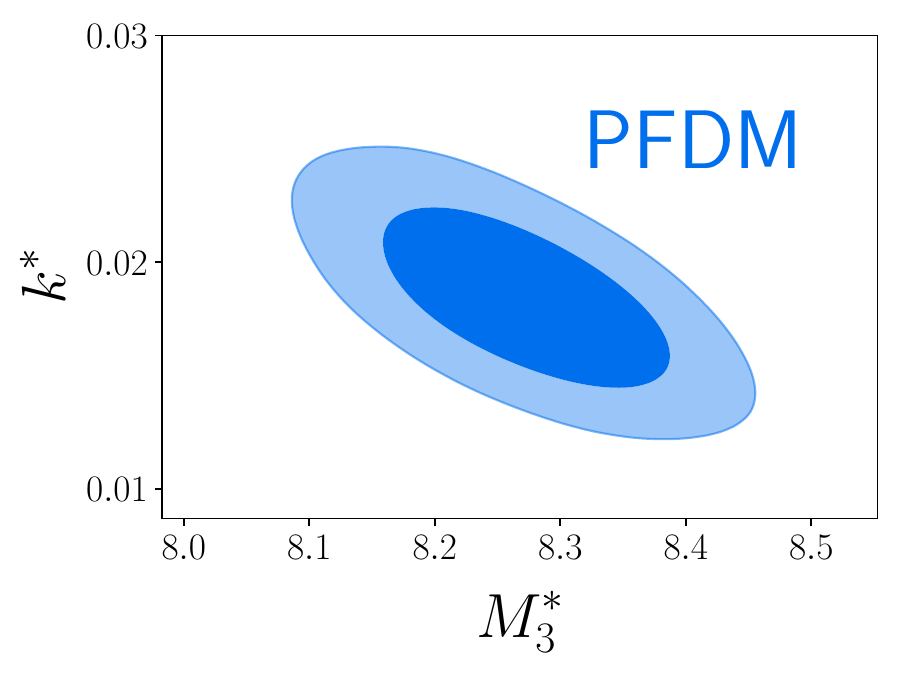}
                \end{minipage}
            }
        }
    \end{tabular}
    \caption{The parameter plots of the modification parameters and the spin parameters $M_p^*$ of the three microquasars within the $68 \%$ and $95 \%$ CL under the single-parameter modified spacetimes of Kerr.}
    \label{fig:4}
\end{figure}

According to the calculations listed in Table \ref{table-2}, we can find some interesting results on the constraints on modification parameters:
(i) The constraints on modification parameters $\Delta^*$ for all axisymmetric spacetimes investigated are strict by using the QPO data, i.e.  at the $68 \%$ CL they have the relative  narrow confidence intervals;
(ii) At the $68 \%$ CL, only the constraint on modification parameter $Q_2^*$ in the KN spacetime admits both negative and positive modifications  (where zero recovers the Kerr solution), while the confidence ranges of modification parameters for other geometries require all strictly  positive;
(iii) According to the constraint results derived from QPO data, the small values of modification parameters are suggested in the KN and PFDM geometries ($Q_2^*=-0.004 \pm 0.078$ and $k^*=$ $0.019_{-0.003}^{+0.024}$). The nonzero constrained value of $k^*$ may provide a signature of dark matter, although its relatively small magnitude might suggest a weak effect;
(iv) The larger value of modification parameter is favored in Kerr-MOG geometry ($\alpha^*=0.795 \pm 0.011$), which indicates that the gravitational field of the Kerr-MOG black hole is stronger than that of the traditional Kerr solution, potentially exhibiting more significant observational features;
(v) The limit results on $\Delta^*$ in all   modified version of Kerr (except  KN model) do not include the value $\Delta^* = 0$ (corresponding to Kerr) at the $68 \%$ CL, which indicates that the investigation into modifications of the Kerr spacetime is valuable;
(vi) Although the modification parameters in the ABG, KN, and Kerr-Sen spacetimes are all related to electric charge, they influence geometry in different ways, which may lead to different constraint results on the parameters. Concretely, the constraints on $Q_1^*=0.360_{-0.011}^{+0.012}$ and $Q_3^*=0.361_{-0.024}^{+0.028}$ are similar, while they differ significantly from that of $Q_2^*=-0.004 \pm 0.078$;
(vii) The modification parameters in Bardeen and KTN geometries are interpreted as the effect originating from the magnetic monopole. It is shown in Table \ref{table-2}  that $b^*=0.229_{-0.034}^{+0.045}$ and $n^*=0.257_{-0.025}^{+0.054}$ exhibit closely matching values.

Additionally, the constraints on some modification parameters in the single-parameter modified spacetimes of Kerr have been discussed in several other studies.
For instance, in the study by \cite{Bardeen-2}, the authors used QPO observational data from microquasars (GRO J1655-40, XTE J1550-564, GRS 1915+105, H1743-322) and the $\mathrm{Sgr~A}^*$ data to constrain the Bardeen spacetime. The best fit, which minimizes $\chi^2$, was found for $b^* \cong 0.25$ under the RP model, which is consistent with our constraint: $b^*=0.229_{-0.034}^{+0.045}$;
In Ref. \cite{modified-Kerr-1}, the authors combined the LIGO-Virgo observational data of the GW150914 event with the KN spacetime and derived a constraint on the charge $Q_2^*$ within the $90 \%$ CL as $Q_2^*<0.33$. This result is consistent with our constraint on $Q_2^*$ in the KN spacetime, which is $-0.004\pm 0.078$;
In the study by \cite{modified-Kerr-2}, the authors combined the KTN spacetime with observational data, including the radiation efficiency and jet power of the microquasar GR0 J1655-40, to determine the NUT parameter $n^*$ range as $0<n^*<0.8$. Our result for the NUT parameter $n^*=0.257_{-0.025}^{+0.054}$ falls within this range, and has a more strict constraint;
In \cite{modified-Kerr-3}, the authors employed the parameter resonance model to combine the BK spacetime with HFQPO observational data from the microquasar GRS 1915+105, resulting in a constraint on the parameter $q^*$ within the range $[-3.258,0.287]$. This range has a more large error than  our result and includes the constraint on $q^*$ derived in this work, which is $0.152 \pm 0.019$;
In \cite{modified-Kerr-4}, the authors considered the spin parameter $a^*$ and the geometric structure of the Kerr-MOG BH, and received a constraint on the deformation parameter $0 \leq \alpha^* \leq 5.250$ in theory when the spin parameter $a^* \approx 0.4$. This range encompasses the constraint on the deformation parameter $\alpha^*$ obtained in our study, which is $0.795 \pm 0.011$;
In \cite{modified-Kerr-5}, the authors incorporated the PFDM spacetime with the Event Horizon Telescope (EHT) observations of M87*, determining a constraint on the parameter $k^*$ within the $1 \sigma$ confidence region, which spans the range $[0,0.0792]$. This result is in agreement with the constraint we derived for the PFDM parameter $k^*$, which is $0.019_{-0.003}^{+0.002}$.

\begin{figure}[htbp]
    \renewcommand{\thefigure}{5}
    \centering
    \subfloat{\includegraphics[width=5.6cm]{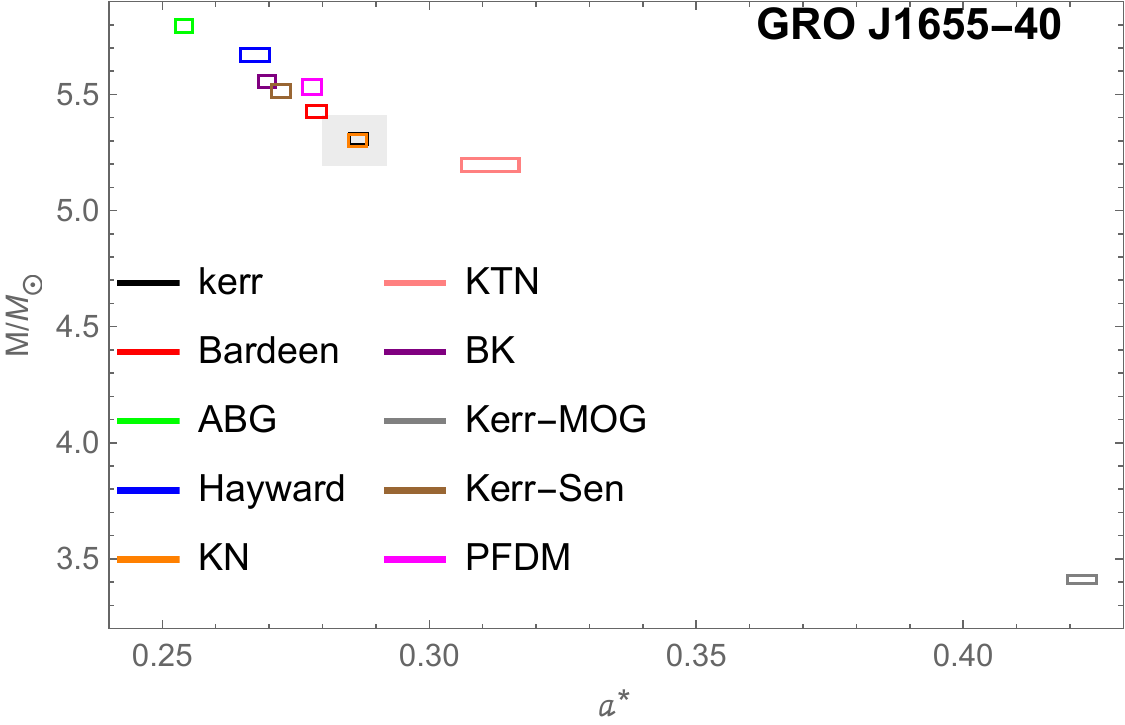}}
    \vspace{2mm}
    \subfloat{\includegraphics[width=5.6cm]{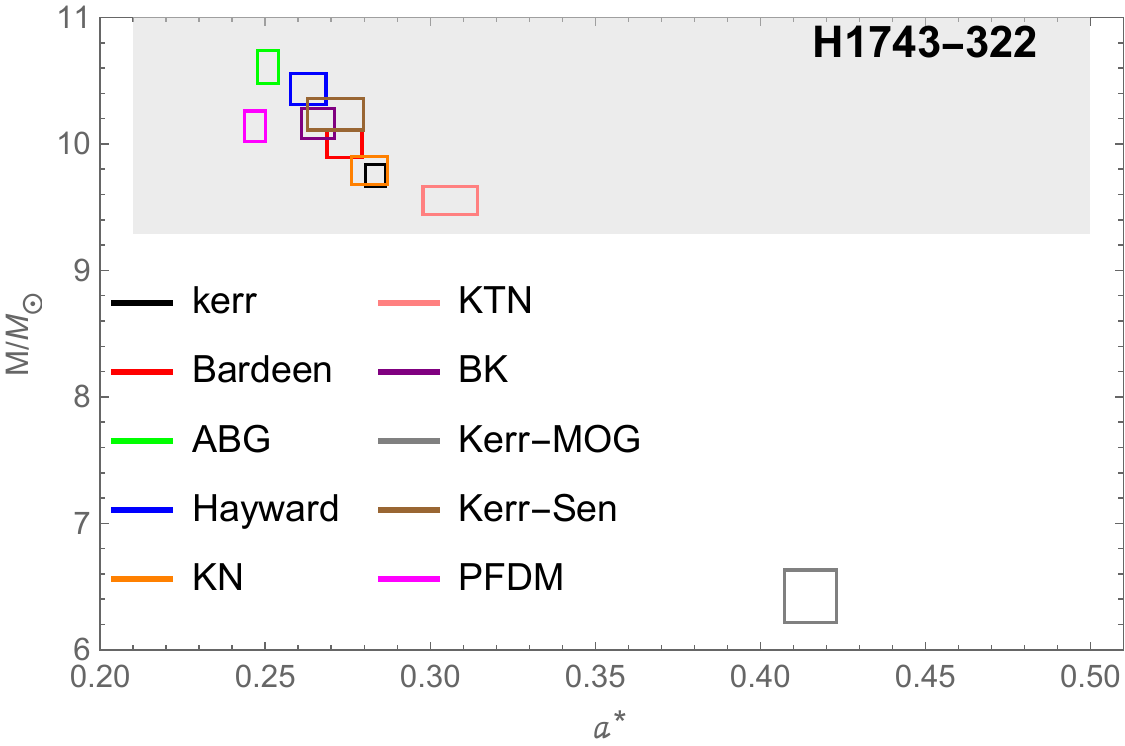}}
    \vspace{2mm}
    \centering
    \subfloat{\includegraphics[width=5.6cm]{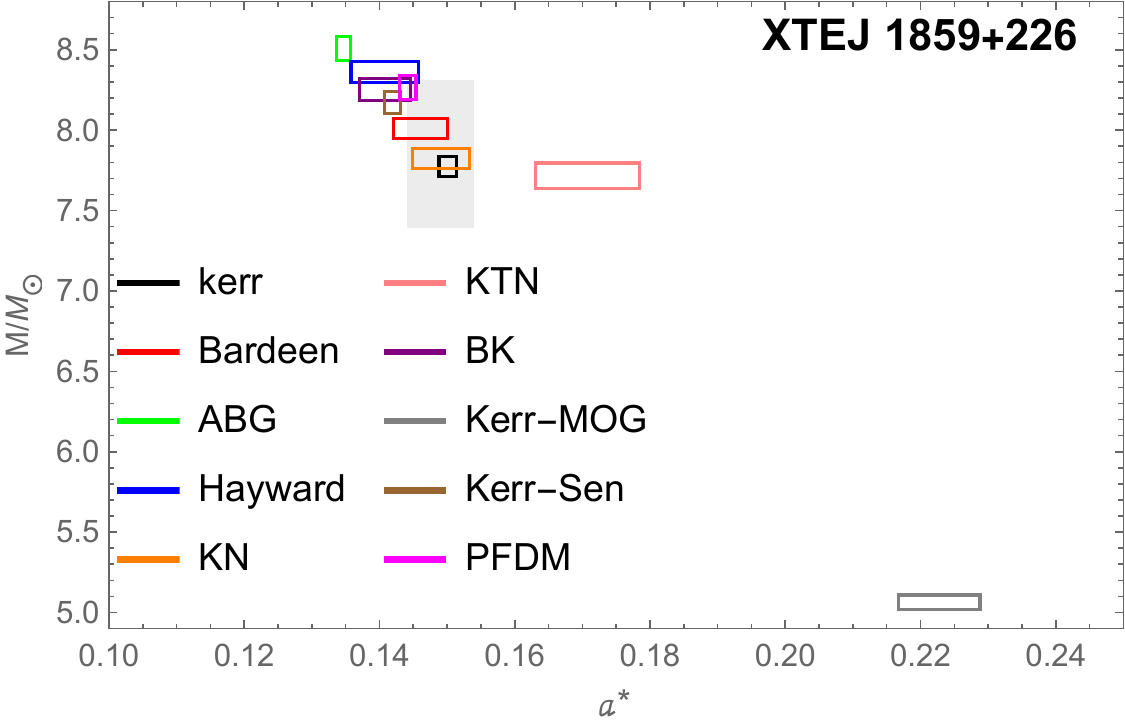}}
    \vspace{2mm}
    \caption{In the Kerr spacetime and its nine single-parameter modified spacetimes, the constraint ranges for the spin $a_p^*$  and mass $M_p^*$  of the three microquasars within the $68 \%$ CL, obtained in this paper, are represented by different colored rectangles. The gray region represents the constraint ranges for the three microquasars with considering the kerr geometry,  reported in the literatures \cite{GRO-1, H1743-1, microquasars-6,XTE-1}.}
    \label{fig:5}
\end{figure}

To provide a clearer comparison for the constraints on the spin parameters $a_p^*$ and masses $M_p^*$ of the three microquasars within the $68 \%$ CL across different spacetime backgrounds, we plotted the constraint results listed in Table \ref{table-2} as a two-dimensional graph in Figure \ref{fig:5}. The horizontal axis represents the spin value $a_p^*$, while the vertical axis represents the mass $M_p^*$.
The left, middle, and right subplots in Figure \ref{fig:5} correspond to the microquasars GRO J1655-40, H1743-322, and XTE J1859+226, respectively. The colored rectangles represent the constraint ranges of $a_p^*$ and $M_p^*$ for various spacetimes considered in this paper.

Additionally, the gray areas in the figure denotes the constraint ranges of $a_p^*$ and $M_p^*$ for these three microquasars in Kerr spacetime using other methods or observational data, as calculated in other references.
 Concretely, (i) for the microquasar GRO J1655-40, the study \cite{GRO-1} used its QPO data and the RP model to derive the spin and mass ranges in Kerr spacetime within the $68.3 \%$ CL, which are $a^*=0.286 \pm 0.006$ and $M^*=5.30 \pm 0.11$, respectively;
(ii) For the microquasar H1743-322, Ref. \cite{H1743-1} performed a kinematic analysis of its radio and X-ray jets, providing a constraint on its spin value in Kerr spacetime within the $68 \%$ CL: $a^*=0.2 \pm 0.3$. Additionally, in \cite{microquasars-6}, the QPO data of H1743-322 were analyzed, yielding constraints on the spin and mass in Kerr spacetime as $0.21 \leq a^* $ and $9.29\lesssim M^*$. Combining the constraints from both references, one has $0.21 \leq a^* \leq 0.5$ and $9.29\lesssim M^*$ for H1743-322;
(iii) For the microquasar XTE J1859+226, Ref. \cite{XTE-1} combined theoretical expressions with its QPO observational data in Kerr spacetime using the RP model, providing constraints on its spin and mass within the $1 \sigma$ range: $a^*=0.149 \pm 0.005$ and $M^*=7.85 \pm 0.46$.

From Figure \ref{fig:5}, we can observe: (i) The constraint results on spin $a_p^*$  and mass $M_p^*$ for the three microquasars (GRO J1655-40, H1743-322, and XTE J1859+226) in this study under Kerr spacetime (black region) are close to the results (orange rectangle) calculated in KN geometry, and both are contained in the constraint ranges reported in the literatures (gray region); (ii) Furthermore, for all the three microquasar cases, Kerr-MOG has the largest value of $a^*$ and the smallest value of $M^*$, while ABG has the largest value of $M^*$ and the smallest $a^*$ (except H1743-322 case).
Specifically, the constraints on the spin and mass values of the three microquasars in Kerr-MOG spacetime show significant deviations from those in the other nine spacetime models.

It is noteworthy that for the mass estimates, the constraint on the microquasar GRO J1655-40 for considered spacetimes in this paper (except for the ABG and Kerr-MOG geometry), are consistent with the optical/near-infrared luminosity measurements \cite{microquasars-2} (where $M^*=5.4 \pm 0.3$).
And  with the exception of the Kerr-MOG spacetime, the constraints from the other nine spacetimes on the microquasar H1743-322 are in agreement with the range reported in \cite{microquasars-6} (where $9.29\lesssim M^*$), and
consistent with the mass range $7.8 \pm 1.9$ gained through the optical dynamical study method for XTE J1859+226 \cite{microquasars-3}.

Different from the consistency between the mass ranges constrained by using the QPO method and those derived from optical/near-infrared luminosity measurements, the spin values for the three microquasars derived using the QPO method across the ten models exhibit substantial discrepancies compared to the spin values obtained through the iron line/continuum-fitting method in Kerr spacetime. For instance, for the microquasar GRO J1655-40, the spin value $a^*$ derived from the continuumfitting method is $0.7 \pm 0.1$ \cite{cf}. In contrast, in all ten models examined in our study, constraints on the spin value for this microquasar, derived using the QPO method within the $68 \%$ CL is below 0.5. This discrepancy has also been found in some other literature, e.g. with possible explanations proposed in \cite{GRO-1, GRO-2, GRO-3, GRO-4}.
As pointed out in reference \cite{GRO-1}, if the RP model is reliable for characterizing the quasi-periodic oscillation behavior of celestial bodies, the inconsistency may stem from the overestimation of the spin parameter by the continuum-fitting method, and may even suggest a violation of the Kerr black hole paradigm. Also, this inconsistency could be attributed to the limitation of the QPO model used, as well as the misunderstanding or insufficient understanding of the physical processes in accretion flows.
In conclusion, this difference remains an area worthy of further investigation and exploration.

Finally, the constraint results for various parameters in ten different rotationally symmetric spacetimes, as presented in this study, indicate that the QPO method could provide stricter constraints on spin values, mass, and modification parameters compared to methods like optical/near-infrared luminosity measurements. Furthermore, differences in the selected model can significantly influence the final results, making the QPO method a powerful tool for testing gravitational theories.
And it is meaningful to assess the status of various spacetimes by combing the observed data with the theoretical models.
\section{$\text{Model camparison for stationary axisymmetric spacetimes}$}

\subsection{{Bayesian analysis}}

In this section, we evaluate the relative superiority of these modified geometries compared to Kerr spacetime using the Bayes factor. Within the framework of Bayesian inference, we focus on the posterior probability $P(D \mid \Theta, \bar{M})$, which represents the posterior probability distribution of the parameters $\Theta$. Here, $\bar{M}$ represents the spacetime model, and $D$ is the data. The posterior probability can be calculated using the following formula \cite{Bys-1, Bys-2}
\begin{equation}
P(\Theta \mid D, \bar{M})=\frac{P(D \mid \Theta, \bar{M}) \pi(\Theta \mid \bar{M})}{P(D \mid \bar{M})}, \label{5.1}
\end{equation}
here, $\pi(\Theta)$ represents the prior distribution of $\Theta$, and $P(D \mid \bar{M})$ is given by
\begin{equation}
P(D \mid \bar{M})=\int P(D \mid \Theta, \bar{M}) \pi(\Theta) d \Theta. \label{5.2}
\end{equation}
And for the model $\bar{M}$, one has
\begin{equation}
P(\bar{M} \mid D) \propto P(D \mid \bar{M}) \pi(\bar{M}) \propto Z_{\bar{M}} \pi_{\bar{M}},\label{5.3}
\end{equation}
here, $Z_{\bar{M}}=P(D \mid \bar{M})$ is the model evidence, and $\pi_{\bar{M}}=\pi(\bar{M})$ is the prior probability of the model. To compare models, we can define the Bayes factor as follows
\begin{equation}
R=\frac{Z_{\bar{M}_1} \pi_{\bar{M}_1}}{Z_{\bar{M}_2} \pi_{\bar{M}_2}}.\label{5.4}
\end{equation}
If $R>1$, it hints that $\bar{M}_1$ is superior to $\bar{M}_2$ in explaining the data, whereas if $R<1$, the opposite is true \cite{Bys-1}.
The strength of evidence when comparing two models, $\bar{M}_1$ and $\bar{M}_2$, is evaluated using an empirical scale known as the ``Jeffreys scale". This scale, as described in \cite{Bys-3,Bys-4,Bys-5}, allows for model preference assessment based on the value of $|\ln R|$. Specifically, if $|\ln R|<1$, the evidence is deemed inconclusive. Weak evidence is indicated by values between $1 \leq|\ln R|<2.5$, while moderate evidence is suggested by $2.5 \leq|\ln R|<5$. Strong evidence corresponds to values in the range $5 \leq|\ln R|<10$, and decisive evidence in favor of one model over the other is indicated by $|\ln R| \geq 10$.
Below, we summarize the values of the Bayes factor $R$ and $|\ln R|$ for the nine types of modified geometries of Kerr relative to Kerr spacetime in Table \ref{table-4}.

\begin{table}[ht]
\setlength{\tabcolsep}{6pt}  
\begin{center}
\begin{tabular}{|c|c|c|c|c|c|c|c|c|c|}
\hline & Bardeen & ABG & Hayward & KN & KTN & BK & Kerr-MOG & Kerr-Sen & PFDM \\
\hline$R$ & 0.971 & 1.057 & 1.029 & 1.114 & 0.971 & 1.029 & 1.071 & 0.929 & 0.914 \\
\hline $|\operatorname{Ln} R|$ & 0.029 & 0.055 & 0.029 & 0.108 & 0.029 & 0.0286 & 0.069 & 0.074 & 0.090 \\
\hline
\end{tabular}
\end{center}
\caption{\label{table-4}
 The values of the Bayes factors and $|\ln R|$ of different types of modified geometries relative to the Kerr spacetime.}
\end{table}

According to the $R$ values in Table \ref{table-4}, we observe that the ABG, Hayward, KN, BK, and Kerr-MOG geometries slightly outperform the Kerr spacetime in explaining the QPO data; While the Bardeen, KTN, Kerr-Sen, and PFDM geometries are somewhat less effective than Kerr spacetime. Furthermore, since all $|\ln R|$ values are less than 1, it indicates that the QPO observational data provide comparable support for both Kerr spacetime and the nine single-parameter modified geometries. This suggests that, on basis of the analysis of Bayes factor and QPO data used in this paper, there is no significant difference in performance between Kerr spacetime and these modified models. It also implies that the Kerr solution predicted by GR remains an effective model for describing the spacetime structure of rotating black holes under the current QPO observational data.

\subsection{{Analysis from Akaike Information Criterion}}

AIC is another statistical method used to evaluate the goodness of a model based on experimental data, and it has been widely applied in various research fields \cite{AIC-1,AIC-2,AIC-3,AIC-4}, such as cosmology \cite{AIC-1,AIC-4}, astrophysics \cite{NS-9}, etc. Its definition is
\begin{equation}
\mathrm{AIC}=-2 \ln \mathcal{L}(\Theta \mid D)_{\max }+2 K.\label{5.5}
\end{equation}
In this formula, $K$ represents the number of free parameters in the model, while $\mathcal{L}_{\max }$ denotes the maximum likelihood obtained using the optimal parameters $\Theta$.
The term $-2 \ln \mathcal{L}(\Theta \mid D)_{\max }$ in Equation \label{5.5} is known as $\chi^2$, which assesses how well the model fits the data, while the term $2 K$ in Equation \label{5.5} reflects the complexity of the model \cite{AIC-5, AIC-6, AIC-7, AIC-8, AIC-9}.

The AIC value of a single model does not have intrinsic meaning; rather, the relative values among different models are what hold practical significance. Thus, the model with the lowest AIC value is considered the best, denoted as $\mathrm{AIC}_{\min }=\min \left\{\mathrm{AIC}_i, i=1, \ldots, N\right\}$, where $\mathrm{AIC}_i$ represents a set of alternative candidate models.
Through the calculation of the model likelihood $\mathcal{L}\left(\bar{M}_i \mid D\right) \propto$ $\exp \left(-\Delta_i / 2\right)$, the relative strength of evidence for each model can be determined, where $\Delta_i=$ $\mathrm{AIC}_i-\mathrm{AIC}_{\min }$ is computed across all candidate models.
The criteria for model selection are as follows: If $0<\Delta_i \leq 2$, model $i$ is almost equally supported by the data as the best model; if $4<\Delta_i \leq 10$, model $i$ is significantly less supported; and if $\Delta_i>10$, model $i$ is essentially irrelevant to the observational data.
Below, we summarize the AIC values of the Kerr spacetime and the nine single-parameter modified forms of the Kerr in Table \ref{table-5}.

\begin{table}[ht]
\setlength{\tabcolsep}{6pt}  
\begin{center}
\begin{tabular}{|l|l|l|l|l|l|l|l|l|l|l|}
\hline & Kerr & Bardeen & ABG & Hayward & KN & KTN & BK & Kerr-MOG & Kerr-Sen & PFDM \\
\hline AIC & 12.976 & 14.572 & 14.428 & 14.442 & 14.429 & 14.424 & 14.497 & 14.480 & 14.699 & 14.712 \\
\hline $\triangle_i$ & 0 & 1.596 & 1.452 & 1.466 & 1.453 & 1.448 & 1.522 & 1.504 & 1.723 & 1.736 \\
\hline
\end{tabular}
\end{center}
\caption{\label{table-5}
 The AIC and $\Delta_i$ values for the Kerr spacetime and its nine single-parameter modified spacetimes.}
\end{table}

From Table \ref{table-5}, it can be observed that the AIC analysis indicates that Kerr spacetime is the best one. Taking the Kerr as the reference model, the $\Delta_i$ values for all other modified spacetime of Kerr are less than 2, suggesting that they also receive almost the same support from the observational QPO data.
Clearly, this paper applies the statistical methods of Bayes factor and AIC model selection to analyze the 10 types of axisymmetric spacetimes explored, and calculations based on QPO observational data show that the support for models differ slightly between the two methods.

\section{$\text{Conclusion}$}

Considering the potential physical issues associated with GR and its prediction of the unique rotating, stationary, axisymmetric Kerr BH, this paper investigates the dynamical effects on particles moving within the Kerr spacetime and its nine single-parameter modified spacetimes. We derive the expressions for the radial, latitudinal, and orbital motions of particles in the background of these ten spacetimes. The calculation results show that as particles approach the black hole, the radial frequency distributions in non-Kerr spacetimes significantly deviate from that in Kerr spacetime; As the particles move farther from the BH, while the radial frequency distributions in the modified spacetimes gradually converge to those in Kerr geometry.

We use the QPO observational data from three microquasars (GRO J1655-40, H1743-322, XTE J1859+226) to put constraints on the Kerr spacetime and the nine single-parameter modified geometries of Kerr within the RP model. Through $\chi^2$ analysis, we provide the limits on the modification parameters in the nine modified spacetimes and the spin and mass parameters of the three microquasars in all ten spacetimes (including Kerr) within the 68\% CL.
The results indicate that, at the $68 \%$ CL free parameters have the relative  small ranges of confidence.
Among them, only the constraint on the modification parameter in the KN spacetime could contain negative, zero, and positive values (zero corresponds to the Kerr), while the confidence ranges of modification parameters for other geometries are all positive, which do not include Kerr at $68 \%$ CL, which indicates that the investigation into modifications of the Kerr spacetime is of significant value.
And according to the constraint results derived from QPO data, the small values of modification parameters are suggested in the KN and PFDM geometries ($Q_2^*=-0.004 \pm 0.078$ and $k^*=$ $0.019_{-0.003}^{+0.024}$), while the larger value of modification parameter is favored in Kerr-MOG geometry ($\alpha^*=0.795 \pm 0.011$).

In the analysis of the spin and mass parameters of the microquasars, we found that, with the exception of the ABG and Kerr-MOG spacetimes, the mass estimates for GRO J1655-40 in the other eight models are consistent with the optical/near-infrared measurements: $5.4 \pm 0.3$.
For H1743-322 and XTE J1859+226, with the exception of the Kerr-MOG spacetime, the mass estimates in the other nine spacetime models are in agreement with the results obtained through dynamical/optical studies: $M^*\gtrsim 9.29$ and $M^*=7.8 \pm 1.9$.
Although the mass estimates are consistent, there is a significant difference between the spin constraints derived from the QPO method and the spin values determined through the iron line method/continuous fitting method in Kerr spacetime.
This issue also can be found in the literature, with some possible explanations proposed in \cite{GRO-1, GRO-2, GRO-3, GRO-4}.
However, this discrepancy still warrants further investigation and exploration. We observed that, for the estimation of free parameter values (modification parameters, mass, and spin parameters), the QPO method provides more stringent constraints compared to optical/near-infrared luminosity measurements, but it depends on the chosen model. Different spacetimes can influence the final physical parameter estimates, thus highlighting the importance of selecting an appropriate geometry to obtain reliable physical constraints.

At the end of the paper, we perform model evaluation for the Kerr spacetime and the nine singleparameter modified spacetimes relative to Kerr using Bayesian analysis and the AIC criterion. The Bayesian analysis indicate that the ABG, Hayward, KN, BK, and Kerr-MOG models show a slight advantage over the Kerr spacetime, while the Bardeen, KTN, Kerr-Sen, and PFDM models have a slightly poorer fit than the Kerr solution in GR. In contrast, the AIC results show that Kerr spacetime remains the optimal model under the current QPO observational data. Clearly, this paper applies the statistical methods of Bayes factor and AIC model selection to analyze the 10 types of axisymmetric spacetimes explored, and calculations based on QPO observational data show that the support for models differ slightly between the two methods.

\textbf{\ Acknowledgments }
 The research work is supported by the National Natural Science Foundation of China (12175095,12075109 and 11865012), and supported by  LiaoNing Revitalization Talents Program (XLYC2007047).

We declare: no new data were created or analysed in this study.

\appendix  
\section{Expressions for the frequencies of particle motion in Kerr spacetime and its nine modified versions}

Below, we present the specific expressions of the orbital frequency $(\omega_{\phi})$, and radial $(\omega_{r})$, latitudinal  $(\omega_{\theta})$ epicyclic frequencies of particles moving on circular orbits in the equatorial plane in the Kerr spacetime and its nine single-parameter modified spacetimes.

(I) In the Kerr spacetime (\ref{2.2}), the concrete formulas for three frequencies: $\omega_{\phi}$, $\omega_{r}$, and $\omega_{\theta}$ are, respectively

\scriptsize
\begin{align*}
\omega_{\phi} =\frac{r^2 \sqrt{M r}-a M r}{r^4-a^2 M r},\tag{A1}
\end{align*}
\begin{align*}
\omega_r^2 =\frac{-4 a^4 M \omega_{\phi}^2+8 a^3 M \omega_{\phi}+a^2 \left(2 M^2 r \omega_{\phi}^2-2 M \left(5 r^2 \omega_{\phi}^2+2\right)+r^3
   \omega_{\phi}^2\right)-4 a M r \omega_{\phi} (M-3 r)+r \left(2 M^2-2 M \left(4 r^3 \omega_{\phi}^2+r\right)+3 r^4 \omega_{\phi}^2\right)}{r^5}
,\tag{A2}
\end{align*}
\begin{align*}
\omega_\theta^2 =\frac{2 a^4 M \omega_{\phi}^2-4 a^3 M \omega_{\phi}+a^2 \left(M \left(4 r^2 \omega_{\phi}^2+2\right)+r^3 \omega_{\phi}^2\right)-4 a M r^2
   \omega_{\phi}+r^5 \omega_{\phi}^2}{r^5}
.\tag{A3}
\end{align*}
\normalsize
In this paper $M$ and $a$ correspond to denote the mass and spin parameter of the BH, respectively.

(II) For a regular  Bardeen BH (\ref{2.3}), the expressions for $\omega_{\phi}$, $\omega_{r}$, and $\omega_{\theta}$ are as follows

\scriptsize
\begin{align*}
\omega_{\phi} =\frac{2 b^2 \left(a M \sqrt{\frac{r^2}{b^2+r^2}}+r^2 \sqrt{\frac{M r \left(r^2-2 b^2\right) \sqrt{\frac{r^2}{b^2+r^2}}}{\left(b^2+r^2\right)^2}}\right)-a M r^2 \sqrt{\frac{r^2}{b^2+r^2}}+r^4 \sqrt{\frac{M r \left(r^2-2 b^2\right)
   \sqrt{\frac{r^2}{b^2+r^2}}}{\left(b^2+r^2\right)^2}}+b^4 \sqrt{\frac{M r \left(r^2-2 b^2\right) \sqrt{\frac{r^2}{b^2+r^2}}}{\left(b^2+r^2\right)^2}}}{-a^2 M r^2 \sqrt{\frac{r^2}{b^2+r^2}}+2 b^2 \left(a^2 M
   \sqrt{\frac{r^2}{b^2+r^2}}+r^3\right)+b^4 r+r^5},\tag{A4}
\end{align*}
\begin{align*}
\omega_r^2 = & \left[ r^3 \left( b^2 + r^2 \right)^5 \right]^{-1} \Big[ a^2 b^{10} r \omega_{\phi}^2 + 5 a^2 b^8 r^3 \omega_{\phi}^2 + 10 a^2 b^6 r^5 \omega_{\phi}^2 - 4 a^2 b^4 M^2 r^5 \omega_{\phi}^2  + 10 a^2 b^4 r^7 \omega_{\phi}^2 -  14 a^2 b^2 M^2 r^7 \omega_{\phi}^2 \nonumber \\
&- \frac{2 M r^{10} \left( \omega_{\phi} \left( 5 a^2 \omega_{\phi} - 6 a + 9 b^2 \omega_{\phi} \right) + 1 \right)}{\sqrt{\frac{r^2}{b^2 + r^2}}}  + M r^8 \left( 2 a \omega_{\phi} \left( 4 a^2 - 9 b^2 \right) - 4 a^2 + \omega_{\phi}^2 \left( -4 a^4 + 9 a^2 b^2 - 12 b^4 \right) + 9 b^2 \right) \nonumber \\
  & + b^4 M r^4 \sqrt{\frac{r^2}{b^2}} \sqrt{\frac{r^2}{b^2} + 1} \left( -6 a \omega_{\phi} \left( 3 a^2 + 7 b^2 \right) + 9 \left( a^2 + b^2 \right) + \omega_{\phi}^2 \left( 9 a^4 + 33 a^2 b^2 - 2 b^4 \right) \right) \nonumber \\
  & + b^8 M (a \omega_{\phi} - 1) \left( \frac{r^2}{b^2} \right)^{3/2} \sqrt{\frac{r^2}{b^2} + 1} \left( 15 a^3 \omega_{\phi} - 15 a^2 + 14 a b^2 \omega_{\phi} + 2 b^2 \right)  + 8 a b^4 M^2 r^5 \omega_{\phi} + 28 a b^2 M^2 r^7 \omega_{\phi} \nonumber \\
  &- 4 a M^2 r^9 \omega_{\phi} + 3 b^{10} r^3 \omega_{\phi}^2  + 15 b^8 r^5 \omega_{\phi}^2 + 30 b^6 r^7 \omega_{\phi}^2 - 4 b^4 M^2 r^5 + 30 b^4 r^9 \omega_{\phi}^2 - 14 b^2 M^2 r^7  - \frac{8 M r^{12} \omega_{\phi}^2}{\sqrt{\frac{r^2}{b^2 + r^2}}}\nonumber \\
  & + 15 b^2 r^{11} \omega_{\phi}^2 + 2 M^2 r^9 + 3 r^{13} \omega_{\phi}^2\Big],\tag{A5}
\end{align*}
\begin{align*}
\omega_{\theta}^{2}=\frac{2 a^4 M \omega_{\phi}^2 \sqrt{\frac{r^2}{b^2+r^2}}+4 a^2 M r^2 \omega_{\phi}^2 \sqrt{\frac{r^2}{b^2+r^2}}-2 a M \sqrt{\frac{r^2}{b^2+r^2}} \left(2 a^2 \omega_{\phi}-a+2 r^2 \omega_{\phi}\right)+r \omega_{\phi}^2 \left(a^2+r^2\right) \left(b^2+r^2\right)}{r^3
   \left(b^2+r^2\right)}.\tag{A6}
\end{align*}
\normalsize
Here, the modification parameter $b$ represents the magnetic monopole charge.

\normalsize
(III) In the framework of general relativity coupled with nonlinear electrodynamics, one can get a singularity-free BH solution (\ref{2.4}), called the ABG. With $Q_{1}$ denoting the charge, in this spacetime  three frequencies can be derived as
\scriptsize
\begin{align*}
\omega_{\phi} =\frac{a M \left(-r^5+Q_1 ^2 r^3+2 Q_1 ^4 r\right)+a Q_1 ^2 r \left(r^2-Q_1
   ^2\right) \sqrt{Q_1 ^2+r^2}+\left(Q_1 ^2+r^2\right)^{7/2} \sqrt{\frac{r^2
   \left(Q_1 ^4-2 Q_1 ^2 M \sqrt{Q_1 ^2+r^2}+r^2 \left(M \sqrt{Q_1 ^2+r^2}-Q_1
   ^2\right)\right)}{\left(Q_1 ^2+r^2\right)^3}}}{r \left(a^2 \left(M \left(2 Q_1
   ^4-r^4+Q_1 ^2 r^2\right)+Q_1 ^2 \left(r^2-Q_1 ^2\right) \sqrt{Q_1
   ^2+r^2}\right)+\left(Q_1 ^2+r^2\right)^{7/2}\right)},\tag{A7}
\end{align*}
\begin{align*}
\omega_r^2 = & \Big[ r^2 \left( r^2 + Q_1^2 \right)^{13/2} \Big]^{-1} \Big[
  - \omega_{\phi}^2 \left( r^2 + Q_1^2 \right)^2 \left( \sqrt{r^2 + Q_1^2} \left( -5 r^4 + 8 Q_1^2 r^2 + Q_1^4 \right) Q_1^2 + M \left( r^2 + Q_1^2 \right) \left( 4 r^4 - 13 Q_1^2 r^2 - 2 Q_1^4 \right) \right) \\
& + a^4 + 2 \omega_{\phi} \left( M \left( 4 r^4 - 13 Q_1^2 r^2 - 2 Q_1^4 \right) \left( r^2 + Q_1^2 \right)^3 + Q_1^2 \left( -5 r^4 + 8 Q_1^2 r^2 + Q_1^4 \right) \left( r^2 + Q_1^2 \right)^{5/2} \right) a^3 \\
& + \left( 2 M^2 \omega_{\phi}^2 \left( r^2 + Q_1^2 \right)^{3/2} \left( r^4 - 7 Q_1^2 r^2 - 2 Q_1^4 \right) r^4 - M \left( r^2 + Q_1^2 \right) \left( 2 \left( 5 r^2 \omega_{\phi}^2 + 2 \right) r^8 + 5 \left( r^2 \omega_{\phi}^2 - 1 \right) Q_1^2 r^6 \right. \right. \\
& \left. \left. - 3 \left( 19 r^2 \omega_{\phi}^2 + 8 \right) Q_1^4 r^4 - 17 \left( 3 r^2 \omega_{\phi}^2 + 1 \right) Q_1^6 r^2 - 2 \left( 7 r^2 \omega_{\phi}^2 + 1 \right) Q_1^8 \right) + \sqrt{r^2 + Q_1^2} \left( \omega_{\phi}^2 r^{12} + \left( 17 r^2 \omega_{\phi}^2 + 5 \right) Q_1^2 r^8 \right. \right. \\
& \left. \left. + 2 \left( 15 r^2 \omega_{\phi}^2 + 1 \right) Q_1^4 r^6 + 4 \left( r^2 \omega_{\phi}^2 - 3 \right) Q_1^6 r^4 - 2 \left( 4 r^2 \omega_{\phi}^2 + 5 \right) Q_1^8 r^2 + \omega_{\phi}^2 Q_1^{12} - \left( r^2 \omega_{\phi}^2 + 1 \right) Q_1^{10} \right) \right) a^2 \\
& + 2 r^2 \omega_{\phi} \left( 2 M^2 \sqrt{r^2 + Q_1^2} \left( - r^6 + 6 Q_1^2 r^4 + 9 Q_1^4 r^2 + 2 Q_1^6 \right) r^2 + M \left( r^2 + Q_1^2 \right) \left( 6 r^8 + Q_1^2 r^6 - 45 Q_1^4 r^4 - 31 Q_1^6 r^2 - 6 Q_1^8 \right) \right. \\
& \left. + Q_1^2 \sqrt{r^2 + Q_1^2} \left( - 7 r^8 - 7 Q_1^2 r^6 + 16 Q_1^4 r^4 + 15 Q_1^6 r^2 + 3 Q_1^8 \right) \right) \\
& + a + r^2 \left( 2 M^2 \left( r^2 + Q_1^2 \right)^{3/2} \left( r^4 - 7 Q_1^2 r^2 - 2 Q_1^4 \right) r^2 + \sqrt{r^2 + Q_1^2} \left( 3 \omega_{\phi}^2 r^{12} + \left( 23 r^2 \omega_{\phi}^2 + 3 \right) Q_1^2 r^8 \right. \right. \\
& \left. \left. + \left( 61 r^2 \omega_{\phi}^2 - 1 \right) Q_1^4 r^6 + 2 \left( 39 r^2 \omega_{\phi}^2 - 8 \right) Q_1^6 r^4 + \left( 53 r^2 \omega_{\phi}^2 - 7 \right) Q_1^8 r^2 + 3 \omega_{\phi}^2 Q_1^{12} + \left( 19 r^2 \omega_{\phi}^2 + 1 \right) Q_1^{10} \right) \right. \\
& \left. - M \left( r^2 + Q_1^2 \right) \left( 8 \omega_{\phi}^2 r^{10} + \left( 26 \omega_{\phi}^2 Q_1^2 + 2 \right) r^8 + \left( 30 \omega_{\phi}^2 Q_1^4 - 3 Q_1^2 \right) r^6 + Q_1^4 \left( 14 \omega_{\phi}^2 Q_1^2 - 33 \right) r^4 \right. \right. \\
& \left. \left. + Q_1^6 \left( 2 \omega_{\phi}^2 Q_1^2 - 11 \right) r^2 + 2 Q_1^8 \right) \right) \Big], \tag{A8}
\end{align*}
\begin{align*}
 \omega_\theta^2=&\Big[r^2 \left(Q_1 ^2+r^2\right)^{5/2}\Big]^{-1}\Big[a^4 \omega_{\phi}^2 \left(2 M \left(Q_1 ^2+r^2\right)-Q_1 ^2 \sqrt{Q_1
   ^2+r^2}\right)+2 a^3 \omega_{\phi} \left(Q_1 ^2 \sqrt{Q_1 ^2+r^2}-2 M \left(Q_1
   ^2+r^2\right)\right)\\
   &+a^2 \left(2 M \left(Q_1 ^2+r^2\right) \left(2 r^2
   \omega_{\phi}^2+1\right)+\sqrt{Q_1 ^2+r^2} \left(-Q_1 ^2+r^4 \omega_{\phi}^2+Q_1 ^4
   \omega_{\phi}^2\right)\right)\\
   &+2 a r^2 \omega_{\phi} \left(Q_1 ^2 \sqrt{Q_1 ^2+r^2}-2 M
   \left(Q_1 ^2+r^2\right)\right)+r^2 \omega_{\phi}^2 \left(Q_1
   ^2+r^2\right)^{5/2}\Big].\tag{A9}
\end{align*}
\normalsize

(IV) For the BH solution described by the  Hayward metric (\ref{2.5}), a model parameter $l$ is introduced to quantify the deviation of the Hayward BH from the Kerr BH. In the Hayward spacetime, the expressions of $\omega_{\phi}$, $\omega_{r}$, and $\omega_{\theta}$ are expressed as

\scriptsize
\begin{align*}
\omega_{\phi} =\frac{\left(l^3+r^3\right)^2 \sqrt{\frac{M r^2 \left(r^3-2 l^3\right)}{\left(l^3+r^3\right)^2}}-a M r \left(r^3-2 l^3\right)}{r \left(2 l^3 \left(a^2 M+r^3\right)-a^2 M r^3+l^6+r^6\right)},\tag{A10}
\end{align*}
 \begin{align*}
 \omega_{r}^{2} =&\Big[r^3 \left(l^2+r^2\right)^5\Big]^{-1}\Big[a^2 l^{10} r \omega_{\phi}^2+5 a^2 l^8 r^3 \omega_{\phi}^2+10 a^2 l^6 r^5 \omega_{\phi}^2-4 a^2 l^4 M^2 r^5 \omega_{\phi}^2+10 a^2 l^4 r^7 \omega_{\phi}^2-14 a^2 l^2 M^2 r^7 \omega_{\phi}^2\\
 &-\frac{2 M r^{10} \left(\omega_{\phi} \left(5 a^2 \omega_{\phi}-6 a+9 l^2
   \omega_{\phi}\right)+1\right)}{\sqrt{\frac{r^2}{l^2+r^2}}}+5 a^2 l^2 r^9 \omega_{\phi}^2+2 a^2 l^8 M (a \omega_{\phi}-1)^2 \sqrt{\frac{r^2}{l^2}} \sqrt{\frac{r^2}{l^2}+1}+2 a^2 M^2 r^9 \omega_{\phi}^2\\
   &+a^2 r^{11} \omega_{\phi}^2+\frac{M r^8 \left(2 a \omega_{\phi}
   \left(4 a^2-9 l^2\right)-4 a^2+\omega_{\phi}^2 \left(-4 a^4+9 a^2 l^2-12 l^4\right)+9 l^2\right)}{\sqrt{\frac{r^2}{l^2+r^2}}}\\
   &+l^4 M r^4 \sqrt{\frac{r^2}{l^2}} \sqrt{\frac{r^2}{l^2}+1} \left(-6 a \omega_{\phi} \left(3 a^2+7 l^2\right)+9
   \left(a^2+l^2\right)+\omega_{\phi}^2 \left(9 a^4+33 a^2 l^2-2 l^4\right)\right)\\
   &+l^8 M (a \omega_{\phi}-1) \left(\frac{r^2}{l^2}\right)^{3/2} \sqrt{\frac{r^2}{l^2}+1} \left(15 a^3 \omega_{\phi}-15 a^2+14 a l^2 \omega_{\phi}+2 l^2\right)+8 a l^4 M^2 r^5\omega_{\phi}+28 a l^2 M^2 r^7 \omega_{\phi}\\
   &-4 a M^2 r^9 \omega_{\phi}+3 l^{10} r^3 \omega_{\phi}^2+15 l^8 r^5 \omega_{\phi}^2+30 l^6 r^7 \omega_{\phi}^2-4 l^4 M^2 r^5+30 l^4 r^9 \omega_{\phi}^2-14 l^2 M^2 r^7-\frac{8 M r^{12} \omega_{\phi}^2}{\sqrt{\frac{r^2}{l^2+r^2}}}\\
   &+15l^2 r^{11} \omega_{\phi}^2+2 M^2 r^9+3 r^{13} \omega_{\phi}^2\Big],\tag{A11}
 \end{align*}
 \begin{align*}
 \omega_\theta^{2}=\frac{2 a^4 M \omega_{\phi}^2-4 a^3 M \omega_{\phi}+a^2 \left(\omega_{\phi}^2 \left(l^3+r^3\right)+M \left(4 r^2 \omega_{\phi}^2+2\right)\right)-4 a M r^2 \omega_{\phi}+r^2 \omega_{\phi}^2 \left(l^3+r^3\right)}{r^2 \left(l^3+r^3\right)}.\tag{A12}
  \end{align*}
\normalsize

(V) For the KN BH (\ref{2.6}), a solution to the Einstein-Maxwell equations, with the charge parameter $Q_{2}$ the expressions for $\omega_{\phi}$, $\omega_{r}$, and $\omega_{\theta}$ can be provided by
\scriptsize
\begin{align*}
\omega_{\phi} = \frac{a \left(Q_2^2-M r\right)+r^3 \sqrt{\frac{M r-Q_2^2}{r^2}}}{a^2 \left(Q_2^2-M r\right)+r^4},\tag{A13}
\end{align*}
\begin{align*}
\omega_{r}^{2} =&r^{-6}\Big[a^4 \omega_{\phi}^2 \left(5 Q_2^2-4 M r\right)+a^3 \left(8 M r \omega_{\phi}-10 Q_2^2 \omega_{\phi}\right)+a^2 \left(r \left(2 M^2 r
   \omega_{\phi}^2-2 M \left(5 r^2 \omega_{\phi}^2+2\right)+r^3 \omega_{\phi}^2\right)\right.\\
   &+\left.Q_2^2 \left(-4 M r \omega_{\phi}^2+11 r^2\omega_{\phi}^2+5\right)+Q_2^4 \omega_{\phi}^2\right)-2 a \omega_{\phi} \left(Q_2^2 r (7 r-4 M)+2 M r^2 (M-3 r)+Q_2^4\right)\\
   &+r^2 \left(2 M^2-2 M
   \left(4 r^3 \omega_{\phi}^2+r\right)+3 r^4 \omega_{\phi}^2\right)+Q_2^2 r \left(-4 M+5 r^3 \omega_{\phi}^2+3 r\right)+Q_2^4\Big],\tag{A14}
\end{align*}
\begin{align*}
\omega_\theta^{2} =&r^{-6}\Big[-a^4 \omega_{\phi}^2 \left(Q_2^2-2 M r\right)+2 a^3 \omega_{\phi} \left(Q_2^2-2 M r\right)+a^2 \left(4 M r^3 \omega_{\phi}^2+2 M r-Q_2^2 \left(2
   r^2 \omega_{\phi}^2+1\right)+r^4 \omega_{\phi}^2\right)\\
   &+2 a r^2 \omega_{\phi} \left(Q_2^2-2 M r\right)+r^6 \omega_{\phi}^2\Big].\tag{A15}
\end{align*}
\normalsize

(VI) In the KTN solution (\ref{2.7}),  we have the forms for $\omega_{\phi}$, $\omega_{r}$, and $\omega_{\theta}$

\scriptsize
\begin{align*}
\omega_{\phi} =\frac{a M \left(n^2-r^2\right)-2 a n^2 r+\left(n^2+r^2\right)^2 \sqrt{\frac{r \left(M \left(r^2-n^2\right)+2 n^2 r\right)}{\left(n^2+r^2\right)^2}}}{a^2 \left(M \left(n^2-r^2\right)-2
   n^2 r\right)+r \left(n^2+r^2\right)^2},\tag{A16}
\end{align*}
\begin{align*}
\omega_{r}^{2} = &\left(n^2+r^2\right)^{-4}\Big[2 a^4 \omega_{\phi}^2 \left(n^2 r (4 M-5 r)-2 M r^3+n^4\right)-4 a^3 \omega_{\phi} \left(n^2 r (4 M-5 r)-2 M r^3+n^4\right)\\
&+a^2 \left(n^4 \left(-2 M^2 \omega_{\phi}^2+6 M r \omega_{\phi}^2-17 r^2
   \omega_{\phi}^2+2\right)+n^2 r \left(-8 M^2 r \omega_{\phi}^2+12 M r^2 \omega_{\phi}^2+8 M-21 r^3 \omega_{\phi}^2-10 r\right)\right.\\
   &\left.+r^3 \left(2 M^2 r \omega_{\phi}^2-2 M \left(5 r^2 \omega_{\phi}^2+2\right)+r^3 \omega_{\phi}^2\right)-3 n^6 \omega_{\phi}^2\right)+4
   a \omega_{\phi} \left(M^2 \left(n^4+4 n^2 r^2-r^4\right)-M r \left(n^4+6 n^2 r^2-3 r^4\right)\right.\\
   &\left.+n^6+4 n^4 r^2+7 n^2 r^4\right)-2 M^2 \left(n^4+4 n^2 r^2-r^4\right)-2 M r
   \left(-2 n^6 \omega_{\phi}^2+n^4+6 n^2 r^2 \left(r^2 \omega_{\phi}^2-1\right)+4 r^6 \omega_{\phi}^2+r^4\right)\\
   &+n^8 \omega_{\phi}^2-6 n^6 r^2 \omega_{\phi}^2-2 n^6-12 n^4 r^4 \omega_{\phi}^2-2 n^2 r^6
   \omega_{\phi}^2-6 n^2 r^4+3 r^8 \omega_{\phi}^2\Big],\tag{A17}
\end{align*}
\begin{align*}
\omega_\theta^{2} = &-\frac{3 a n \omega_{\phi}^2 \left(2 a n \left(\left(a^2+n^2+r^2\right)^2-a^2 \left(a^2+r (r-2 M)-n^2\right)\right)-4 a n \left(n^2+r^2\right) \left(a^2+r (r-2
   M)-n^2\right)\right)}{\left(n^2+r^2\right)^4}\\
   &-\left(2 \left(n^2+r^2\right)\right)^{-3}\Big[\omega_{\phi}^2 \left(\left(n^2+r^2\right) \left(4 a^2 \left(a^2+r (r-2 M)-n^2\right)-8 n^2 \left(a^2+r (r-2 M)-n^2\right)-2
   \left(a^2+n^2+r^2\right)^2\right)\right.\\
   &\left.-2 a^2 \left(\left(a^2+n^2+r^2\right)^2-a^2 \left(a^2+r (r-2 M)-n^2\right)\right)\right)\Big]-\frac{4 a \omega_{\phi} \left(a^2+n^2+r^2\right) \left(M
   r+n^2\right)}{\left(n^2+r^2\right)^3}\\
   &+\frac{6 a n^2 \omega_{\phi} \left(a^2+n^2+r^2\right) \left(4 M r+2 n^2-2 r^2\right)}{\left(n^2+r^2\right)^4}+\frac{2 a^2 \left(M
   r+n^2\right)}{\left(n^2+r^2\right)^3}-\frac{3 a^2 n^2 \left(4 M r+2 n^2-2 r^2\right)}{\left(n^2+r^2\right)^4}.\tag{A18}
\end{align*}
\normalsize
The  gravitational magnetic monopole  parameter $n$  denotes a unique dual mass effect.

(VII) In the BK spacetime (\ref{2.8}), we have the forms for $\omega_{\phi}$, $\omega_{r}$, and $\omega_{\theta}$,

\scriptsize
\begin{align*}
\omega_{\phi} =\frac{a (q-M r)+r^3 \sqrt{\frac{M r-q}{r^2}}}{a^2 (q-M r)+r^4},\tag{A19}
\end{align*}
\begin{align*}
\omega_{r}^{2} =&r^{-6}\Big[a^4 \omega_{\phi}^2 (5 q-4 M r)+a^3 (8 M r \omega_{\phi}-10 q \omega_{\phi})+a^2 \left(q^2 \omega_{\phi}^2+q \left(-4 M r \omega_{\phi}^2+11 r^2
   \omega_{\phi}^2+5\right)\right.\\
   &\left.+r \left(2 M^2 r \omega_{\phi}^2-2 M \left(5 r^2 \omega_{\phi}^2+2\right)+r^3 \omega_{\phi}^2\right)\right)-2 a \omega_{\phi}
   \left(q^2+q r (7 r-4 M)+2 M r^2 (M-3 r)\right)\\
   &+q^2+q r \left(-4 M+5 r^3 \omega_{\phi}^2+3 r\right)+r^2 \left(2 M^2-2 M \left(4 r^3
   \omega_{\phi}^2+r\right)+3 r^4 \omega_{\phi}^2\right)\Big],\tag{A20}
\end{align*}
\begin{align*}
\omega_\theta^{2} =&r^{-6}\Big[-a^4 \omega_{\phi}^2 (q-2 M r)+2 a^3 \omega_{\phi} (q-2 M r)+a^2 \left(-q \left(2 r^2 \omega_{\phi}^2+1\right)+4 M r^3 \omega_{\phi}^2+2 M r+r^4
   \omega_{\phi}^2\right)+2 a r^2 \omega_{\phi} (q-2 M r)+r^6 \omega_{\phi}^2\Big].\tag{A21}
\end{align*}
\normalsize
Here the "tidal charge" parameter $q$ encodes the effect of the extra dimensions on the gravitational field.

(VIII) In the Kerr-MOG spacetime (\ref{2.9}), the modification parameter is defined as $\alpha=\frac{G_{K}}{G}-1$ with $G_{K}$ denoting the enhanced gravitational constant.
And the frequencies $\omega_{\phi}$, $\omega_{r}$, and $\omega_{\theta}$ have the forms, respectively

\scriptsize
\begin{align*}
\omega_{\phi} =\frac{a (\alpha +1) M (\alpha  M-r)+r^3 \sqrt{\frac{(\alpha +1) M (r-\alpha
   M)}{r^2}}}{a^2 (\alpha +1) M (\alpha  M-r)+r^4},
\tag{A22}
\end{align*}
\begin{align*}
 \omega_{r}^{2} =&r^{-6}\Big[a^4 (\alpha +1) M \omega_{\phi}^2 (5 \alpha  M-4 r)-2 a^3 (\alpha +1) M \omega_{\phi}
   (5 \alpha  M-4 r)+a^2 \left(\alpha ^2 (\alpha +1)^2 M^4 \omega_{\phi}^2-4 \alpha\right.\\
   &\left.(\alpha +1)^2 M^3 r \omega_{\phi}^2+(\alpha +1) M^2 \left(5 \alpha +(13 \alpha +2)
   r^2 \omega_{\phi}^2\right)-2 (\alpha +1) M r \left(5 r^2 \omega_{\phi}^2+2\right)+r^4
   \omega_{\phi}^2\right)\\
   &-2 a (\alpha +1) M \omega_{\phi} \left(\alpha ^2 (\alpha +1) M^3-4
   \alpha  (\alpha +1) M^2 r+(9 \alpha +2) M r^2-6 r^3\right)+\alpha ^2 (\alpha
   +1)^2 M^4\\
   &-4 \alpha  (\alpha +1)^2 M^3 r+(\alpha +1) M^2 r^2 \left(5 \alpha
   \left(r^2 \omega_{\phi}^2+1\right)+2\right)-2 (\alpha +1) M r^3 \left(4 r^2
   \omega_{\phi}^2+1\right)+3 r^6 \omega_{\phi}^2\Big],
\tag{A23}
\end{align*}
\begin{align*}
 \omega_\theta^{2} =&r^{-6}\Big[-a^4 \omega_{\phi}^2 (\alpha-2 M r)+2 a^3 \omega_{\phi} (\alpha-2 M r)+a^2 \left(-\alpha \left(2 r^2 \omega_{\phi}^2+1\right)+4 M r^3 \omega_{\phi}^2+2 M r+r^4
   \omega_{\phi}^2\right)+2 a r^2 \omega_{\phi} (\alpha-2 M r)+r^6 \omega_{\phi}^2\Big].
\tag{A24}
\end{align*}
\normalsize

(IX) In the Kerr-Sen spacetime (\ref{2.10}), the explicit expressions for $\omega_{\phi}$, $\omega_{r}$, and $\omega_{\theta}$ can be written as

\scriptsize
\begin{align*}
\omega_{\phi} =-\frac{M \left(\sqrt{2} \left(M r+Q_3^2\right)^2 \sqrt{\frac{M^2 \left(2 M
   r+Q_3^2\right)}{\left(M r+Q_3^2\right)^2}}-2 a M^3\right)}{2 a^2 M^4-\left(M
   r+Q_3^2\right)^2 \left(2 M r+Q_3^2\right)},\tag{A25}
\end{align*}
\begin{align*}
 \omega_{r}^{2} =&\left(2 M r^2 \left(M r+Q_3^2\right)^4\right)^{-1}\Big[-2 a^4 M^5 \omega_{\phi}^2 \left(4 M r+Q_3^2\right)+4 a^3 M^5 \omega_{\phi} \left(4 M
   r+Q_3^2\right)+a^2 M \left(4 M^6 r^2 \omega_{\phi}^2-4 M^5 r \left(5 r^2
   \omega_{\phi}^2+2\right)\right.\\
   &\left.-2 M^4 \left(14 Q_3^2 r^2 \omega_{\phi}^2+Q_3^2-r^4
   \omega_{\phi}^2\right)+2 M^3 Q_3^2 r \omega_{\phi}^2 \left(3 r^2-4
   Q_3^2\right)+7 M^2 Q_3^4 r^2 \omega_{\phi}^2+4 M Q_3^6 r \omega_{\phi}^2+Q_3^8
   \omega_{\phi}^2\right)\\
   &+8 a M^4 r \omega_{\phi} \left(-M^2 r (M-3 r)+4 M Q_3^2
   r+Q_3^4\right)+2 r \left(2 M^7 r-2 M^6 \left(4 r^4 \omega_{\phi}^2+r^2\right)+M^5
   \left(3 r^5 \omega_{\phi}^2-Q_3^2 r \left(23 r^2 \omega_{\phi}^2+2\right)\right)\right.\\
   &\left.+12 M^4 Q_3^2r^2 \omega_{\phi}^2 \left(r^2-2 Q_3^2\right)+M^3 Q_3^4 r \omega_{\phi}^2 \left(19 r^2-11
   Q_3^2\right)+M^2 Q_3^6 \omega_{\phi}^2 \left(15 r^2-2 Q_3^2\right)+6 M Q_3^8 r
   \omega_{\phi}^2+Q_3^{10} \omega_{\phi}^2\right)\Big]
,\tag{A26}
\end{align*}
\begin{align*}
\omega_\theta^{2} =&\left(r^2 \left(M r+Q_3^2\right)^3\right)^{-1}\Big[2 a^4 M^4 \omega_{\phi}^2-4 a^3 M^4 \omega_{\phi}+a^2 M \left(M^3 \left(4 r^2
   \omega_{\phi}^2+2\right)+M^2 r \omega_{\phi}^2 \left(4 Q_3^2+r^2\right)+2 M Q_3^2 r^2
   \omega_{\phi}^2+Q_3^4 r \omega_{\phi}^2\right)\\
   &-4 a M^3 r \omega_{\phi} \left(M r+Q_3^2\right)+r^2
   \omega_{\phi}^2 \left(M r+Q_3^2\right)^3\Big].\tag{A27}
\end{align*}
\normalsize
The modification parameter is written as $Q_3=Q^2 / M$, where $Q$ denotes the electric charge.

(X) In the PFDM spacetime (\ref{2.11}), the expressions for $\omega_{\phi}$, $\omega_{r}$, and $\omega_{\theta}$ are

\scriptsize
\begin{align*}
\omega_{\phi} =\frac{-a k \log \left(\frac{r}{| k| }\right)+a (2 M+k)-\sqrt{2} r^2 \sqrt{\frac{-k \log \left(\frac{r}{| k| }\right)+2 M+k}{r}}}{-a^2
   k \log \left(\frac{r}{| k| }\right)+a^2 (2 M+k)-2 r^3},\tag{A28}
\end{align*}
\begin{align*}
 \omega_{r}^{2} =&(2 r^5)^{-1}\Big[-\left(a^4 \omega_{\phi}^2 (8 M+5 k)\right)+2 a^3 \omega_{\phi} (8 M+5 k)-a^2 \left(-4 M^2 r \omega_{\phi}^2+M \left(-2 k r \omega_{\phi}^2+20
   r^2 \omega_{\phi}^2+8\right)+k^2 r \omega_{\phi}^2\right.\\
   &\left.+11 k r^2 \omega_{\phi}^2+5 k-2 r^3 \omega_{\phi}^2\right)+k \log \left(\frac{r}{| k| }\right)
   \left(4 a^4 \omega_{\phi}^2-8 a^3 \omega_{\phi}+a^2 \left(-4 M r \omega_{\phi}^2-k r \omega_{\phi}^2+10 r^2 \omega_{\phi}^2+4\right)\right.\\
   &\left.+2 a r \omega_{\phi} (4
   M+k-6 r)+r \left(-4 M-k+8 r^3 \omega_{\phi}^2+2 r\right)\right)+k^2 r (a \omega_{\phi}-1)^2 \log ^2\left(\frac{r}{| k| }\right)\\
   &+2 a r
   \omega_{\phi} \left(-4 M^2-2 M (k-6 r)+k (k+7 r)\right)+r \left(4 M^2+2 M \left(k-8 r^3 \omega_{\phi}^2-2 r\right)-k^2-2 k r^3
   \omega_{\phi}^2-3 k r+6 r^4 \omega_{\phi}^2\right)\Big],\tag{A29}
\end{align*}
\begin{align*}
\omega_\theta^{2} =r^{-5}\Big[2 a^4 M \omega_{\phi}^2-4 a^3 M \omega_{\phi}-a k (a \omega_{\phi}-1) \left(a^2 \omega_{\phi}-a+2 r^2 \omega_{\phi}\right) \log \left(\frac{r}{| k|
   }\right)+a^2 \left(4 M r^2 \omega_{\phi}^2+2 M+r^3 \omega_{\phi}^2\right)-4 a M r^2 \omega_{\phi}+r^5 \omega_{\phi}^2\Big].\tag{A30}
\end{align*}
\normalsize
The modification parameter $k$ characterizes the intensity of dark matter.

\end{document}